\ifx\texorpdfstring\undefined\newcommand\texorpdfstring[2]{#1}\fi
\documentclass[aps,prd,nofootinbib,floatfix,preprintnumbers,tightenlines,11pt,longbibliography,superscriptaddress,altaffilletter
]{revtex4-1}

\usepackage{comment}

\usepackage{amsfonts}
\usepackage{amsmath}
\usepackage{amssymb}
\usepackage{bbm}
\usepackage{bm}
\usepackage{graphicx}
\usepackage{cancel}
\usepackage[pdftex]{xcolor}
\usepackage{pagecolor}
\usepackage{afterpage}
\usepackage[sort&compress]{natbib}
\usepackage[colorlinks=true,linkcolor=blue,filecolor=blue,urlcolor=blue,citecolor=blue,pdftex,plainpages=false]{hyperref}
\usepackage{appendix}

\usepackage{xcolor}
\renewcommand*\thesection{\arabic{section}}
\usepackage[explicit]{titlesec}
\definecolor{Blue1}{HTML}{0098FF}

\titleformat{\section}[hang]{\Large\bfseries\sffamily}%
{\rlap{\color{Blue1}\rule[-9pt]{\textwidth}{1.2pt}}\colorbox{Blue1}{%
           \raisebox{0pt}[13pt][6pt]{ \makebox[60pt]{
                \fontfamily{phv}\selectfont\color{white}{\thesection}}
            }}}%
{15pt}%
{ \color{Blue1}#1
}
\titlespacing*{\section}{0pt}{3mm}{5mm}

\mathchardef\mhyphen="2D

\begin{document}
	\count\footins = 1000


\title{
Quantum Simulation for High Energy Physics}

\author{Christian~W.~Bauer}
\email{Editor and corresponding author: cwbauer@lbl.gov}
\affiliation{
	Physics Division, Lawrence Berkeley National Laboratory, Berkeley, CA 94720, USA}
	
\author{Zohreh~Davoudi}\email{Editor and corresponding author: davoudi@umd.edu}
\affiliation{
 	Department of Physics and Maryland Center for Fundamental Physics, 
University of Maryland, College Park, MD 20742, USA}

\author{A.~Baha~Balantekin}
\affiliation{
 	Department of Physics, University of Wisconsin, Madison, WI 53706, USA}
\author{Tanmoy~Bhattacharya}
\affiliation{
    T-2, Los Alamos National Laboratory, Los Alamos, NM 87545, USA}
    
\author{Marcela~Carena}
\affiliation{Fermi National Accelerator Laboratory, Batavia,  IL 60510, USA}
\affiliation{Enrico Fermi Institute, University of Chicago, Chicago, IL 60637, USA}
\affiliation{Kavli Institute for Cosmological Physics, University of Chicago, Chicago, IL 60637, USA}
\affiliation{Department of Physics, University of Chicago, Chicago, IL 60637, USA)}

\author{Wibe~A.~de~Jong}
\affiliation{
	Physics Division, Lawrence Berkeley National Laboratory, Berkeley, CA 94720, USA}

\author{Patrick Draper}
\affiliation{Department of Physics and Illinois Center for the Advanced Study of the Universe and Illinois Quantum Information Science and Technology Center, University of Illinois, Urbana, IL 61801, USA}

\author{Aida~El-Khadra}
\affiliation{Department of Physics and Illinois Center for the Advanced Study of the Universe and Illinois Quantum Information Science and Technology Center, University of Illinois, Urbana, IL 61801, USA}

\author{Nate~Gemelke}
\affiliation{QuEra Computing Inc, Boston, MA 02135, USA}

\author{Masanori~Hanada}
\affiliation{Department of Mathematics, University of Surrey, Guildford, Surrey, GU2 7XH, UK}

\author{Dmitri~Kharzeev}
\affiliation{Center for Nuclear Theory, Department of Physics and Astronomy, Stony Brook University, NY 11794-3800, USA}
\affiliation{Department of Physics, Brookhaven National Laboratory, Upton New York 11973, USA}

\author{Henry~Lamm}
\affiliation{Fermi National Accelerator Laboratory, Batavia,  IL 60510, USA}

\author{Ying-Ying~Li}
\affiliation{Fermi National Accelerator Laboratory, Batavia,  IL 60510, USA}

\author{Junyu~Liu}
\affiliation{Pritzker School of Molecular Engineering, Chicago Quantum Exchange, and Kadanoff Center for Theoretical Physics, University of Chicago, IL 60637, USA}
\affiliation{qBraid Co., Harper Court 5235, Chicago, IL 60615, USA}

\author{Mikhail Lukin}
\affiliation{Department of Physics, Harvard University, Cambridge, MA 02138, USA}

\author{Yannick~Meurice}
\affiliation{Department of Physics and Astronomy, University of Iowa, Iowa City, IA 52242, USA}

\author{Christopher~Monroe}
\affiliation{Duke Quantum Center, Duke University, Durham, NC 27701, USA}
\affiliation{Department of Electrical and Computer Engineering, Duke University, Durham, NC 27708, USA}
\affiliation{Department of Physics, Duke University, Durham, NC 27708}
\affiliation{IonQ, Inc., College Park, MD 20740, USA}

\author{Benjamin~Nachman}
\affiliation{
	Physics Division, Lawrence Berkeley National Laboratory, Berkeley, CA 94720, USA}

\author{Guido~Pagano}
\affiliation{Physics and Astronomy Department, Rice University, Houston, TX 77005, USA}

\author{John~Preskill}
\affiliation{Institute for Quantum Information and Matter, California Institute of Technology, Pasadena, CA 91125, USA}

\author{Enrico~Rinaldi}
\affiliation{Physics Department, University of Michigan, Ann Arbor, MI 48109, USA}
\affiliation{Theoretical Quantum Physics Laboratory, Center for Pioneering Research, RIKEN, Wako, Saitama 351-0198, Japan}
\affiliation{
Interdisciplinary Theoretical and Mathematical Sciences Program (iTHEMS), RIKEN, Wako, Saitama 351-0198, Japan}

\author{Alessandro~Roggero}
\affiliation{Dipartimento di Fisica, University of Trento, via Sommarive 14, I–38123, Povo, Trento, Italy}
\affiliation{INFN-TIFPA Trento Institute of Fundamental Physics and Applications,  Trento, Italy}

\author{David~I.~Santiago}
\affiliation{Quantum Nanoelectronics Laboratory, Department of Physics,
University of California at Berkeley, Berkeley, CA 94720, USA}
\affiliation{Computational Research Division, Lawrence Berkeley National Lab, Berkeley, CA 94720, USA}

\author{Martin~J.~Savage}
\affiliation{InQubator for Quantum Simulation (IQuS), Department of Physics,
University of Washington, Seattle, WA 98195, USA}

\author{Irfan~Siddiqi}
\affiliation{Quantum Nanoelectronics Laboratory, Department of Physics,
University of California at Berkeley, Berkeley, CA 94720, USA}

\affiliation{Computational Research Division, Lawrence Berkeley National Lab, Berkeley, CA 94720, USA}
\affiliation{
	Materials Sciences Division, Lawrence Berkeley National Lab, Berkeley, CA 94720, USA}

\author{George~Siopsis}
\affiliation{Department of Physics and Astronomy, University of Tennessee, Knoxville, TN 37996-1200, USA}

\author{David~Van~Zanten}
\affiliation{Fermi National Accelerator Laboratory, Batavia,  IL 60510, USA}

\author{Nathan~Wiebe}
\affiliation{Department of Computer Science, University of Toronto, Toronto, ON M5S 2E4, Canada}
\affiliation{Pacific Northwest National Laboratory, Richland, WA 99354, USA}

\author{Yukari~Yamauchi}
\affiliation{
 	Department of Physics and Maryland Center for Fundamental Physics, 
University of Maryland, College Park, MD 20742, USA}

\author{K\"ubra~Yeter-Aydeniz}
\affiliation{Emerging Technologies and Physical Sciences Department, The MITRE Corporation, 7515 Colshire Drive, McLean, VA 22102-7539, USA}

\author{Silvia~Zorzetti}
\affiliation{Fermi National Accelerator Laboratory, Batavia,  IL 60510, USA}

\newcommand\snowmass{\begin{center}\rule[-0.2in]{\hsize}{0.01in}\\\rule{\hsize}{0.01in}\\
\vskip 0.1in Submitted to the  Proceedings of the US Community Study\\ 
on the Future of Particle Physics (Snowmass 2021)\\ 
\rule{\hsize}{0.01in}\\\rule[+0.2in]{\hsize}{0.01in} \end{center}}

\snowmass

\begin{abstract}
\noindent
\textbf{Preprint Report No.} UMD-PP-022-04, LA-UR-22-22100, RIKEN-iTHEMS-Report-22, FERMILAB-PUB-22-249-SQMS-T, IQuS@UW-21-027, MITRE-21-03848-2.

\vspace{0.5 cm}

\noindent
\textbf{Abstract.} It is for the first time that \emph{quantum simulation for High Energy Physics (HEP)} is studied in the U.S. decadal particle-physics community planning, and in fact until recently, this was not considered a mainstream topic in the community. This fact speaks of a remarkable rate of growth of this subfield over the past few years, stimulated by the impressive advancements in Quantum Information Sciences (QIS) and associated technologies over the past decade, and the significant investment in this area by the government and private sectors in the U.S. and other countries. High-energy physicists have quickly identified problems of importance to our understanding of nature at the most fundamental level, from tiniest distances to cosmological extents, that are intractable with classical computers but may benefit from \emph{quantum advantage}. They have initiated, and continue to carry out, a vigorous program in theory, algorithm, and hardware co-design for simulations of relevance to the HEP mission. This community whitepaper is an attempt to bring this exciting and yet challenging area of research to the spotlight, and to elaborate on what the promises, requirements, challenges, and potential solutions are over the next decade and beyond.

\vspace{0.2 cm}

\noindent
This whitepaper is prepared for the topical groups CompF6 (Quantum computing), TF05 (Lattice Gauge Theory), and TF10 (Quantum Information Science) within the Computational Frontier and Theory Frontier of the U.S. Community Study on the Future of Particle Physics (Snowmass 2021).

\end{abstract}

\maketitle

\nopagebreak[4]

  \tableofcontents
\newpage
\phantomsection
\addcontentsline{toc}{section}{[Theory Frontier] Executive Summary}
\section*{[Theory Frontier] Executive Summary}
\label{sec:th-exec}
\noindent
HEP is full of important theoretical questions whose answers seem to be intractable using classical-computing techniques, spanning many different subfields, including collider physics, neutrino (astro)physics, cosmology and early-universe physics, and quantum gravity. Quantum simulation has the potential to provide computationally-feasible approaches to many of these problems. Explicitly, quantum simulation promises: i) computations of full scattering processes, as well as \emph{ab initio} calculations of parton distribution functions and other non-perturbative matrix elements, ii) simulations of collective neutrino oscillations in core-collapse supernovae, and of neutrino-nucleus scattering cross sections crucial for the Deep Underground Neutrino Experiment (DUNE), iii) studies of non-equilibrium dynamics in high-energy particle collisions, in inflationary phase of the universe, in CP-violating scenarios, and for models of dark matter, iv) describing bulk gravitational phenomena via simulating dual quantum field theories on the boundary and accessing quantum gravity in table-top experiments. While progress is being made in investigating the applications of quantum simulation in these problems, it is anticipated that over time, more problems will be identified in HEP that may benefit from quantum advantage.

Advances in many fields are needed to make this wide range of problems accessible to quantum simulators: i) Fundamentally new formulations of the problems within the framework of quantum field theories may be required. One of the main requirements is to develop efficient ways to turn the infinite-dimensional Hilbert space of quantum field theories into a finite-dimensional one. This often requires discretizing the space by going to a lattice formulation and digitizing the field values. While much progress has been made over the past years to devise efficient formulations, particularly for gauge theories of the Standard Model, more research is required to determine the pros and cons of each formulation in the context of concrete quantum-simulation algorithms, or even find entirely new formulations. Moreover, finding optimal ways to protect or utilize the symmetries of the underlying theories, in particular gauge symmetries, is an active area of research that most likely will benefit from new theoretical ideas. Finally, systematic uncertainties of quantum simulating field theories given the truncations imposed, and issues related to renormalization, need to be investigated carefully in the context of Hamiltonian simulation. ii) Extensive algorithmic research for both digital and analog simulations needs to occur rooted in physics inputs and theoretical understandings. For digital schemes, low-overhead and efficient encodings of degrees of freedom to qubits, and algorithms with tight and rigorous error bounds are needed. Furthermore, concrete protocols are needed for preparing non-trivial initial states and for measuring the outcome relevant for a range of observables, from scattering amplitudes and structure functions, to identifying phases of matter and phase transitions, potentially through quantum-information measures. iii) Since certain problems might be more natural to implement on analog quantum simulators, which often exhibit intrinsic fermionic and bosonic degrees of freedom, algorithms need to be developed to utilize analog quantum simulators for HEP problems. This amounts to understanding how one can map given quantum field theories onto a variety of analog quantum simulators each with distinct intrinsic degrees of freedom, native interactions, and connectivity properties.

A tight collaboration between experimentalists developing quantum simulators and HEP theorists, as well as effective communications among scientists in both HEP and QIS, are required for a program in quantum simulation in HEP to succeed. Given the multidisciplinary nature of this field, proper education, training, and retention of the next generation of scientists is critical, and must involve universities, national laboratories, and private industry. As it is true in most fields, successfully establishing a new area of research requires many parts to work together in harmony, and significant investments in education, scientific ingenuity, and resources need to be made to reach the era where quantum simulation can lead to solving problems that are currently unimaginable.\looseness-1

\newpage
\phantomsection
\addcontentsline{toc}{section}{[Computational Frontier] Executive Summary}
\section*{[Computational Frontier] Executive Summary}
\label{sec:comp-exec}
\noindent
Research and discovery in HEP has relied on a major computational pursuit, and high-performance computing has entered, evolved, and exploited to the full in many HEP disciplines. As a result, high-energy physicists are in a prime position to recognize the potential benefits of ever-advancing quantum-simulation hardware and its growing algorithm/software/compiler/user ecosystem. They have, indeed, identified a range of problems outside the capabilities of emerging Exascale classical computers and beyond, spanning subfields such as collider physics, neutrino (astro)physics, cosmology and early-universe physics, and quantum gravity.

The simulations of relevance to HEP involve quantum field theories at their core. A theoretical campaign has gained intensity in recent years to frame quantum field theories in the language of Hamiltonian simulation, which is the natural language of quantum simulators, and to quantify their systematic uncertainties when truncations are imposed to fit the simulations on finite simulating platforms. Such theoretical investigations must be developed in tight connections with algorithmic and hardware developments. For one, the number of general state preparation, time evolution, and measurement and tomography algorithms have been dramatically growing in the QIS community. It is not yet known what the best theoretical formulations of the HEP problems are coupled with these algorithms, or if newly-devised algorithms tailored to field-theory simulations are needed. The algorithms need to be concrete, allow for realistic estimates of quantum resources required to execute them given desired accuracy, and take advantage of physics input such as locality, symmetries, and gauge and scale invariances when present. To operate simulations in an analog or hybrid analog-digital mode, and engineer quantum field theories of increasing complexity and relevance, enhanced modalities and advanced quantum-control capabilities must be developed and added to current state-of-the-art platforms, and a range of physical architectures from atomic, molecular, optical, and solid-state systems must be carefully considered.

To ensure the feasibility of the simulation approaches, prevent a wide gap between theorists' proposals and experimental realities, and tighten the theoretical algorithmic scalings by supplementing empirical observations, implementation and benchmarking using the near-term noisy intermediate-scale quantum hardware will continue to be a critical endeavor for the HEP community. General-purpose error-correction and noise-mitigation methods, or those tailored to HEP simulations or inspired by HEP developments, are a necessity, as are hybrid classical-quantum algorithms with clear cost-saving benefits. Furthermore, to enable the broader HEP community to readily participate, models for programming quantum computers need to mature to a level in which abstractions and library-based methods can be adopted to expedite programming and ensure portability. Efficient compilers that can take the programs written in high-level programming languages to efficient low-level code are needed for HEP applications. The quantum-computing paradigm may also require that users effectively engage across the quantum-computing stack, and to have access to software layers that generate pulses needed to drive the interactions of relevance to HEP applications, such as those of gauge theories.
 
Since quantum technology is advancing rapidly in universities, national laboratories, and private sector, it is important that collaboration and cooperation is ensured between all these institutions. A careful study on whether industry-developed quantum hardware satisfies the needs of the HEP community is required. Most likely, the HEP community needs to play a crucial role in the development of quantum simulators, while at the same time needs to have access to those platforms developed in industry and other partners. Such  multidisciplinary research requires training, retention, and empowering the workforce, and in some cases, retooling and reorienting the talent in HEP to enter this rapidly advancing field. Importantly, diversity and inclusivity will be crucial in ensuring an intellectually open and strong program at the intersection of HEP and QIS.

\newpage
\section{Introduction}
\label{sec:intro}
\noindent
Quantum simulation refers to simulating a complex quantum system using another quantum system that can be fabricated in a laboratory, and be sufficiently isolated and controlled to maintain quantum coherence. The simulator can closely resemble the degrees of freedom of the target system, and its dynamics can be engineered to closely follow those of the simulated theory, in which case it is said to operate in an analog mode. The simulator can also be developed in such a way that the simulation will be agnostic to the underlying physical architecture, and is engineered to implement a set of universal and elementary operations on an array of few-state quantum units, prototypically two-state `qubits',
in which case it is said to operate as a digital quantum computer. It can also combine features of the two modes for more adaptability and efficiency. The simulated theory often describes a quantum many-body system that is hard to simulate classically as the Hilbert spaces are exponentially large in particle number or system size, prohibiting exact or approximate classical Hamiltonian-simulation methods, or other  methods that are stochastic in nature and may suffer from sign problems. Not only can the Hilbert spaces be encoded exponentially more compactly on a quantum simulator, but also quantum operations retain quantum correlations and effectively perform massively parallel computations on the encoded wavefunction. Harnessing such a significant computing resource amounts to finding a mapping between the dynamics of the target system and those which can be encoded in the simulator via quantum algorithms. One desires such algorithms to require resources that scale sub-exponentially with the system size and the accuracy desired. Furthermore, efficient error-correction algorithms in the far term, and effective error-mitigation techniques in the near term, are needed to overcome the effect of quantum decoherence in realistic implementations. Finally, an integrated hardware-software-compiler stack is desired such that the users with applications can easily interact with the simulator, without significant barriers in translating between the simulated theory's language and the machine's language.

The development and advancement of the quantum-simulation program in HEP requires all the above topics to be considered in the context of the target problems in HEP. Perhaps more importantly in the short term is to identify such target problems, i.e., those that are likely to remain intractable with current theoretical and computational methods, but have the potential to be solved with quantum simulators. Once such problems are identified, the underlying theoretical framework must be adopted properly and the corresponding quantum-simulations algorithms need to be developed. Since the underlying theoretical framework in HEP is generally quantum field theories (QFTs), the aim becomes quantum simulating quantum fields and their interactions. For many of the questions of relevance to HEP research, quantum simulating the Standard Model (SM) and particularly the strong force is a primary objective, but simulating a range of effective field theories (EFTs), conformal field theories, or prototype models of quantum gravity, will also be a critical component of the program for their applications in low-energy and high-energy regimes in nature. Finally, given the underlying simulation to be performed, the proper choice of the quantum simulator becomes essential and, importantly, it will likely be the case that special-purpose quantum simulators for HEP will be required. Such dedicated simulators may come in different varieties given different existing physical architectures, and their development can be facilitated by a co-design process involving HEP scientists and quantum scientists and developers. Given such a multi-pronged and multi-disciplinary area of research and development in quantum simulation of HEP, a new and skilled generation of workforce will need to be trained and empowered, and strong partnerships among universities, national laboratories, and technology companies to be pursued. This will allow theoretical, algorithmic, and experimental lines of this research to be advanced quickly and simultaneously.

\begin{figure}[t!]
\includegraphics[scale=0.585]{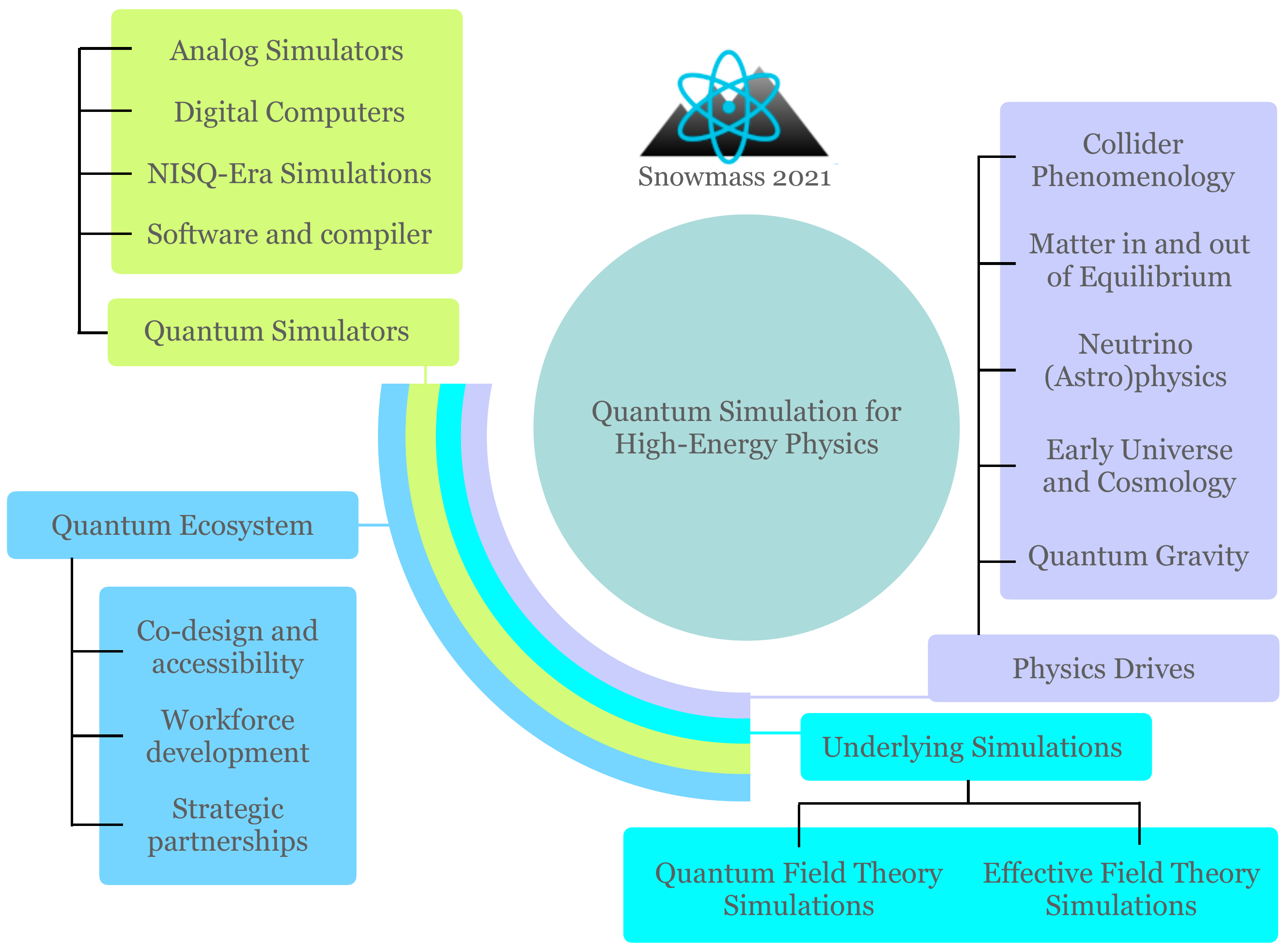}
\caption{Aspects of a quantum-simulation program for HEP that are studied in this whitepaper.
}
\label{fig:chart}
\end{figure}

This whitepaper is an attempt to bring this exciting and yet challenging area of research to the spotlight, and to elaborate on what the promises, requirements, challenges, and potential solutions are over the next decade and beyond. Figure~\ref{fig:chart} presents an schematic overview of the aspects of quantum-simulation program for HEP over the next decade that are studied in this whitepaper.  The abundant supplemental materials contain more in-depth discussions on several topics along with relevant references. It should be noted that this document only concerns research directions that directly impact the mission of the field of HEP as defined within the U.S., that is to ``explore the fundamental constituents of matter and energy, and to reveal the profound connections underlying everything we see, including the smallest and the largest structures in the universe''~\cite{Ritz2014}. The whitepaper, therefore, does not discuss problems of relevance in other closely-related fields such as Nuclear and Hadronic Physics~\cite{Geesaman2015} unless progress in those problems are expected to impact theoretical and experimental research in HEP.
\vspace{0.5 cm}

\section{Physics drives}
\label{sec:drives}
\noindent
HEP is full of important theoretical questions whose answers seem to be intractable using traditional classical-computing techniques. Addressing these problems, which span many different subfields, including collider physics, neutrino (astro)physics, cosmology and early-universe physics, and quantum gravity, will have a major impact on our understating of nature and its underlying working principles at the most fundamental level. High-energy physicists are actively identifying concrete areas that quantum-simulation methods and technologies can advance in the coming decade. 

Probing the SM at the highest possible energies is usually achieved using high-energy particle colliders, and particle collisions are notoriously difficult to describe. Strong interaction, being asymptotically free~\cite{gross1973ultraviolet,politzer1973reliable}, is described in terms of weakly-interacting quarks and gluons at high energies. However, since the final and, in many cases, initial, states of the collisions involve hadrons, non-perturbative physics is required for any full description of the events. Since the simulation of collider physics requires understanding real-time dynamics, normal lattice-field-theory techniques based on a Monte Carlo approach is hampered by sign problems~\cite{troyer2005computational,alexandru2016monte}.
For this reason, particle collisions are nowadays described theoretically using various types of approximations~\cite{particle2020review, campbell2022event}. Processes at the highest energies are computed using a perturbative evaluation
of the full quantum-mechanical amplitudes~\cite{gehrmann2021precision}. The production of a large number of additional partons is traditionally described by a parton-shower algorithm~\cite{sjostrand2006pythia,corcella2001herwig,gleisberg2009event}, which is based on classical emission
probabilities. Finally, phenomenological models are used to describe how the resulting partons hadronize to form color-neutral hadrons. Quantum computers hold the promise to simulate scattering processes from first principles, and, in principle, without any uncontrolled approximations.
The basic idea is to approximate the continuous model by a set of discretized formulations, which can systematically approximate the original model with increased precision. It is believed that all relevant ingredients required to compute the S-matrix can be calculated on a quantum computer using resources that scale only polynomially with the number of lattice sites~\cite{jordan2012quantum}. This was demonstrated explicitly for a scalar field theory~\cite{jordan2011quantum}, but is believed to also remain true for more complicated field theories, such as the gauge theories of the SM. Understanding the behavior of the strong interaction in the non-perturbative regime can allow the calculation of long-distance quantities that are important ingredients to collider observables. Such non-perturbative ingredients often arise in EFT approaches, in which short-distance physics, which is calculable in perturbation theory, is separated from long-distance physics, which is often non-perturbative. One of the well-known example of such non-perturbative quantities is the parton distribution function, but other ingredients such as jet and soft functions arise, for example, in soft collinear effective theory (SCET)~\cite{bauer2000summing,bauer2001effective,bauer2001invariant,bauer2002soft}. 

Quantum computing is also required to understand strongly-coupled matter at high density or far from equilibrium. In the coming decade, results from  heavy-ion and proton collisions at the Relativistic Heavy Ion Collider (RHIC) and the Large Hadron Collider (LHC), as well as gravitational-wave searches from the Laser Interferometer Gravitational-Wave Observatory (LIGO), will lead to an unprecedented level of experimental data probing strongly-interacting matter. These problems are again inaccessible to standard Monte Carlo lattice-gauge-theory techniques due to a sign problem. Quantum computers offer the promise of reducing the computational complexity in this problem from exponential to polynomial by naturally incorporating entanglement as quantum states~\cite{temme2011quantum,desai2021z3}. Such simulations should lead to precision theoretical results for the QCD equation of state and the behavior of phase transitions in strongly-interacting matter~\cite{czajka2021quantum}. The ability of quantum simulations to perform real-time evolution will also provide insight into the approach to equilibrium in strongly-interacting systems~\cite{riera2012thermalization}. There is tantalizing evidence from heavy-ion and proton-collision experiments at RHIC and LHC that the strongly-interacting matter thermalizes in a remarkably short period of time, in disagreement with na\"{\i}ve scaling arguments.  One possible solution to this puzzle is that the dynamics of quantum entanglement plays a role in the equilibration process~\cite{kharzeev2005color,kharzeev2017deep,baker2018thermal,berges2021qcd}. Furthermore, it is expected that entanglement Hamiltonians and associated entanglement spectrum~\cite{li2008entanglement} can reveal unique properties of the quantum many-body system such as whether and how the system thermalizes and if phase transitions occur~\cite{geraedts2016many, kaufman2016quantum, yang2015two,Mueller:2021gxd}.  Studying such behavior in large, far-from-equilibrium strongly-interacting quantum matter is prohibitively difficult with classical computers. Quantum simulators represent a natural tool to create such conditions and track the real-time evolution of entanglement in strongly-interacting matter.  

Another area of HEP where classically intractable problems are present is neutrino (astro)physics. A prime example is core-collapse supernovae and neutron-star mergers, where the very large number of neutrinos present require taking neutrino-neutrino interactions, both within the SM and beyond, into account. Collective neutrino oscillations have been shown to potentially have an important impact in supernova environment, both in the neutrino-driven explosion mechanism and in the ensuing nucleosynthesis in the ejected material~\cite{qian1993connection,qian1995neutrino,fogli2007collective}. This results in a full many-body problem, hence an exact solution to the dynamical evolution of flavor from a general initial configuration amounts to a computational cost which is exponential in the number of neutrinos involved.
Quantum computing can be used to simulate the propagation of neutrinos, in particular collective neutrino oscillations, with resources that are polynomial in the number of neutrinos. Due to the simple structure of the neutrino-neutrino Hamiltonian, it is also possible, at least for simple geometries and energy distributions, to map  neutrino systems directly into the degrees of freedom and interactions of quantum simulators.~\cite{davoudi2020towards,bermudez2017long}.
In a rather different direction, the measurement of fundamental properties of neutrinos such as their absolute masses, mixing angles, and the CP-violation parameters depends upon accurate determination of neutrino cross sections with the material, often nuclear isotopes, used in the detector in experiments such as in the Deep Underground Neutrino Experiment (DUNE). A full description of the neutrino-nucleus scattering cross section is a formidable theoretical challenge due to the wide energy range required for the analysis of the experiment. 
Future quantum simulations can help these efforts by allowing for both more efficient representations of the nuclear target's wavefunction, and the extraction of inclusive and semi-exclusive reaction cross sections with controllable uncertainties~\cite{roggero2019dynamic,roggero2020quantum}. Early attempts with current digital quantum devices are encouraging (see e.g., Refs.~\cite{dumitrescu2018cloud,lu2019simulations,roggero2020quantum,stetcu2021variational,baroni2021nuclear}) but to achieve the required accuracy, large-scale error-corrected quantum simulators are likely required.

Since the universe is inherently quantum, predictions about earlier epochs of the universe may need to include quantum effects. Multiple phenomena across cosmology and the early universe, including inflation, baryon asymmetry, phase transitions, and dark matter, may need to be described by the non-equilibrium dynamics of non-perturbative quantum fields, which again is not possible using standard lattice-field-theory methods and requires approximations that lead to uncontrolled systematics. Simulating such phenomena {\it ab initio} provides opportunities for demonstrating practical quantum advantage.
The inflationary phase is terminated by non-perturbatively transferring energy to particle degrees of freedom through reheating and preheating. These far-from-equilibrium and non-perturbative processes could leave imprints on the sky today~\cite{Kofman:1994rk,kofman1997towards,greene1997structure}. Though in the long term, large-scale quantum simulations of quantum inflationary fields are desired~\cite{chatrchyan2021analog, liu2021quantum,butera2021particle,barata2021single},  near-term studies can be useful as well~\cite{tsamis1997quantum,chatrchyan2021analog2,Sakharov:1967dj,lagnese2021false, pirvu2022bubble,milsted2020collisions,neuenhahn2015quantum,arrighi2018quantum,rossini2020dynamics,boettcher2020quantum,xu2019probing,nie2019experimental,Heyl_2019,berkowitz2015lattice,wantz2010axion,turner1986cosmic}.
Non-equilibrium dynamics are also required for generating the observed baryon asymmetry~\cite{sakharov1998violation}. 
Standard treatments rely upon the dynamics when the interaction rate is close to the Hubble rate, which is assuming nearly equilibrium behavior~\cite{wainwright2012cosmotransitions}, but simplistic comparisons can be insufficient to ensure an adiabatic evolution~\cite{polkovnikov2008breakdown}. 
Serving as inputs to the time evolution of classical fields, the properties of nucleating bubbles are usually extracted from an effective potential, but the phase-transition history defined solely by the conventional critical-temperature calculation can be misleading~\cite{baum2021nucleation}. As a result, it is necessary to perform more detailed calculations of the nucleation mechanism. A first priority for quantum research in this area is to rigorously define matrix elements for bubble properties and exploratory calculations of them~\cite{lagnese2021false, pirvu2022bubble,milsted2020collisions}. Additionally, the interplay between phase transitions and curved spacetime can be investigated with the Noisy Intermediate-Scale Quantum (NISQ) simulators for low-dimensional toy models~\cite{neuenhahn2015quantum,arrighi2018quantum,rossini2020dynamics,boettcher2020quantum}.
Another mystery of the early universe is what accounts for around 85\% of the matter observed only via its gravitational effects. Light dark matter such as axions typically require a more complex and non-equilibrium history to be produced. To extract the axion-mass information, QCD free energy as a function of the CP-violating phase and temperature has to be calculated, a problem that is hindered by a sign problem classically. With real-time simulations of the early universe, quantum computers can also simulate the non-equilibrium histories of various types of dark matter, such as light dark matter, topological defects~\cite{vilenkin1994cosmic}, and primordial black holes~\cite{zel1967hypothesis}.

Besides the important phenomenological questions raised above, constructing a complete and convincing quantum theory of gravity is a grand challenge facing fundamental physics. The main goals are to develop a deeper understanding of the fundamental laws of nature, to resolve long-standing puzzles about what happens inside black holes, and to explain the initial conditions in the history of the universe. Despite limited guidance from experiments, remarkable progress is being achieved, examples being the AdS/CFT duality~\cite{maldacena1999large}, augmented by the realization that bulk geometry emerges from boundary entanglement~\cite{ryu2006holographic}.
Despite this progress by melding insights from the holographic correspondence with ideas from QIS~\cite{almheiri2015bulk,pastawski2015holographic,harlow2019constraints,hayden2007black,sekino2008fast,Maldacena2016bound,susskind2016computational,stanford2014complexity,brown2016holographic,Sachdev1993gapless,Kitaev_talk,engelhardt2015quantum,penington2020entanglement,almheiri2019entropy,penington2019replica,almheiri2021entropy}, much is still missing from our current understanding of quantum gravity. Thanks to holographic duality, bulk gravitational phenomena can be described by a ``dual'' boundary quantum system consisting of many particles strongly interacting with one another. In principle, this boundary system can be simulated using a quantum computer, opening opportunities for exploring quantum gravity in table-top experiments. One needs to find a (non-gravitational) quantum system that has a gravitational dual, realize this quantum system in a feasible experiment, and develop a dictionary relating gravitational phenomena of interest to measurable observables in the
quantum system. More tractable, but still ambitious, targets with good gravitational duals would be certain matrix models~\cite{banks1997m,itzhaki1998supergravity,berenstein2002strings}. The information of the gravitational geometry is encoded in matrix degrees of freedom~\cite{banks1997m,witten1996bound,hanada2021bulk}. Although Monte Carlo simulations on classical computers provided non-trivial test of this conjecture (see e.g., Refs.~\cite{berkowitz2016precision,bergner2021confinement}), the details of the encoding of geometry into matrices have been out of reach. Quantum simulation would provide a practical tool in this problem.
Though spectacular insights into quantum gravity may not be expected in the next ten years, the community can develop tools, methods, and insights that will lay foundations for quantum technology, enabling profound advances in fundamental physics.

While the physics drives of the program as enumerated in this whitepaper are the most immediate motivations for using quantum technologies in simulation over the coming decades, it is likely that as high-energy physicists discover the power of quantum simulation, and enjoy access to the current and next generations of quantum simulators, new problems and additional physics goals will be identified and explored.

\vspace{0.5 cm}

\section{Underlying simulations}
\label{sec:simulation}
\noindent
Quantum field theories are the underlying mathematical description of three of the four fundamental forces in nature, constituting the SM of particle physics. They are also closely connected to theories of quantum gravity via the bulk/boundary duality. QFTs in form of EFTs also provide a systematic mechanism to organize interactions in nature at low energies where there exists a separation of scales, and coarse-grained descriptions of degrees of freedom are applicable, as well at high energies where the unknown physics beyond the SM can be introduced via higher-dimension operators of SM fields. Various perturbative and non-perturbative approaches to solving QFTs have been developed and successfully applied over the years. In the realm of SM predictions in the non-perturbative regime of the strong interactions, lattice-field-theory methods have proven the most reliable approach to date, providing critical input to many experimental programs in HEP, see e.g., Refs.~\cite{aoki2021flag,lehner2019opportunities,cirigliano2019role,detmold2019hadrons,davoudi2021nuclear,bazavov2019hot,ding2015thermodynamics,detar2009qcd} for recent reviews. Nonetheless, the conventional lattice-field-theory program based on the path-integral formalism of QFTs relies on Monte Carlo importance-sampling methods, and this statistical feature has halted progress in several problems. These include finite-density systems where a signal-to-noise or a fermionic sign problem demands an exponential increase in generated samples and measurements with the system size. The same issue has prevented meaningful progress in real-time problems such as scattering processes, except for those at low energies and low inelasticities that can be addressed by indirect methods~\cite{luscher1986volume,luscher1991two,bulava2022hadron,briceno2018scattering,davoudi2018path,hansen2019lattice}. That is because the gauge-field importance sampling is enabled by a Wick rotation to Euclidean spacetime. A Hamiltonian-simulation approach does not  encounter such issues but the size of the required Hilbert space scales exponentially with the system size. This  has limited classical computing methods, including tensor-network approaches that have so far mostly targeted simpler low-dimensional theories and in systems without volume-law entanglement~\cite{banuls2020review,banuls2020simulating} (see Refs.~\cite{magnifico2021lattice,felser2020two} for examples of progress in higher-dimensional simulations). Quantum simulation is a natural mechanism to implement Hamiltonian simulation of QFTs, however, similar to the development of the conventional lattice-field-theory program, it will need a dedicated program of continuous research and innovation in theory, algorithm, hardware implementation, and co-design to reach the level of maturity that is needed for addressing the physics drives of the HEP program.

The Hilbert space of QFTs is infinite dimensional as fields are defined on an infinite and continuum spacetime. For bosonic theories, the local Hilbert space of the fields is also infinite dimensional. Simulating QFTs on any finite-capacity quantum hardware requires truncating the extent of the volume, discretizing the space, and digitizing and/or truncating the on-site Hilbert space of the bosons. In the context of Hamiltonian simulation, various approaches are proposed to make QFTs finite dimensional in a systematic way. Furthermore, evolution in a quantum simulator occurs in a digital or discrete manner leading to time digitization errors or in an analog or continuous manner but with potentially approximate Hamiltonians implemented. As with the development of the conventional lattice-field-theory program, quantifying the systematic uncertainties in quantum simulation of lattice field theories, including renormalization and continuum limit, finite boundary effects, boson-field digitization/truncation errors, and inaccuracy in the time evolution stage is crucial to achieving realistic results and  resource estimates. Effective field theories may provide a pathway to such characterizations as with the conventional program. While initial efforts along these lines have started (see Suppl.~\ref{suppl:QFT}), these questions will continue to be at the center of theoretical studies in the upcoming years.  

The basis chosen to span the Hilbert space informs the truncation scheme, and proper choices can result in faster convergence to the desired asymptotic limits and/or lower computational cost, classically or quantumly. For gauge theories, the manifestation of gauge invariance and of local Gauss' law constraints are another determining factor in adopting the proper formulation. The Hamiltonian formulation of lattice gauge theories (LGTs) provides the starting point for a range of frameworks: i) irreducible representation (irrep) or electric-field bases that diagonalizes the electric Hamiltonian, which are a natural bases for expressing the Gauss' laws, and can be truncated to a finite number of irreps without breaking gauge invariance~\cite{kogut1975hamiltonian,klco2018quantum,klco20202,byrnes2006simulating,ciavarella2021trailhead}, ii) prepotential and loop-string-hadron bases which build the Hilbert space that satisfies the non-Abelian Gauss' laws via constructing gauge-invariant operators locally, and impose only an Abelian constraint on the link~\cite{mathur2005harmonic,anishetty20142,anishetty2009irreducible,mathur2010n,raychowdhury2020loop}, iii) group-element bases that simplify the implementation of gauge-matter and the magnetic Hamiltonians, and are a more economical choice in the weak-coupling regime relevant for approaching the continuum limit~\cite{zohar2015formulation,lamm2019general}, iv) dual or magnetic basis which diagonalizes the magnetic Hamiltonian, is obtained via discrete group Fourier transforms when applicable, or via tensor-renormalization group methods in given systems~\cite{kaplan2020gauss,haase2021resource,bauer2021efficient,mathur2016lattice,liu2013exact,xie2012coarse,meurice2020tensor}, v) light-cone quantization of fields leading to basis states in the light-cone momentum occupation-number basis with encoding-cost benefits over other formulations but with subtleties regarding the zero modes and UV-IR mixing during renormalization~\cite{kreshchuk2021light,kreshchuk2021simulating,kreshchuk2022quantum}, vi) quantum link model~\cite{brower1999qcd,brower2004d} or qubitization~\cite{buser2020state,bhattacharya2021qubit,alexandru2021universality} formulations of gauge theories that assume a finite-dimensional on-site Hilbert space for the gauge fields, but introduce an extra dimension that needs to be taken to infinity, or alternatively that take advantage of universality classes near the critical points to 
approach the continuum limit without the need to introduce the extra dimension, vii) matrix models which provide a way to map gauge theories to quantum mechanical models via dimensional reduction, while preserving some of the interesting non-perturbative dynamics and structure of the parent QFT~\cite{Luscher:1982ma,van2001qcd,banks1997m,banks2006towards,anninos2020notes}, among a few other choices. We do not know what the most optimal formulation is for QCD, that is a SU(3) LGT in 3+1 D is with multiple flavors of quarks in the fundamental representation, as this theory has not yet been fully studied in the context of the formulations mentioned above. The next decade of developments will shed light on this important theoretical question.

An important ongoing investigation is to what extent preserving symmetries, and, particularly, local gauge symmetries and Gauss' law, matter in a quantum simulator. Questions regarding the robustness of simulations to gauge-symmetry-breaking perturbations in the dynamics are starting to be explored in simpler gauge theories and quantum link models~\cite{halimeh2020fate,van2021reliability}. Furthermore, several proposals and algorithms are put forward in recent years for detecting and discarding Gauss's law violations~\cite{stryker2019oracles,raychowdhury2020solving}, and for suppressing coherent gauge-symmetry-violating noise~\cite{banerjee2012atomic,hauke2013quantum,banerjee2013atomic,marcos2013superconducting,rajput2021hybridized,tran2021faster,lamm2020suppressing,halimeh2021gauge,halimeh2022gauge,halimeh2021stabilizing,van2021suppressing,stannigel2014constrained,kasper2020non,ciavarella2021trailhead}, taking advantage of features like introduction of energy penalties, classical noise and Zeno effect, quantum control, dynamical decoupling, random rotations of the state throughout the evolution via unitaries generated by the symmetry (or pseudo symmetry) operator, and controlled operations in digital circuits. The value of these strategies and their limitations must be confirmed in realistic experiments~\cite{nguyen2021digital} and in the context of each of the Hamiltonian formulations discussed above. Furthermore, more targeted strategies need to be developed to suppress the incoherent noise and, more generally, to represent gauge-theory simulations as dynamics of open quantum systems coupled to the environment. 

Digital quantum algorithms for QFTs for both near- and far-term era of quantum computing need to be developed in the context of various Hamiltonian formulations and the particular quantities that are aimed to be extracted. Near-term era refers to at least the next decade of hardware developments, where the qubit resources are scarce and the gate fidelities, particularly those of the entangling gates, are not high enough to enable reliable error-corrected simulations. In the far-term era, when fault tolerance is achieved, auxiliary qubit registers can be used at ease but error-corrected synthesis of non-Clifford single-qubit gates such as T gates becomes costly. Each scenario, therefore, requires different resource optimizations. For non-Abelian gauge theories, for example, the non-commuting algebra of the group generally introduces costly arithmetic in terms of both near- and far-term resource measures. Additionally, the choice of simulation algorithms that amounts to the method of approximating the unitary time evolution, as well as qubit encoding that amounts to storing fermionic and bosonic quantum numbers (and the Fermi statistics of fermions), greatly impacts resource requirements of the simulation, and full or partial retainment of symmetries and conservation laws at each stage of the evolution. While algorithmic progress for Abelian and non-Abelian gauge theories of relevance to the SM has been significant in recent years~\cite{byrnes2006simulating,shaw2020quantum,ciavarella2021trailhead,kan2021lattice}, as with classical computing, the most efficient algorithms are likely not going to be the early algorithms. Furthermore, the asymptotic scaling with simulation accuracy must be greatly improved in the coming years by taking advantage of physics inputs~\cite{su2021nearly,csahinouglu2021hamiltonian,hatomura2022state,yi2021spectral,zhao2021hamiltonian} and insights from empirical analyses~\cite{childs2019nearly,stetina2020simulating,nguyen2021digital}. The algorithms necessary for preparation of non-trivial initial states such as hadrons and nuclei, quantities such as scattering and transition amplitudes, structure functions,  equal- and out-of-time correlation functions, and entanglement measures, as well as efficient state-tomography methods, need to be also advanced. Furthermore, proper definitions of entanglement in the context of lattice gauge theories (which represent non-separable Hilbert spaces due to local constraints) must be developed within various Hamiltonian formulations~\cite{donnelly2012decomposition,casini2014remarks,aoki2015definition,soni2016aspects}.    

Analog quantum simulators provide another potentially promising path to quantum simulation of QFTs. Analog simulators may naturally offer fermionic and bosonic degrees of freedom removing the need for expensive qubit encodings, or provide tunable interactions between larger local Hilbert spaces like qudits. On the other hand, one needs to engineer the interactions of these degrees of freedom to represent the dynamics of the QFT Hamiltonian of interest. Analog simulators may come in larger sizes with sizable Hilbert spaces and with two- and three-dimensional geometries, and offer the possibility of engineering interaction graphs representing curved spacetime~\cite{kollar2019hyperbolic,kollar2019hyperbolic,periwal2021programmable}, hence enabling studies pertinent to black-hole physics and quantum information~\cite{boettcher2020quantum,swingle2016measuring}. These exciting possibilities are faced with the challenge of finding the proper platform for a given problem both in terms of similarities of Hilbert spaces, the mapping of the symmetries, and the ability to engineer complex interactions with the knobs available in the simulator. For example, simulating the dynamics of both Abelian and non-Abelian gauge theories in 2+1 and higher dimensions have proven hard
~\cite{zohar2012simulating,zohar2011confinement,zohar2013simulating,tagliacozzo2013optical,ott2021scalable,gonzalez2022hardware} given the higher-body interactions present when working in the irrep basis, and non-local interactions when working in dual basis. Some progress has been reported in recent years, but a first implementation of complete building blocks of a lattice gauge theory with high  fidelity remains a critical goal of the program in the coming decade. If such simulations become possible and can be scaled, one needs to face the question of rigorous error-bound analysis, error corrections, and verifiability that are less developed for analog simulations compared with their digital counterparts. In the same spirit, continuous-variable quantum computing presents another opportunity for quantum simulating QFTs, as it may provide a more natural way of treating bosonic fields in the simulation~\cite{marshall2015quantum,yeter2018quantum}.

It is important to emphasize the role of hardware implementation and benchmarks in guiding the course of developments  in quantum simulation. Many recent experiments and implementations of a variety of QFT problems on a range of quantum architecture in both analog and digital modes (see e.g., Refs.~\cite{martinez2016real,klco2018quantum,nguyen2021digital,kokail2019self,lu2019simulations,alam2021quantum,klco20202,atas20212,ciavarella2021trailhead,ciavarella2021preparation,xu20213+,mildenberger2022probing,schweizer2019floquet,gorg2019realization,mil2019realizing,yang2020observation,zhou2021thermalization} have generated an exciting platform for communications and collaborations with leading experts in quantum hardware and software technology.  It has also generated proposals and experiments dedicated to developing simulators suitable for QFTs on a range of platforms, see e.g., Refs.~\cite{hauke2013quantum,davoudi2020towards,andrade2021engineering,surace2020lattice,marcos2013superconducting,davoudi2021towards,casanova2011quantum,zohar2013cold,banerjee2013atomic,dasgupta2022cold,zohar2013quantum}. Continuing this trend is essential as otherwise theory and algorithmic developments will be disconnected from the reality of hardware. Furthermore, implementations on various platforms assist in verifying the outcome of quantum simulation once the limits of classical computing are challenged in the forthcoming simulations. To achieve meaningful progress, over the next decade a series of models from low-dimensional theories and simpler gauge groups need to be identified and progressively made more complex to follow, or ideally, guide,  hardware developments.

Finally, hybrid classical-quantum approaches to quantum simulation should be taken advantage of to a full degree in both near and far terms. On one hand, variational-based algorithms can be proven useful in spectroscopy studies as they rely on classical optimizers to find the ground state of a given Hamiltonian given a non-trivial wavefunction prepared on a quantum computer, see Refs.~\cite{klco2018quantum,lu2019simulations,kokail2019self,atas20212} for several gauge-theory examples. They can also be taken advantage of in scattering problems as recently proposed~\cite{liu2021towards}. On the other hand, classical computers may
prepare non-trivial states through the use of conventional lattice-field-theory methods (as long as sign and signal-to-noise problems are not encountered) or tensor-network-inspired methods (as long as states are area-law entangled). Quantum computers can then use this initial-state input to perform the classically-intractable time evolution, hence saving quantum resources needed for state preparation~\cite{harmalkar2020quantum,gustafson2021toward}. Furthermore, various costly stages of a conventional lattice-field-theory calculation may be expedited by quantum processors. These include generation of gauge-field configurations, matrix inversions~\cite{chang2019quantum}, Wick contractions, and design of suitable interpolating-operator sets for the states~\cite{avkhadiev2020accelerating}. Over the next decade, the community will need to continue to identify and advance such hybrid classical-quantum approaches to simulation.

The framework discussed so far can be used to simulate the full S-matrix of an underlying QFT. However, in the case of QCD theory, quantities at short distances (high energies) can, in most cases, be reliably calculated perturbatively, and it is the physics at long distances (low energies) that is non-perturbative in nature. EFTs can be used to separate the physics at long and short distances, with the EFT describing the long-distance dynamics, while short-distance effects can be included through coefficients that are obtained by perturbative renormalization-group matching, via non-perturbative methods, or by matching to experiment in regimes where the EFTs are applicable. The EFT setup then defines the long-distance ingredients that can, in principle, be computed on a quantum computer, together with a prescription of how to combine it with the short-distance physics to obtain a physical observable. This strategy saves quantum resources as simulating the full underlying theory is often costly.

A well-known non-perturbative quantity in hadronic physics is the parton distribution function (PDF). It involves a matrix element of two quark fields separated by a light-like direction, which can not be calculated using traditional lattice-field-theory techniques directly due to a sign problem. While several techniques have been put forward to make  calculations of PDFs possible with notable success~\cite{aglietti1998model,liu2000parton,ji2013parton,chambers2017nucleon,ma2018extracting}, quantum computers give rise to the possibility to compute the matrix element relevant for the PDFs directly and from first principles. Several proposals have been put forward in recent years to demonstrate how PDFs can be accessed on a quantum computer~\cite{lamm2020parton,echevarria2021quantum,li2021partonic,mueller2020deeply,qian2021solving}. One can calculate the hadronic tensor on a quantum computer and then extract the PDF using perturbative information for the partonic scattering~\cite{lamm2020parton}. Alternatively, the Wilson line can be explicitly constructed using plaquette operators or fermion hopping terms, allowing for an estimate of the full quantum computation of the PDF~\cite{lamm2020parton,echevarria2021quantum}. Finally, a PDF calculation in the NJL model was performed in Ref.~\cite{li2021partonic} using a variational ansatz for the proton state, and following Ref.~\cite{pedernales2014efficient} for the correlation function. At this stage, it is not clear what the realistic computational-resource requirements are for computing PDFs and hadronic tensor to given accuracy, as the complete algorithms, including that needed for preparation of hadronic states in QCD on a quantum computer, are either non-existing or premature. Over the next decade, theoretical and algorithmic research will improve upon these initial analyses.

An EFT describing the collinear and soft physics in jet-like collider events is SCET~\cite{bauer2000summing,bauer2001effective,bauer2001invariant,bauer2002soft}. The PDFs mentioned above are collinear matrix elements in the theory, however soft matrix elements are required as well to make predictions with jets in the final state. Since these soft matrix elements are again non-perturbative in nature, quantum algorithms might be suited to compute them from first principles. This has been studied in a simplified theory, namely a scalar field theory interacting with static sources. All necessary quantum circuits to compute the soft function in this example were worked out in Ref.~\cite{bauer2021simulating}, along with necessary quantum circuits and a small-scale simulation on an IBM Quantum device. The progress in this problem is correlated with that in simulating lattice gauge theories in 3+1 dimensions. 

Another example of important long-distance effects in collider events is collinear radiation, which traditionally is described by parton-shower algorithms~\cite{particle2020review,sjostrand2006pythia,bahr2008herwig++,gleisberg2009event,bauer2007event}. The very nature of a probabilistic Markov-Chain algorithm makes including quantum-interference effects challenging, since collider events typically contain a very large number of final-state particles. Quantum-interference effects that can be present are effects at subleading orders in the inverse number of colors, and interference between amplitudes with different intermediate particles and different internal kinematics. This problem, therefore, is a suitable candidate for quantum simulation. A quantum algorithm has been developed in Ref.~\cite{nachman2021quantum} that reproduces the regular parton shower, while by computing all possible amplitudes at the same time, it also includes quantum-interference effects. More work is required to develop a full parton-shower algorithm for the SM, and to find ways to include the most relevant quantum-interference effects, such as color interference.

Finally, low-energy EFTs of nuclear interactions are important for the HEP mission as they provide a consistent framework to describe the nuclear targets used in high-energy experiments such as in long-baseline neutrino experiments, neutrinoless double-$\beta$ decay experiments~\cite{engel2017status}, and 
direct dark-matter detections~\cite{baudis2013signatures,andreoli2019quantum}. Dynamical properties of nuclei like inelastic cross sections are challenging to compute classically, especially for semi-exclusive scattering in medium- and large-mass nuclei, and quantum simulations have the potential of being impactful in such problems. Simulation of these theories is similar to a quantum-chemistry simulation, given the nonrelativistic nature of interactions, thus techniques developed in quantum chemistry can be ported to the nuclear simulations. 
A few main differences are the presence of additional fermionic species, the presence of three- and higher-nucleon interactions, and the presence of long-range pion-exchange interactions. Progress has been made in applying variational techniques to obtaining the ground-state properties of light nuclei~\cite{dumitrescu2018cloud,lu2019simulations,shehab2019toward}, and in algorithms for nuclear-reaction cross sections and neutrino-nucleus response functions~\cite{roggero2020preparation,roggero2019dynamic}. Further progress is needed to optimize and advance such algorithms by applying new developments in simulation algorithms and low-weight fermionic encodings, see Suppl.~\ref{suppl:EFT-Nuclear}. In a similar vein, \emph{ab~initio} many-body calculations based in the underlying few-body hadronic interactions are needed to address questions regarding the composition of the interior of neutron stars~\cite{burgio2021neutron,kruger2013neutron,gandolfi2015neutron,lonardoni2020nuclear}, which is relevant for the analysis gravitational-wave emission from merging neutron stars~\cite{tews2019confronting}. Such calculations are computationally challenging, but are crucial in discerning the role of non-nucleonic degrees of freedom in the equation of state of neutron stars~\cite{vidana2018hyperons,tolos2020strangeness}. Quantum-computing the many-baryon problem is similar in nature to the computations described above for many-nucleon systems. To constrain the many unknown low-energy constants of the effective hadronic description, it is anticipated that the lattice-QCD program in the few-hadron sector could be matched to a nuclear and hypernuclear structure program~\cite{lu2019simulations}, similar to what is currently promoted using conventional tools~\cite{barnea2015effective,contessi2017ground,bansal2018pion}. The output of such efforts will subsequently impact research in HEP, including problems in the intensity and cosmic frontiers.
\vspace{0.5 cm}

\section{Simulator requirements}
\label{sec:simulator}
\noindent
Currently, quantum hardware is being developed in a variety of forms. This includes, on one end of the universality spectrum, special-purpose simulators whose intrinsic or engineered Hamiltonian emulates that of the simulated theory, and on the opposite end, gate-based universal systems with, in principle, the ability to accurately implement any unitary operation. The choice of the suitable architecture, and the mode of operation of the simulator given the simulation problem, impacts the efficiency of the algorithms, and the accuracy achieved with finite resources. It is important to understand the capabilities and limitations of current leading simulating hardware and their prospects for simulating HEP models, how peculiarities of the NISQ era of computing may provide exploratory opportunities and open the door to potential co-design efforts, and what form of software and compiler developments is required for HEP applications in the NISQ era and beyond.

Today, functional analog quantum simulators exist based on atomic, molecular, optical, and solid-state technologies, and are being constantly optimized for programmable quantum simulation. Systems of neutral atoms have been used as analog quantum simulators since the first production of quantum degenerate Bose and Fermi gases~\cite{anderson1995observation,demarco1999onset,ketterle2008making,hadzibabic2002two}. They nowadays come in the form of purpose-built optical lattices with the addition of quantum gas microscope~\cite{bakr2009quantum,sherson2010single,cheuk2015quantum}, allowing very large systems of neutral atoms to be placed in a single many-body state. These systems are can be made in various configurations, with a range of optical lattices with controllable geometry, dynamical couplings, and Floquet engineering~\cite{schafer2020tools} available. Furthermore, cold atoms held in independently movable optical tweezers and driven by laser light into Rydberg states have been developed as a platform for quantum information processing, with successful demonstration of simulating classically intractable quantum many-body systems~\cite{semeghini2021probing,bluvstein2021controlling}. Trapped-ion systems are also among popular quantum-simulation platforms in which atomic ions, that are naturally interacting via long-range Coulomb interactions, are confined through electromagnetic fields, forming 1D or 2D crystals in space. Each atomic ion stores an effective spin with a long idle coherence time. A tunable long-range Ising interaction between all ion pairs is achieved via coupling off-resonantly the spin and the motional degrees of freedom, a feature that had led to many interesting quantum-simulation experiments of spin systems with long-range interactions~\cite{Monroe2021}. Trapped-ion systems can be utilized to simulate three- and multi-spin Hamiltonians as well~\cite{hauke2013quantum,Bermudez2009competing,andrade2021engineering,katz2022n}. They can also operate with qudits where extra internal atomic levels are addressed~\cite{low2020practical,senko2015realization}. Furthermore, analog and hybrid analog-digital implementations are plausible using phonons as dynamical degrees of freedom~\cite{yang2016analog,davoudi2021towards,casanova2011quantum}. Among other notable platforms~\cite{altman2021quantum} with potential prospect for analog simulation of models of relevance to HEP are laser-cooled polar molecules, cavity quantum electrodynamics, superconducting quantum circuits, and dopants in semiconductors such as in Silicon, each presenting unique opportunities but are currently restricted in scalability, controllablity, and/or coherence times, see Suppl.~\ref{suppl:analog}.

HEP simulations, particularly those rooted in QFTs, demand the development of simulators of large scale that exhibit good quantum coherence and high-quality readout, but perhaps more importantly, significant amount of control beyond what has been customary in the quantum simulation of simpler spin systems in the past. Since the analog simulator and target system often do not share the same fundamental symmetries or connectivity of interactions, computing resources are inevitably lost to encoding overheads. Additionally, these restrictions might make it difficult or impossible to engineer certain Hamiltonians. With simulators exhibiting fermionic and bosonic degrees of freedom, and newer modalities involving efficient multiqubit operations, new opportunities to mitigate these overheads will likely arise, and the issues with encoding fermionic statistics and the need for low truncation of bosonic modes may be circumvented. The existing analog simulators are, unfortunately, still far away from presenting the essential capabilities for simulating gauge theories of the SM. A hybrid approach can be explored, where native and more versatile sets of operations are used, but the evolution is digitized to avoid the need for the challenging task of simultaneously applying an increasingly large number of terms in the local Hamiltonian of gauge-field theories. Finally, while trapped-ion and cold-atom platforms have been explored more extensively in the context of quantum simulation of gauge theories, it is not yet known if a range of other platforms will provide unique opportunities for this task. This is a question that only combined theoretical and experimental research can illuminate in the coming decade.

Digital quantum computing implements algorithms as sequences of universal gates on the underlying qubit architecture, based on superconducting systems, cavity QED, neutral atoms, or trapped ions, among other technologies. Current technologies are limited in scale below the 100-qubit level but roadmaps exist for scaling us these system considerably.
Trapped-ion quantum computers feature qubits with long coherence times (minutes)~\cite{Wang2017}, single-qubit with $>99.9999\%$ average fidelity~\cite{Harty2014high} and two-qubit gates with >98\% fidelities~\cite{Pino2021demonstration,Wright2019benchmarking}, and reported readout fidelity of $\sim$99.97\% \cite{Christensen2020high}. Ion chains with tens of functional qubits are achieved in current systems, and importantly, are all mutually connected. This feature greatly reduces circuit depths as the qubits do not need to be swapped and placed in proximity of each other to enable a gate between them. The control of large ion chains are limited by the dense motional (phonon) modes, demanding that ion crystals be broken into spatially separated modules and connected via either via photonic links \cite{monroe2013scaling, Monroe2014} or shuttling~\cite{Pino2021demonstration}. 
Another popular digital quantum-computing platform is superconducting electronics-based Quantum Processing Units (SC QPUs). Being semiconductor systems, SC QPUs can leverage extremely high purity solid-state materials and sophisticated material-processing techniques to produce QPUs with coherence times $\sim 100$’s $\mu$s, high single-qubit, 99.8--99.96\%, and two-qubit, 98\%--99.6\%, gate fidelities~\cite{qubit-aqt}, and to realize chips of varied qubit counts from a few to close to 100.
Imperfections in materials, control systems, QPU design, and electromagnetic environment are among the culprits of QPU operation performance---areas that will be constantly improved. Other emerging platforms, such as highly-scalable neutral atoms, are likely to introduce another candidate architecture for digital quantum computing over the next decades.

New programming paradigms, compilation, and transcription processes will be necessary to understand how best to utilize digital quantum computers in HEP applications, given their enormous qubit and gate requirements, see e.g., initial estimates in Refs.~\cite{jordan2011quantum,byrnes2006simulating,shaw2020quantum,ciavarella2021trailhead,kan2021lattice}. For the purpose of simulating gauge theories of the SM, for example, it may be best to work with the formulations and encodings that retain the locality of interactions, both among bosons and among fermions and bosons, depending on the connectivity pattern inherent to the hardware architecture used, in order to minimize costly swap operations. For the purpose of quantum simulation, any model Hamiltonian with local and semi-local interactions can be decomposed into smaller units of time evolution through the Trotter-Suzuki expansion~\cite{lloyd1996universal} and efficiently decomposed into universal gate operations. However, other simulation algorithms that are costly for present-day hardware may be of value in the fault-tolerant and large-scale era of quantum computing for HEP. While proof-of-concept studies of error-correction with different encodings have been completed on various platforms~\cite{egan2021fault,ryan2021realization,krinner2021realizing,heeres2017implementing,fluhmann2019encoding,nguyen2021demonstration}, no system to date has sufficient resources to meaningfully utilize it in real-world algorithms. Through deep user interaction, co-design, algorithm innovation, and continued improvement in digital-computing hardware, the field is moving toward accelerating the timeline toward universal fault-tolerant quantum computation. As such, HEP physicists need to keep an eye on developing optimal fault-tolerant algorithms, and study the interplay between error-correction protocols and gauge constraints in gauge-theory simulations~\cite{rajput2021quantum}.

NISQ era refers to an era of quantum computing where noisy non-error-corrected operations are performed on devices with $\sim$50--100 qubits~\cite{preskill2018quantum}. It is imperative to not dismiss the possibilities provided by the NISQ hardware, and to take advantage of available devices with various capabilities and capacities, see Suppl.~\ref{suppl:NISQ}. Developing, optimizing, and testing the building blocks of QFT simulations from simpler (Abelian) low-dimensional  models to more complex (non-Abelian) models in higher dimensions is a near-term task that can be started with existing NISQ devices~\cite{martinez2016real,klco2018quantum,nguyen2021digital,kokail2019self,lu2019simulations,alam2021quantum,klco20202,atas20212,ciavarella2021trailhead,ciavarella2021preparation}. In this context, error mitigation, device comparisons, and benchmarking are essential for making progress. The HEP community will continue to develop and apply new NISQ-tailored simulation algorithms in the coming years, and to determine if any quantum advantage can be anticipated in the HEP applications in the NISQ era.

There are a few promising approaches that can produce meaningful results even with shallow circuits and without active error-correction sequences in the NISQ era. These include hybrid algorithms, such as variational quantum eigensolver (VQE)~\cite{peruzzo2014variational,mcclean2016theory} and quantum approximate optimization algorithms (QAOA)~\cite{farhi2014quantum}, that divide classical and quantum resources such that only steps which require probing a large combinatorial space are executed on qubits~\cite{tilly2021variational}. Many early demonstrations of these approaches arose in the context of lattice gauge theories~\cite{klco2018quantum,lu2019simulations,kokail2019self,atas20212}. Quantum annealers, that are non-universal quantum optimizer machines~\cite{johnson2011quantum}, have also proven useful in this context, with recent implementation of a number of quantum-simulation problems of importance to HEP~\cite{ARahman:2021ktn,Illa:2022jqb}. Another approach involves increasing the size of the operational Hilbert space by using three- and higher-level systems in place of qubits. On another front, in the NISQ era of computing, one could encode the portion the problem that is harder to compute with classical methods on a quantum processor (e.g., real-time dynamics), and combine the results when possible with the components that can be evaluated with more ease on a classical computer (e.g., preparation of certain states). Finally, the translation of a quantum algorithm into specific gates must be done with an eye on minimizing quantum-resource requirements in the NISQ era. For example, the original quantum algorithm for computing scattering amplitudes in a scalar field theory~\cite{jordan2011quantum,jordan2011quantum,jordan2012quantum} is costly for present-day quantum computers, but several improvements have been suggested since then to bring down the cost considerably~\cite{moosavian2019site,klco2020minimally,bauer2021practical}. Furthermore,
strategies for fully or partially removing the redundant gauge degrees of freedom may reduce the encoding overhead, but such formulations can change the local nature of interactions inherent in the original theory and may result in higher gate complexity~\cite{martinez2016real,nguyen2021digital,davoudi2021search,mathur2016lattice}. As a result, careful analysis of quantum-resource costs and the susceptibility to noise will be essential in simulating gauge theories in the NISQ era. 

Error mitigation is another important ingredient to make sense of the measurements in the NISQ era. There are two general types of errors that can affect a quantum simulation: readout errors and the gate errors. Currently, readout errors have error rates between $<$1 to 10\%. The largest gate errors occur in entangling operators, such as the CNOT gate, with a typical error rate slightly below 1\%. However, gate errors accumulate and become dominant for longer circuits. One way the readout errors can be mitigated is by preparing a given state and recording the measurement of this state, hence providing a conversion matrix to apply to any other measurement, a technique that can take advantage of developments in collider physics~\cite{maciejewski2020mitigation}. Gate errors include stochastic and coherent errors. Stochastic errors can be understood as either coherent errors with randomly varying control parameters or as processes that entangle the system with its environment. Coherent errors, i.e., those arising from collective couplings and including unitary noise, can be turned into incoherent errors, i.e., those having a stochastic and hence non-unitary nature, via certain methods~\cite{wallman2016noise,li2017efficient,cai2019constructing,cai2020mitigating}. A common method to reduce stochastic gate errors is zero-noise extrapolation, which measures a given circuit at different noise levels (by changing the entangling-gate operation time or by replacing each gate or unitary block by a larger number of gates/operations), and then extrapolates it to zero noise~\cite{li2017efficient,temme2017error,kandala2019error,klco2018quantum,he2020zero,sun2021mitigating,bauer2021computationally}. Among other noise-mitigation proposals~\cite{endo2018practical,endo2019mitigating,otten2019recovering,otten2019accounting,strikis2021learning,urbanek2021mitigating}, one with special relevance to HEP applications is to use the symmetries of the simulated system, such as gauge invariance, to detect and mitigate errors~\cite{bonet2018low,otten2019noise, raychowdhury2020solving}. Further noise-suppression techniques at the level of the hardware, and noise-mitigation schemes at the level of algorithms, may emerge from communications and collaborations among the HEP and QIS communities in the coming decade(s), given the high-energy physicists' expertise in suppressing and mitigating errors and background noise in precision experimental settings.

Last but not least, there is a need for expanding the programmability of quantum-computing devices for testing new and creative quantum algorithms. A variety of low-level programming languages akin to assembler-level programming models in classical computing have been developed over the past decade. Transpilers and early low-level compilers are being developed to optimize the user generated operations---generally referred to as circuits---into a shorter and more efficient set that can be translated to pulses and other fundamental operations on quantum hardware. Many of these tools also incorporate hardware-specific knowledge, such as topologies and error rates, to deliver the best performance possible. To ease the programmability of quantum computers, higher-level programming frameworks will be needed. Within the Department of Energy (DOE), efforts are underway in the Office of Advanced Scientific Computing Research (ASCR) to develop a software toolkit that includes higher-level programming models~\cite{nguyen2021composable}, but no such programming models suitable for HEP are currently available, and must be developed in the coming decade.

To advance HEP applications, and enable the broader HEP community to readily participate, models and approaches for programming quantum computers need to mature to a level in which abstractions and library-based methods can be adopted to expedite programming and ensure portability. This will require the development of programming languages, potentially domain-specific languages, that can readily express the discretization and complex interactions, such as the need to describe the coupling of fermions and bosons and open quantum systems, provides key building blocks, such as non-trivial arithmetic subroutines that are prevalent in quantum simulating the dynamics of non-Abelian lattice gauge theories, or large (sparse) matrix inversion for expediting Monte-Carlo-based routines present in the conventional lattice-gauge-theory program. In addition to programming languages and libraries, the HEP research community would benefit from efficient compilers that can take the programs written in high-level programming languages to efficient low-level code that can be run efficiently on quantum-computing hardware. Open-source software and tools, such as those that enable debugging and validating quantum-computer results, will be needed. One could also utilize quantum computers as well-controlled analog systems. This requires the users to effectively engage across the quantum-computing stack, and to have access to software layers that generate pulses needed to drive the interactions of relevance to HEP applications.

\vspace{0.5 cm}

\section{Quantum ecosystem}
\label{sec:ecosystem}
\noindent
While it is in principle possible to construct gate-based quantum-computing algorithms without detailed knowledge of the underlying hardware, the resource constraints of today’s systems imply severe limitations on the scale of computations, and imply a need for hardware-aware efficient implementations and designs that take advantage of a detailed understanding of system interactions and connectivity, as well as error and decoherence mechanisms. This favors tightly-coupled collaborative models for executing science programs, in which hardware providers and domain experts in HEP applications co-develop scientific agendas. Among successful examples in the present day are the DOE-funded Advanced Quantum Testbed (AQT)~\cite{aqt} at the Lawrence Berkeley National Laboratory (LBNL) and the Superconducting Quantum Materials and Systems Center (SQMS)at Fermi National Accelerator Laboratory (FNAL), where internal and external collaborators as well as users are engaged in developing quantum simulators in tight collaboration with scientists developing algorithms. In addition to DOE programs at the national laboratories, several commercial hardware service providers now offer in-house services to help design efficient hardware-cognizant implementations on demand~\cite{IBM,Quera,Xanadu,QCI,Dwave}. This will remove the wall between the users and cloud-based services with preset features, and will allow the needs of the domain scientists to be communicated directly to the hardware and software developers. At the university setting, experimental efforts have long been in harmony with theoretical efforts in quantum simulation at the same or nearby institutions, and such a model will potentially be the key to success in arrangements outside academia as well. For HEP applications, resources and opportunities in co-design efforts will need to be directed toward systematic evaluation of the best methods to simulate Hamiltonians of relevance to HEP efficiently and accurately, design and perform state preparation and tomography in strongly-interacting QFTs, understand the role of decoherence mechanisms, and how best to incorporate symmetries and gauge and scale invariance, or take advantage of them to mitigate or correct the errors. 

This co-design requirement also motivates the need for accessible devices  that HEP and other domain scientists can use to benchmark algorithms and test new ideas at ease. Therefore, it is important for the HEP community to identify optimal ways to have access to state-of-the-art programmable quantum computers or quantum-simulation experiments. In the coming years, the ability to access multiple platforms with a variety of architectures would lead to rapid progress on target problems in HEP, as different platforms may be suitable for different problems. Furthermore, it is often necessary to run relatively large simulations on classical computers, as numerical tests and benchmarks in the quantum-simulation problems are a necessity. It is conceivable that these computations start to require non-negligible time at High Performance Computing facilities, and this requirement should be recognized ahead of time. One successful model for how accessibility of resources in HEP can be ensured is the successful USQCD-collaboration model~\cite{usqcd}. One could similarly envisage a meta collaboration of US HEP-QIS theorists whose main goal would be to strengthen individual efforts, and coordinate access to different hardware platforms available at national laboratories or commercial sites. It would also facilitate communications between researchers developing quantum-computing hardware and the theorists developing algorithms and software for HEP applications. Similar to the USQCD model, such a meta collaboration can bring up an algorithm-development effort feeding into independent and basically competing individual groups. Direct and broad access to leading hardware and software technologies,  via a community-approved mechanism that can unify, advocate for, and coordinate the accessibility needs of HEP scientists may, therefore, be an important component of a future quantum ecosystem in HEP.

Since quantum simulation for HEP requires a considerably different and much more multidisciplinary skill set than other established HEP areas, collaborations with other disciplines such as atomic-molecular-optical physics, solid-state and condensed-matter physics, computer science, and electrical and material engineering are anticipated to achieve the QIS goals relevant to the HEP science mission. This is because the types of problems that require quantum simulation in HEP are similar in form to those in the other domain sciences. Furthermore, the underlying simulating hardware is itself a physical system which could be taken advantage of in co-designing algorithms and protocols. It is also the case that the techniques that are being developed for the simulation of QFTs may be of relevance for quantum sensing and quantum communications. It would, therefore, be useful for funding agencies to investigate ways for funding to flow across the different domain-sciences areas to flourish essential coordinations and collaborations. The formation of the National QIS Research Centers~\cite{NQI} has already stimulated interdisciplinary research, but funding models that can establish long-term centers with permanent scientists will be important to the continuity of the efforts and completion of long-term projects. Ensuring a strong HEP involvement in these centers can ensure that the HEP-specific goals are being met via the proximity and accessibility of hardware and software expertise in QIS.

An important aspect of quantum simulation is that there exists a significant and  growing expertise  in the private sector. In order for HEP to be at the forefront of quantum simulation, engagement with technology companies is critical, and will be mutually beneficial. However, all developments of importance to HEP research from engagements with technology companies need to be future-proofed. That is to say that any advance that enables accomplishing one or more objectives of the HEP mission must be able to reside in the community and not be lost behind an IP barrier if a company decides that this line of research is no longer a priority. Given the large investment of the private sector in the development of quantum computers, it is important to carefully evaluate how the need of the HEP program is served by the devices that will likely be developed by technology companies. Important considerations are that these devices are developed with an eye on problems that have a more direct monetary payoff, such as quantum chemistry. While many of the problems between different science areas are quite similar, the HEP community might have unique needs that might not be fully addressed by the private sector. This would be similar to the situation in the lattice-QCD program, where building dedicated hardware proved to be important to fully realize the science goals, as well as contributed to a development and flourishing of parallel architectures. A careful study over the next decade will be needed to assess the hardware needs of the HEP community, which will inform how many HEP resources will need to be allocated for hardware development. 

Finally but crucially, advancing quantum simulations of processes essential to HEP research objectives requires a diverse and inclusive quantum-ready workforce with skills that extend significantly beyond those traditionally in HEP, making workforce development a key component of the program. The skills required for quantum simulations of HEP processes~\cite{hughes2021assessing} include HEP phenomenology, quantum field theory and quantum mechanics, lattice field theory, high-performance computing, statistical analysis, experimental design and optimization, machine learning and artificial intelligence, software-stack development, quantum- and classical-computing algorithms, quantum-circuit design, implementation and optimization, and more. Integration with quantum-hardware development through HEP co-design efforts will further broaden this skill-set. Developing a skilled workforce has to be achieved through the natural re-alignment of some of the existing HEP workforce, but also through the recruitment and training of junior scientists. The intrinsically interdisciplinary nature of quantum simulation for HEP requires collaborating with expertise in QIS, computer science, applied mathematics, statistics, material science, nuclear physics, and more, with scientists, engineers, and developers residing in national laboratories, universities, and technology companies. Historically, HEP scientists have contributed to, or moved into, other areas of research, such as computing, ``big data", and device fabrication, and have shaped the development of those areas in substantial ways. The same is anticipated in the emerging quantum era.

Education and training programs for the skills mentioned above will have to be distributed between universities, national laboratories, and the private sector. The educational and training pipelines need to be both strengthened and expanded in the area of QIS, and coordination between the different sectors is required, since all have an important role to play. Given the scope of the anticipated quantum ecosystem, the pipeline for quantum education and training should begin even before students enter university. New and creative ways of educating and training junior, mid-career, and senior scientists, engineers, and developers need be encouraged, keeping an eye on inclusivity and diversity so that under-represented sectors of the population have unimpeded and equal access to QIS education and training. It is also crucial to ensure retention of talent such that a well-trained new generation of scientists in quantum simulation for HEP have attractive permanent positions to look forward to, since otherwise they will leave the field for lucrative positions in the private sector. This can come through faculty positions at universities, permanent scientist positions at national laboratories, and ideally joint positions between the two, following a successful model in joint faculty/staff positions in nuclear physics over the past decades.

\vspace{0.5 cm}
\clearpage

\phantomsection
\addcontentsline{toc}{section}{---Supplemental Material---}
\section*{------------------------------Supplemental Material------------------------------}
\noindent
\clearpage

\renewcommand*\thesection{\arabic{section}}
\definecolor{Blue2}{HTML}{0044FF}

\titleformat{\section}[hang]{\large\bfseries\sffamily}%
{\rlap{\color{Blue2}\rule[-9pt]{\textwidth}{1.2pt}}\colorbox{Blue2}{%
           \raisebox{0pt}[13pt][6pt]{ \makebox[60pt]{
                \fontfamily{phv}\selectfont\color{white}{\thesection}}
            }}}%
{15pt}%
{ \color{Blue2}#1
}
\titlespacing*{\section}{0pt}{3mm}{5mm}

\appendix


\renewcommand{\thesection}{\Roman{section}}
\section{
Physics drive: Collider phenomenology}
\label{suppl:collider}
\noindent
Probing the Standard Model at the highest possible energies is usually achieved using high-energy particle colliders. In such experiments, two initial-state particles (typically electrons, positrons, protons or anti-protons) are accelerated to very high energies and then collide with one another. In this collision, the initial-state particles can scatter inelastically, and the large kinetic energy in the colliding particles can be used to create intermediate, massive particles. By studying the decay products of these collisions, one can infer what kind of intermediate particles were produced. A main use of particle colliders is to look for deviations from the SM predictions, which are expected at some scale due to the inability of the SM to describe well-established  facts about nature, such as the existence of dark matter and the matter-antimatter asymmetry. 

Particle collisions are notoriously difficult to describe. On one hand, this is due to the fact that particle collisions are governed by physics at widely different length scales. While particle collisions are often aimed at discovering processes happening at the highest energies (shortest distances), the observed distribution of final states is affected by physics at energies ranging from the large kinetic energy in the colliding partons (quark and gluon constituents) all the way to low energies describing the binding of partons into hadrons. On the other hand, high-energy collisions typically give rise to a large number of final-state partons (even before hadronization and decay), making them too complicated to be calculable using perturbative techniques based on e.g., Feynman diagrams.

For these reasons, particle collisions are nowadays described theoretically using various types of approximations, each valid for a certain energy range in the process. Processes at the highest energies typically involve only a small set of final-state particles, allowing a perturbative evaluation of the full quantum-mechanical amplitudes. The production of the large number of additional partons is traditionally described by a parton-shower algorithm~\cite{sjostrand2006pythia,corcella2001herwig,gleisberg2009event}, which is based on classical emission probabilities rooted in a collinear approximation in the limit of infinite number of colors ($N_C \to \infty$). Finally, one uses phenomenological models to describe how the resulting partons hadronize to form color neutral hadrons. The parton-shower algorithm is typically combined with some hadronization model to allow so-called exclusive event generators, which take a partonic state produced in a short-distance process and turns it into a fully exclusive final state containing only stable or long-lived particles which can be observed in a particle detector. For a review and discussion on event generation in particle physics, see Refs.~\cite{particle2020review, campbell2022event}. The short-distance process is computed to a given order in perturbation theory. Much work has been devoted to calculate short-distance processes as precisely as possible, and many processes are available at the second order in perturbation theory, and some even at the third order. For a recent review, see Ref.~\cite{gehrmann2021precision} and references therein. Care needs to also be taken to avoid double counting when combining parton-shower algorithms with short-distance calculations at higher orders in perturbation theory. While the combination of these ingredients  allows to simulate fully-exclusive scattering processes, the presence of the various approximations made imply that the resulting distributions can only be trusted for certain observables, and it is difficult to estimate the uncertainty in the obtained predictions. 

In order to reduce the sensitivity to the details of the modeling of hadronization effects and to the approximations made in the parton shower, comparisons of the obtained simulations to experimental measurements are limited to observables that are ``sufficiently inclusive''. While parton showers allow one to calculate less inclusive observables, the results are much more dependent on the particular choices made in the modeling, and therefore, will have significant uncertainties. 

Quantum computers hold the promise to simulate scattering processes from first principles, and in principle, without any uncontrolled approximations. The basic idea, which is explained in more detail in Suppl.~\ref{suppl:QFT}, is to discretize the continuous spatial dimensions and to digitize the continuous field values~\cite{kogut1979introduction,kogut1983lattice,kogut1975hamiltonian}. This turns the infinite-dimensional Hilbert space of the field theory describing the SM into a finite-dimensional one, which can be described by the rules of regular quantum mechanics. Since the Hilbert space is exponential in the number of lattice points required, it is far too large to allow such a computation using classical computers. However, it is believed that all relevant ingredients required  to compute the S-matrix can be calculated on a quantum computer using resources that scale only polynomially with the number of lattice sites~\cite{jordan2012quantum}. This was demonstrated explicitly for a scalar field theory~\cite{jordan2011quantum}, but is believed to also remain true for more complicated field theories, such as gauge theories of the SM. 

As an alternative to the simulation of the full S-matrix using quantum algorithms, one can also attempt to devise quantum algorithms for the parton shower and the hadronization process. The idea of a quantum parton shower is to still work in the collinear approximation that underlies a classical parton shower, but to include quantum interference effects that are not possible in a traditional approach using classical probability distributions. For example, a quantum parton shower was shown to be able to include quantum interference effects arising from amplitudes with the same final-state particles, but different intermediate-particle flavors~\cite{nachman2021quantum}. One can also hope to go beyond the $N_c \to \infty$ limit of regular parton shower (with $N_c$ being the number of colors), including the quantum interference between different color structures. 

Another approach that might be suitable for collider physics is to simulate the physics at a particular energy scale on a quantum computer, while maintaining more traditional approaches for physics at other scales. For example, one could try to develop quantum algorithms for hadronization processes, which might allow one to go beyond the relatively simple models used in traditional approaches. However, in order for such an approach to be meaningful, a proper separation of the various energy scales present in a collider process is required. This is typically achieved using EFTs, as discussed in Suppl~\ref{suppl:EFT}.
\vspace{0.5 cm}

\section{Physics drive: Matter in and out of equilibrium}
\label{suppl:matter}
\noindent
Many open problems in HEP deal with the behavior of strongly-coupled fermionic matter at high density or far from equilibrium. In the coming decade, results from  heavy-ion and proton collisions at the Relativistic Heavy Ion Collider (RHIC) and the Large Hadron Collider (LHC) as well as gravitational-wave searches from the Laser Interferometer Gravitational-Wave Observatory (LIGO) will lead to an unprecedented level of experimental data probing strongly-interacting matter. Because of the strong coupling, one is typically forced to rely on the numerical method of lattice QCD for first-principles predictions of the state of matter. However, both in finite-density systems and for real-time dynamics, the use of familiar Monte Carlo approach to lattice gauge theories is hampered by sign problems~\cite{troyer2005computational,alexandru2016monte}. The quantum simulation of strongly-interacting matter in and out of equilibrium holds the promise of avoiding such problems and can be performed efficiently, i.e., with resources that scale only polynomially in the size of the system by encoding the entanglement of quantum states~\cite{temme2011quantum,desai2021z3}. In finite-density systems, such simulations should lead to precision theoretical results for the QCD equation of state and the behavior of phase transitions in strongly-interacting matter~\cite{czajka2021quantum}. Beyond their intrinsic value for better understanding QCD, the current theoretical uncertainty on these predictions are anticipated to be the limiting factor in precision physics and searches for new physics in the coming decade. 

One area of real-time dynamics accessible to quantum computers is the direct simulation of the collision of leptons, hadrons, and nuclei, thus ``solving'' hadronization in the sense of providing quantitative, testable predictions, as discussed in Suppl.~\ref{suppl:collider}. Current resource estimates suggest such computations require millions of logical qubits and thus represent targets on a longer time scale. Instead, one can study these collision experiments with phenomenological models composed of effective theories. A quantum computer provides a non-perturbative calculation of the low-energy or long-distance observables in effective theories with less resources compared to a quantum simulation involving the entire collision. Examples of such observables are diffusion coefficients, exclusive decay rates~\cite{ciavarella2020algorithm}, parton distribution functions~\cite{lamm2020parton,kreshchuk2020quantum,echevarria2021quantum,kreshchuk2021simulating,li2021partonic,perez2021determining}, hadronic tensor~\cite{lamm2020parton}, transport coefficient~\cite{cohen2021quantum}, and jet functions. These observables require substantially fewer quantum resources because they avoid manipulating late-time asymptotic states, and therefore, represent interesting targets for simulation in the coming decade.  Current estimates suggest that evaluation of the low-order transport coefficients for use in relativistic hydrodynamics require the smallest quantum resources, but the qubit and gate estimates are still substantial for QCD~\cite{cohen2021quantum,kan2021lattice}. Obtaining such results for phenomenological models such as the $2+1$ dimensional quantum Ising model will be a physically interesting near-term target of quantum computation.  Other interesting questions in the development of quantum algorithms are the effect of thermal fluctuations in hydrodynamics and finite-volume effects in the quantum simulation of transport coefficients. With even fewer quantum resources, the dynamics of confinement and string breaking can be investigated in low-dimensional models~\cite{banerjee2012atomic,verdel2020real,nachman2021quantum} to improve phenomenological models of parton showers~\cite{hoche2016introduction} and hadronization, e.g the Lund model~\cite{andersson1983parton}.

Beyond computing non-perturbative inputs, the ability of quantum simulations to perform real-time evolution will provide insight into the approach to equilibrium in strongly-interacting systems~\cite{riera2012thermalization}. There is tantalizing evidence from heavy-ion and proton collision experiments at the RHIC and LHC that the strongly-interacting matter thermalizes in a remarkably short period of time, a fraction of a Fermi. This runs counter to naive scaling arguments and suggests that our current understanding of the dynamics of strongly-interacting systems in their approach to equilibrium is missing fundamental insights. One possible solution to this puzzle is that the dynamics of quantum entanglement plays a role in the equilibration process~\cite{kharzeev2005color,kharzeev2017deep,baker2018thermal,berges2021qcd}. Indeed, it has been proposed some time ago that a quench in an entangled system can lead to apparently thermal behavior if only a part of the system is observed~\cite{deutsch1991quantum,srednicki1994chaos}. Studying such behavior in large, far-from-equilibrium strongly-interacting quantum matter is prohibitively difficult with classical computers. Therefore, the use of quantum computing represents a necessary tool in order to study the real-time evolution of entanglement in strongly interacting matter.

The insights from such research would enhance our understanding of the role of quantum information in HEP. In particular, it is known that entanglement spectrum~\cite{li2008entanglement}, that is the spectrum of eigenvalues of the (negative of the logarithm of) reduced density matrix, contains more complete information about the system than the entanglement entropy~\cite{amico2008entanglement,eisert2010colloquium}. The distribution of level spacing in the entanglement spectrum can reveal whether and how the system thermalizes, and the evolution of entanglement-spectral gaps can signal phase transitions~\cite{geraedts2016many, kaufman2016quantum, yang2015two,Mueller:2021gxd}. Furthermore, elastic and inelastic processes are shown to represent different rates of entanglement spreading in the final state~\cite{milsted2020collisions}, as do the evolution of confined vs. deconfined phases of matter after a quench~\cite{pichler2016real}. While it is only recently that thermalization and associated questions in gauge theories have started to be explored~\cite{Mueller:2021gxd,halimeh2022robust,banerjee2021quantum,brenes2018many,dalmonte2022entanglement} using entanglement measures in classical and quantum Hamiltonian-simulation studies, it is conceivable that such explorations will gain a considerable boost as large-scale programmable simulators become available~\cite{kokail2021entanglement,pichler2016measurement,yirka2021qubit,kokail2021quantum}. Quench experiments may reveal interesting out-of-equilibrium features of the physical system~\cite{mitra2018quantum,canovi2011quantum} while being relatively straightforward to set up, as demonstrated in quantum-simulation experiments of spin systems that start to push the limits of classical computing~\cite{bernien2017probing,zhang2017observation}. Nonetheless, entanglement spectroscopy could be relatively costly and recent ideas in shadow tomography and related protocols~\cite{aaronson2019shadow,huang2020predicting} may provide economical ways to learn about entanglement structure of the final state, see e.g., Refs.~\cite{kokail2021entanglement,kokail2021quantum}.

\vspace{0.5 cm}

\section{Physics drive: Neutrino (astro)physics}
\label{suppl:neutrino}
\noindent

In certain astrophysical environments, such as core-collapse supernovae and neutron-star mergers, the very large number of neutrinos present require taking neutrino-neutrino interactions, both within the Standard Model and beyond, into account. Proper description of neutrino propagation in such environments includes the effects of neutrino mixing, forward scattering from the background particles, i.e., the Mikheyev–Smirnov–Wolfenstein (MSW) effect, forward scattering of neutrinos off each other and other collisions of neutrinos. The first three processes can give rise to collective neutrino oscillations which have been shown to potentially have an important impact in supernova environment, both in the neutrino-driven explosion mechanism and in the ensuing nucleosynthesis in the ejected material~\cite{qian1993connection,qian1995neutrino,fogli2007collective}.

Correlations caused by neutrino-neutrino scattering processes result in a full many-body problem and an exact solution to the dynamical evolution of flavor from a general initial configuration will, therefore, require a computational cost which is exponential in the number of neutrinos involved. A peculiar property of the neutrino-neutrino interactions, arising from the point-like nature of weak interactions in coordinate space, is that they are extremely long ranged when represented in momentum space, resulting in large numbers of neutrinos' momentum modes to be coupled together.

A simple approach would be to use a mean-field approximation in which a test neutrino interacts with a ``mean field'' representing the influence of all the other neutrinos. Thanks to the ``infinite'' range of the neutrino-neutrino interaction in momentum space, this approximation can be rigorously justified when computing equilibrium properties at low energies (see, e.g., Ref.~\cite{brandao2016product}) but its validity in a more general out-of-equilibrium setting is yet not completely understood. Nevertheless, the mean-field approach has been extensively utilized, revealing a large variety of interesting physics such as synchronization, splits in the neutrino energy spectra, and early-stage fast flavor oscillations. An important direction of current research is to clarify the range of validity of this approximation and to understand whether there exist situations where neutrino-neutrino correlation could change the mean-field predictions in a qualitative way. 

QIS tools could be utilized in the study of neutrino astrophysics in two ways: One is using quantum computing to simulate the propagation of neutrinos, in particular collective neutrino oscillations. Indeed early attempts with available digital quantum computers are encouraging~\cite{hall2021simulation,yeter2021collective}. However, a full digital simulation of the quantum many-neutrino systems is still in the far future. Due to the simple structure of the neutrino-neutrino Hamiltonian, it is also possible, at least for simple geometries and energy distributions, to map a neutrino system directly into the degrees of freedom and interactions of quantum simulators. Trapped-ion quantum devices are an ideal candidate for these near-term explorations due to the possibility of exactly mapping the neutrino interactions (see, e.g., Refs.~\cite{davoudi2020towards,bermudez2017long}). Further development of quantum simulators to use qudits as fundamental degrees of freedom will also open the possibility of extending these studies beyond the simple two-flavor approximation.

In the near term, a more immediate application of QIS is to use tools such as entanglement measures to assess the limits of applicability of the mean-field approximation. Since in the mean-field approximation, the entropy of entanglement for each neutrino vanishes, obtaining non-zero values of entanglement entropy in exactly-solvable simplified models would suggest how the mean-field approximation may be improved~\cite{cervia2019entanglement}. Indeed recent work shows that away from spectral split energies, mean field may be a good description, suggesting a hybrid approach where many-body effects are explored for neutrinos with energies around the spectral split energies~ \cite{patwardhan2021spectral}. It is also important to understand what observables would be affected by quantum correlations, since the presence of entanglement in the many-body neutrino state does not necessarily imply an error in simple observables like the individual flavor polarizations~\cite{martin2021classical}. A better understanding of the evolution of entanglement with the size of the neutrino system being simulated is also critical to understand the crossover between the semi-classical regime, which can be explored using, for example, tensor-network methods capable of describing weakly-entangled many-neutrino states and a fully-quantum regime requiring full-scale quantum simulations (see, e.g., Refs.~\cite{roggero2021entanglement,roggero2021dynamical,cervia2022collective}).

The measurement of fundamental properties of neutrinos such as their absolute masses, mixing angles, and the presence of CP violation is also a major goal of the HEP community both in the U.S. and worldwide, and large-scale experiments, such as the Deep Underground Neutrino Experiment, have been commissioned to achieve this goal. Besides its importance at a fundamental level, an accurate determination of these parameters would also inform simulation of neutrino interactions in astrophysical settings. In order to extract neutrino properties, long-baseline experiments rely on accurate determination of neutrino cross sections with the material used in their detector. For the DUNE experiment, as well as the currently-operating ones like MicroBoone, this is the $^{40}$Ar nucleus. A full description of the neutrino-Argon scattering cross section is a formidable theoretical challenge due to the wide energy range required for the analysis of the experiment, the necessity to have information about semi-exclusive processes (such as neutron emission), as well as the open-shell structure of the $^{40}$Ar nucleus which makes it challenging to describe accurately the properties of the target. 

Classical techniques including Quantum Monte Carlo, Coupled Cluster and various Green's Function based schemes are currently being extended to address in part this challenge (see~Ref.\cite{rocco2020ab} for a recent review of these theoretical efforts). Future quantum simulations can help these efforts by allowing for both more efficient representations of the nuclear target's wavefunction, and the extraction of both inclusive and semi-exclusive reaction cross sections with controllable uncertainties~\cite{roggero2019dynamic,roggero2020quantum}. Early attempts with current digital quantum devices are encouraging (see, e.g., Refs.~\cite{dumitrescu2018cloud,lu2019simulations,roggero2020quantum,stetcu2021variational,baroni2021nuclear}) but to achieve the required accuracy, large-scale error-corrected quantum simulators are likely required. In the near term, it will then be important, together with improving the scalability of accurate algorithms, to understand in more detail what type of semi-exclusive data would be available by quantum simulations and how to integrate this information in the event-generator codes employed to analyze the experiment (see, e.g., Refs.~\cite{isaacson2021novel, campbell2022event} for recent work in this direction for classical simulations). Another important aspect that merits consideration in the future is how to properly account for relativistic corrections which become important at large energy/momentum transfer (see, e.g., Ref.\cite{rocco2018relativistic}). These directions will likely form the basis of an active research in quantum simulations of relevance to the neutrino program.
\vspace{0.5 cm}

\section{Physics drive: Cosmology and early universe}
\label{suppl:cosmology}
\noindent
The universe is inherently quantum, therefore predictions about earlier epochs of the universe should include quantum effects. Multiple phenomena across cosmology and the early universe, including inflation, baryon asymmetry, phase transitions, and dark matter need to be described by the non-equilibrium dynamics of non-perturbative quantum fields. At their core, these problems demand tools capable of non-perturbatively time evolving the quantum fields, which is a challenging task.  Alas, the only first-principles and systematic method to date, that is lattice field theory, is impeded in this endeavour by seemingly intractable sign problems. To sidestep this obstacle, state-of-the-art calculations assume adiabatic or near-equilibrium evolution, and/or perturbative field theory. Such approximations may receive large corrections from far-from-equilibrium or non-perturbative effects that are difficult to quantify. Obtaining these corrections or simulating \emph{ab initio} such phenomena constitute opportunities for practical quantum advantage. Alongside the development of quantum hardware, theoretical developments are required in connecting classical- or perturbative-physics intuition with well-defined renormalized matrix elements and algorithms to compute them. It is anticipated that addressing the problems discussed below with quantum hardware will, in the long run, change our understanding of the early universe. 

In the inflationary paradigm, the universe experienced a period of accelerated expansion brought on by quantum fluctuations, evolving as nearly classical fields, before terminating by non-perturbatively transferring energy to particle degrees of freedom through \emph{reheating} and \emph{preheating}. These far-from-equilibrium and non-perturbative processes could leave imprints on the sky today~\cite{Kofman:1994rk,kofman1997towards,greene1997structure}. While in the long term, large-scale quantum simulations of quantum inflationary fields are desired~\cite{chatrchyan2021analog, liu2021quantum,butera2021particle,barata2021single}, near-term studies could improve calculations of the quantum back-reactions~\cite{tsamis1997quantum} onto classical inflation fields and scalar/tensor perturbations directly from non-perturbative quantum effects. Further near-term opportunities exist in using analog quantum devices to simulate the dynamics of reheating with ultracold Bose gases~\cite{chatrchyan2021analog2}.

Generating the observed baryon asymmetry requires non-equilibrium dynamics~\cite{Sakharov:1967dj}. Potential sources of this non-equilibrium behavior include heavy-particle decays~\cite{weinberg1979cosmological}, the Affleck-Dine mechanism~\cite{affleck1985new}, and first-order phase transitions~\cite{quiros1998finite,hindmarsh2021phase}. Standard treatments rely upon the dynamics when the interaction rate is close to the Hubble rate, which is essentially assuming near-equilibrium behavior~\cite{wainwright2012cosmotransitions}. Condensed-matter studies suggest that such simplistic comparisons can be insufficient to ensure adiabatic evolution~\cite{polkovnikov2008breakdown}. The state-of-the-art perturbative calculations of the effective potential at finite temperature may also suffer from non-negligible higher-order corrections~\cite{curtin2018thermal}. For a first-order phase transition, interesting phenomena such as particle production and gravitational-waves generation are yet to be fully understood. Serving as inputs to the time evolution of classical fields, the properties of nucleating bubbles are usually extracted from the effective potential. As recently stressed in the literature~\cite{baum2021nucleation}, the phase-transition history defined solely by the conventional critical-temperature calculation can be misleading, and it is necessary to perform more detailed calculations of the nucleation mechanism. Early quantum research should focus on rigorously defining matrix elements for bubble properties and exploratory calculations of them~\cite{lagnese2021false, pirvu2022bubble,milsted2020collisions}. Additionally, the interplay between phase transitions and curved spacetime can be investigated with the NISQ-era simulators to manifest non-equilibrium behavior in low-dimensional toy models~\cite{neuenhahn2015quantum,arrighi2018quantum,rossini2020dynamics,boettcher2020quantum}. Studies of bubble nucleation and the growth of entanglement with energy and with the number of collisions have been carried out using non-integrable Ising spin-chains systems \cite{milsted2020collisions}. Given the limited digital quantum resources, analog quantum simulations of similar processes~\cite{ng2021fate} would also be beneficial. Moving beyond the near-equilibrium phase transitions, the quantum devices will also allow for the study of dynamical phase transitions which arise in non-equilibrium statistical mechanics~\cite{xu2019probing,nie2019experimental,Heyl_2019}. 

Another mystery of the early universe is what accounts for around 85\% of the matter observed only via its gravitational effects. Light dark matter such as axions typically require a more complex and non-equilibrium history to be produced. For example, the strong-CP $\theta$ information and the temperature-dependent axion mass~\cite{berkowitz2015lattice} are required to predict the relic abundance from misalignment mechanism. To extract the mass information, QCD free energy as a function of the CP-violating phase and temperature has to be calculated. However, due to the presence of the CP-violating phase, classical calculations suffer from a sign problem. Though dilute instanton gas model (DIGM) at high temperature or the interacting instanton liquid model (IILM) around the QCD phase transition has been explored in detail~\cite{wantz2010axion,turner1986cosmic}, it is unclear to what extent the DIGM and IILM are valid and how to control their uncertainties. Quantum computers could be used to compute the QCD free energy at finite temperature with a finite $\theta$-term ~\cite{chakraborty2020digital,kan2021investigating}. With real-time simulations of the early universe, quantum computers can also simulate the non-equilibrium histories of light dark matter as well as reduce the systematic uncertainties involved. Quantum simulation may allow probing the dynamics of other types of dark-matter-like topological defects~\cite{vilenkin1994cosmic}, or primordial black holes~\cite{zel1967hypothesis} that are difficult to analyze because of their relation to the strong-field theory.

\vspace{0.5 cm}

\section{Physics drive: Nonperturbative quantum gravity}
\label{suppl:gravity}
\noindent
Constructing a complete and convincing quantum theory of gravity is a grand challenge facing fundamental physics. There are compelling reasons to pursue this quest. We crave a deeper understanding of the fundamental laws of nature. We hope to resolve long-standing puzzles about what happens inside black holes, and about how black holes process information. We desire general principles that can explain the initial conditions in the history of the universe. We anticipate that progress in quantum gravity will teach us broader lessons applicable to other areas of physics, and finally, studying quantum gravity is a fun intellectual endeavor! 

Quantum gravitational phenomena are so elusive that one is forced to develop the theory with limited guidance from experiments. It takes hubris even to try, and yet, remarkable progress is being achieved, particularly in the past few years. Many valuable lessons flow from the discovery of AdS/CFT duality twenty four years ago~\cite{maldacena1999large}, augmented fifteen years ago by the realization that, in the context of AdS/CFT, bulk geometry emerges from boundary entanglement~\cite{ryu2006holographic}. In a sense, quantum entanglement is what holds space together. 

Progress has been fueled by melding insights from the holographic correspondence with ideas from QIS. It is discovered that the dictionary mapping bulk to boundary physics can be viewed as the encoding map of a quantum error-correcting code~\cite{almheiri2015bulk,pastawski2015holographic}. Among other consequences, this viewpoint has sharpened our understanding of why exact global symmetries are disallowed in bulk quantum gravity~\cite{harlow2019constraints}. We have learned that black holes are the most efficient possible scramblers of quantum information~\cite{hayden2007black,sekino2008fast}, and have leveraged that insight to deepen our understanding of quantum chaos more broadly~\cite{Maldacena2016bound}. 

It is further seen that computational complexity of a boundary theory can be related to geometrical properties in the bulk~\cite{susskind2016computational,stanford2014complexity,brown2016holographic}, and that a surprisingly simple quantum system can have a holographic dual which is helpful for understanding the system's behavior~\cite{Sachdev1993gapless,Kitaev_talk}. More recently, by extending the connection between geometry and entanglement to include quantum effects in the bulk~\cite{engelhardt2015quantum}, it is learnt how to compute the so-called Page curve, which describes how quantum information escapes as a black hole evaporates~\cite{penington2020entanglement,almheiri2019entropy}. Remarkably, thanks to the discovery of replica wormhole contributions to the Euclidean path integral~\cite{penington2019replica,almheiri2021entropy}, semiclassical computations validate the unitarity of black-hole evaporation, without invoking any explicit description of the black hole's microscopic degrees of freedom.  

Despite this encouraging progress, much is still missing from our current understanding of quantum gravity. While the Euclidean path integral seems to be a surprisingly powerful tool, it is not known how to formulate it precisely in a theory of gravity. The knowledge about the quantum code relating bulk and boundary degrees of freedom is yet incomplete. We do not fully understand why bulk physics is local on distance scales small compared to the AdS curvature scale, which is related to our unsatisfying grasp of how quantum gravity works in asymptotically flat spacetime or in de Sitter space. It is not yet known what happens at the singularity inside a black hole, or even how to describe the experience of someone who falls through the event horizon. The computational complexity of the dictionary that maps the region deep inside a black hole to the region outside cannot yet be characterized with confidence. Finally, there does not exist a systematic way to identify quantum systems that admit useful holographic dual descriptions. 

Quantum simulations, in both the near term and the longer term, can help to fill these gaps in our current understanding. Thanks to holographic duality, bulk gravitational phenomena can be described in a completely different language that does not involve gravity at all. Instead, the ``dual'' boundary quantum system consists of many particles strongly interacting with one another. In principle, this boundary system can be simulated using a quantum computer, opening opportunities for exploring quantum gravity in laboratory experiments. Furthermore, duality is a two-way street: On the one hand, experiments with quantum devices might illuminate properties of quantum gravity that are analytically intractable. On the other hand, by interpreting the behavior of many strongly-interacting particles in terms of gravitational phenomena, one might better understand and control that behavior.

For example, properties of emergent geometry and gravitational back reaction in the bulk can be accessed by exploring the entanglement structure of the dual boundary theory.  A particular challenge is understanding why the bulk gravitational theory is (approximately) local, given that operators which are spacelike separated in the bulk correspond to operators in the boundary dual that act on overlapping regions. Bulk operator commutation relations can, in principle, be studied in the boundary theory via experiments that probe transport or linear response. Quantum corrections to semiclassical gravity, especially non-perturbative ones, are difficult to compute analytically. Eventually, it may be possible to measure such corrections in studies of the dual theory.

To realize this vision of investigating quantum gravity through laboratory experiments, three ingredients are needed: i) A (non-gravitational) quantum system that has a gravitational dual, at least approximately. ii) A proposal for realizing the quantum system in a feasible experiment. iii) A dictionary relating gravitational phenomena of interest to measurable observables in the quantum system. The best understood case is conformally-invariant four-dimensional maximally supersymmetry SU($N$) Yang-Mills theory, such that the bulk curvature radius is large compared to the string scale when the the boundary theory is strongly coupled, and bulk effects higher order in the gravitational constant $G_N$ are suppressed when $N$ is large ~\cite{maldacena1999large}. Admittedly, simulating this boundary theory accurately will require a large-scale fault-tolerant quantum computer which might not be available for decades, but this eventual goal provides one strong incentive (among many) for advancing the tools needed to simulate dynamics in conformal field theories. 

More tractable, but still ambitious, targets with good gravitational duals would be  Banks, Fischler, Shenker, Susskind (BFSS) and Berenstein, Maldacena, Nastase (BMN) matrix models~\cite{banks1997m,itzhaki1998supergravity,berenstein2002strings}. The information of the gravitational geometry is encoded in matrix degrees of freedom~\cite{banks1997m,witten1996bound,hanada2021bulk}. For example, the low-energy states are expected to describe M-theory black holes or black zero-branes~\cite{itzhaki1998supergravity}. Although Monte Carlo simulations on classical computers provided non-trivial test of this conjecture (see e.g., Refs.~\cite{berkowitz2016precision,bergner2021confinement} for state-of-the-art results), the details of the encoding of geometry into matrices have been out of reach. Quantum simulation would provide a practical tool in this problem. For example, a quantum state describing both a black hole and a probe D0-brane can be obtained by constraining some matrix entries~\cite{hanada2021bulk}, and the motion of such a probe should be described by the black-hole geometry created by other degrees of freedom~\cite{maldacena1999large}. Classical machine learning and hybrid quantum-classical algorithms have been successfully applied to simple matrix models~\cite{han2020deep,rinaldi2022matrix}, and quantum simulations including quantum machine learning would enable studies of the full BFSS and BMN models. Real-time evolution associated with the motion of D0-branes in the black-hole geometry, or simpler processes such as the scattering of a small number of D0-branes~\cite{becker1997two,becker1997higher,okawa1999multi}, would provide valuable clues to understand quantum gravity. Quantum entanglement between color degrees of freedom can also be studied, and it may lead to a generalization of Ryu-Takayanagi approach that is based on the splitting of spatial regions. In principle, such approaches that focus on matrix degrees of freedom can also be applied to 4D supersymmetric Yang-Mills theory. Furthermore, various supersymmetric systems including 4D supersymmetric Yang-Mills theory can be regularized using matrix models, hence the quantum simulation of matrix models can be the first step toward quantum simulating supersymmetric QFTs, see Ref.~\cite{gharibyan2021toward} and references therein.

Meanwhile, studies of the Sachdev-Ye-Kitaev (SYK) model ~\cite{Sachdev1993gapless,Kitaev_talk}, describing many fermions with strong all-to-all couplings, have alerted us that more accessible models can have useful gravitational duals worthy of further investigation. This development invites us to contemplate table-top experiments in which quantum information residing in a complex many-particle system is first scrambled and then refocused, a phenomenon best understood in the dual bulk picture as transmission of quantum information through a wormhole in space~\cite{gao2017traversable, maldacena2017diving}. Thus, gravitational intuition may guide our interpretation of dynamics in strongly-coupled many-particle systems even in the relatively near term, especially for systems with long-range couplings~\cite{brown2019quantum,nezami2021quantum,schuster2021many}.  
Realistically, the goal in the near term should be to light the way toward progress in the more distant future. Though spectacular insights into quantum gravity may not be expected in the next ten years, the community can develop tools, methods, and insights that will lay foundations for quantum technology, enabling profound advances in fundamental physics.

\clearpage

\section{Underlying simulations:  Simulating quantum field theories}
\label{suppl:QFT}
\noindent
\addtocontents{toc}{\protect\setcounter{tocdepth}{2}}
\noindent
Quantum field theory is an elegant mathematical framework that combines quantum mechanics and special relativity. The development of gauge field theories through various stages of conceptual and mathematical progress, along with abundant experimental verifications, marked the birth of the Standard Model of particle physics in the 20th century.  Given that nature has chosen gauge theories as the mechanism governing subatomic particles up to length scales probed by experiments to date, physicists continue to believe that such theories, in one form or another, are strong candidates for physics beyond the SM~\cite{csaki2016beyond,willenbrock2014effective}, including that of the dark sector~\cite{petraki2013review,kribs2016review,albert2017towards,urena2019brief}. Furthermore, effective descriptions developed to describe certain limits of the SM, i.e., effective field theories, are a form of QFTs. Moreover, there exist intriguing connections between quantum gravity and conformal field theories. QFTs are, therefore, the backbone of HEP, and any attempt at simulating nature from first-principles amounts to simulating quantum fields and their interactions. Perturbative methods have proven powerful in accurate predictions of the behavior of subatomic particles, e.g., at particle colliders. However, for strong interactions in the low-energy regime, where features such as confinement and hadronization arise, one needs to apply non-perturbative methods as the interaction strength becomes sizable. Furthermore, in systems where electroweak and strong interactions are both in play, such as in hadrons and nuclei, consistent inclusion of both interactions is required when handling the strong interactions non-perturbatively

\subsubsection{Conventional lattice field theory and the case for quantum simulation}
A reliable non-perturbative tool to simulate QFTs is lattice field theory~\cite{wilson2004origins,kogut1983lattice,montvay1997quantum,rothe2012lattice,rebbi1983lattice}, which when applied to the theory of strong force is called lattice quantum chromodynamics (QCD). Lattice QCD is a numerical technique that systematically estimates, via Monte Carlo sampling, the quantum-mechanical correlation functions of hadrons, nuclei, and finite matter from interactions among constituent quarks and gluons. It has led to some of the most impressive computations in theoretical physics, from the determination of muon’s anomalous magnetic moment to studies of light and heavy meson decays for testing the SM of particle physics and searching for violations of fundamental symmetries in nature~\cite{aoki2021flag,lehner2019opportunities,cirigliano2019role}. It has also led to progress in illuminating spectral, structure, and reaction properties of nucleons and light nuclei~\cite{beane2011nuclear,davoudi2018light,detmold2019hadrons,davoudi2021nuclear} to inform astrophysical models and high-energy collider experiments, and in understanding the dilute QCD matter at finite temperatures~\cite{bazavov2019hot,ding2015thermodynamics,detar2009qcd} to shed light on the phases of strongly-interacting matter. Beyond QCD, lattice field theory has been used to explore BSM theories in the non-perturbative regime~\cite{brower2019lattice}.

Beside the exponential growth of the Hilbert space of QCD as a function of system’s size, the statistical nature of the lattice-QCD method means that only a finite sample of (infinite) quantum configurations are produced and processed to estimate expectation values. This approach suffers a significant drawback: with finite statistics, if contribution
s to the system’s partition function with oscillating signs arise, statistical averages cannot be estimated reliably, leading to an infamous sign problem. This means that lattice-QCD calculations are bound to be performed in imaginary time so as to allow a sign-problem-free sampling using an Euclidean action. Moreover, in finite-density systems with a fermionic chemical potential, the probability distribution used in the Monte Carlo sampling of quantum configurations is oscillatory even in Euclidean spacetime and introduces a sign problem. A closely related problem is an exponential signal-to-noise degradation in nuclear correlation functions, challenging precision lattice-QCD calculations of nuclei. This limits the accuracy and the precision of theoretical predictions for a range of HEP experiments that use hadron and nuclei as target.  Such limitations also mean that many fundamental questions regarding equilibrium and non-equilibrium phases of QCD, including the mechanism of thermalization, hydrodynamization, fragmentation, and hadronization in hadron collisions and early universe will remain unexplored, as are a range of other critical questions in HEP, as detailed in this whitepaper. With this understanding, it is essential to seek alternative computational paradigms that approach these problems fundamentally differently.

A primary question is the following: Can a  lattice-field-theory program based on quantum simulation be fully developed to complement and expand the conventional program? To answer this question, one must recall the course of developments in lattice QCD over multiple decades. It consisted of, first of all, formally defining the QCD path integral and observables in a finite discrete spacetime in such a way that as many symmetries as possible are kept, or systematically recovered, in the continuum infinite-volume limit, starting from the pioneering work of K. Wilson~\cite{wilson1974confinement}. It is now proceeding to connect Euclidean finite-volume quantities to Minkowski infinite-volume amplitudes in the few-hadron sector, starting from the pioneering work of M. L\"uscher~\cite{luscher1986volume,luscher1991two}, along with many other theoretical advances. It also consisted of devising algorithms that, over time, scaled better with system’s parameters and took advantage not only of advances in applied mathematics and computer science but importantly of physics input, such as expression of symmetries and constraints, and renormalization group and scale separation, to make seemingly impossible computations possible. Furthermore, it relied on adjusting algorithms and compilations to the hardware architecture, and remarkably in instances, impacted the development of computing architecture itself via a co-design process, see Suppl.~\ref{suppl:codevelopment}.

A quantum-simulation-based lattice-gauge-theory program, similarly, should require developments in all these three areas: (i) theoretical foundation, (ii) algorithmic research, and (iii) hardware awareness, motivating the case for co-design of dedicated QFT simulators.

\subsubsection{Theoretical developments for quantum simulation of QFTs}
The most common framework to compute static and dynamical observables on quantum hardware is the Hamiltonian framework. This is because the unitary time evolution can be naturally implemented on a quantum device, either in a continuous (analog) manner or a digital (gate-based) manner. While Kogut and Susskind laid the ground for a Hamiltonian formulation of lattice gauge theories in the 1970s~\cite{kogut1975hamiltonian}, for the sake of quantum simulation, more considerations are in play. The infinite-dimensional Hilbert space of gauge bosons must be truncated and one may wonder whether electric-field, magnetic-field, or some dual representation will lead to faster convergence to the exact theory toward the continuum limit. Furthermore, local Gauss’s laws must be imposed on the Hilbert space (or more generally, gauge invariance should be preserved either directly upon the Hilbert space or dynamically), or else the simulation may explore a vast unphysical Hilbert space due to algorithmic or hardware imperfections. Last but not least, the continuum and infinite-volume limits of observables, static or dynamical, in the Hamiltonian framework, must be understood. We will briefly discuss each of these theoretical directions, and will enumerate avenues for progress in the coming decade. 

The Hilbert space of local QFTss is infinite dimensional since, ignoring mathematical subtleties, the states can be described by functions from physical space into a target field space. The physical space is a continuum and of infinite extent. For fermionic theories, the target space is finite dimensional locally, while for bosonic theories like gauge theories, the target space is also continuous. On the other hand, quantum computers realized as discrete systems can simulate only finite-dimensional systems, and this requires discretizing/digitizing and bounding both physical space and the target space. The original field theory is then realized as a double limit of removing both of these discretizations and bounds. To estimate resources required to simulate a theory to a desired precision, one needs to understand these limits.

There are a number of frameworks developed for systematically converting infinite-dimensional field theories to a finite-dimensional counterpart. As an example of the choices encountered, consider a scalar field theory discretized on a spatial lattice. One may proceed by digitizing and bounding the field and its conjugate variable~\cite{jordan2011quantum}, or alternatively can quantize the theory in terms of harmonic-oscillator excitations and bound the allowed occupation of the oscillator modes~\cite{klco2019digitization}. Additionally, for simulations in the low particle-number sector, a single-particle digitization may prove more economical~\cite{barata2021single}. Each of these approaches may result in different rates of convergence to the predictions of the infinite-dimensional theory, as well as different resource requirements in simulation. The situation for gauge theories is more involved given the presence of local gauge symmetries and their expression in basis states, and a number of leading ideas are currently being explored. These include:
\begin{itemize}
\item[--] \textit{Global and local irreducible-representation bases.} The starting point of these approaches is the Hamiltonian formulation of lattice gauge theories by Kogut and Susskind~\cite{kogut1975hamiltonian}. The time variable is continuous and a partial gauge fixing is performed by choosing $A_0=0$, where $A_0$ is the temporal component of the gauge field. $A_0$ is non-dynamical in the Kogut-Susskind Hamiltonian. Therefore, it appears as a Lagrange multiplier for the Gauss's law operator, which is then required to vanish when acting on the physical states. Since Gauss's law is a statement on the divergence of (color) electric field, the electric field or irreducible representation (irrep) basis comes with advantages in applying Gauss's law. For example, by analytically solving Gauss's law at every lattice site, only Casimir eigenvalues are left as dynamical variables, from which a new Hamiltonian matrix can be formed that has a lower dimension and does not involve all or some of the unphysical transitions. This process can be done globally, with a cost that scales exponentially with the size of the system and is impractical for sizable simulations. It can also be done locally or semi-locally, in which case the classical pre-processing is scalable with the system's size, but unphysical transitions among various (semi-) local blocks are still plausible and must be eliminated in the simulation algorithm at a cost. Applications of this approach to U(1), SU(2), and SU(3) pure gauge theories in the context of quantum simulation have appeared in recent years~\cite{klco2018quantum,klco20202,byrnes2006simulating,ciavarella2021trailhead}.
\item[--] \textit{Prepotential and loop-string-hadron formulations.} 
The starting point of these formulations is still Kogut-Susskind Hamiltonian of non-Abelian LGTs, and one works in the irrep basis, in which the electric Hamiltonian is diagonal, while the gauge-matter coupling and magnetic Hamiltonians are non-diagonal. Prepotential formulation amounts to breaking the representation of the gauge link operator to left and right Schwinger bosons of the SU(N) theory, and building SU(N)-invariant operators from these at each site~\cite{mathur2005harmonic,anishetty20142,anishetty2009irreducible,mathur2010n}. By coupling prepotentials to fundamental fermions, one can construct gauge-invariant bosonic and fermionic operators, the so-called loops, strings, and hadrons~\cite{raychowdhury2020loop}. The loop-string-hadron formulation, therefore, expresses the non-Abelian dynamics in terms of strictly charge-conserving underlying operators.
 The ordinary non-Abelian Gauss's law constraints are made automatic, though auxiliary U(1) constraints are introduced and imposed to ensure the equality of the group Casimir on the broken links. The loop-string-hadron Hamiltonian is naturally expressed as a sum of one-sparse terms, which could benefit time-evolution subroutines~\cite{davoudi2021search}. The development of the loop-string-hadron formulation for QCD is among the immediate next goals of this program.
\item[--] \textit{Group-element basis and discrete subgroups.} The magnetic Hamiltonian on the lattice is defined with semi-local operators involving the product of links along a closed path such as a plaquette. These have a non-trivial action on the states expressed in the irrep basis. Furthermore, the magnetic Hamiltonian dominates the dynamics of U(1) and SU(N) gauge theories toward the continuum limit, and choosing an irrep basis requires retaining a large number electric-field excitations in this limit, hence increasing the computing-resource requirement. One may, therefore, desire to work in the group-element basis~\cite{zohar2015formulation}, which simplifies the simulation of both the gauge-matter coupling and the magnetic Hamiltonians. Furthermore, by formulating in this basis, one maintains a close relation to standard lattice-field-theory methods which simplifies analysis ~\cite{ji2020gluon,ji2022gluon} and development of algorithms~\cite{lamm2019general,harmalkar2020quantum,carena2021lattice,carena2022improved}. However, quantizing and truncating the group elements in the SU(N) LGT is not straightforward and may violate the group symmetry. One approach around this is to approximate continuous gauge groups by their crystal-like subgroups. This crystallization reduces qubit costs ~\cite{hackett2019digitizing,alexandru2019gluon,ji2020gluon,alam2021quantum,alexandru2021spectrum,hartung2022digitising} with realistic estimates for SU(3) LGT being $\sim 10$ qubits per gauge link, and is agnostic to the particular Hamiltonian chosen~\cite{carena2022improved}. This remnant gauge symmetry simplifies renormalization issues, in particular from gauge-symmetry violation. By proper choices of Hamiltonians and actions, it has been demonstrated that for U(1) and SU(N), systematic errors from this approximation should remain negligible for quantum simulations for the forseeable future where qubit counts remain below $\mathcal{O}(10^7)$ and the lattice spacings $a\gtrsim 0.06$ fm~\cite{alexandru2019gluon,alexandru2021spectrum}. As quantum resources improve, systematic improvements to the Hamiltonians are possible~\cite{ji2020gluon,ji2022gluon,haase2021resource,carena2022improved}.
\item[--] \textit{Magnetic or dual representations.}
A formulation in the magnetic basis, where the magnetic Hamiltonian is diagonal, requires much fewer basis states to be retained in the truncation, yielding a much more efficient representation at weak couplings relevant for the continuum limit. The dual nature of the electric and magnetic basis~\cite{kaplan2020gauss} has allowed a magnetic basis to be constructed for a compact Abelian gauge theory~\cite{haase2021resource,paulson2021simulating}. Starting from an electric basis which keeps a large number of electric basis states, and converting this to a magnetic basis using a Fourier transform, the resulting magnetic basis can then be truncated, giving a much better description at small couplings than an electric basis of the same dimension, while performing much worse at large couplings. Another basis for the same compact U(1) theory exists~\cite{bauer2021efficient}, in which the magnetic and electric basis are related to each other by a simple Fourier transform,  and the scheme is shown to work at both small and large couplings alike. The development of dual bases for non-Abelian gauge theories will be an important next step for the field, but early efforts indicate that such dual formulations often lead to more complex and generally non-local electric Hamiltonians~\cite{mathur2016lattice}. The benefits of such dual formulations, therefore, must be thoroughly examined in the context of quantum resources required to achieve given accuracy.

\item[--] \textit{Tensor renormalization group.} A complementary approach starts with the standard Lagrangian formulation used in lattice gauge theory and uses character expansions (for instance Fourier series) developed in the context of strong coupling expansions to rewrite partition functions and average observables in terms of products of traced tensors. In most situations of interest, this provides a {\it discrete} reformulation that can be exploited for quantum computing and can be verified at small volume with conventional methods. The tensors are local objects that contain all the information about the model and its symmetries. They can be seen as the building blocks of various type of computations. When continuous field variables are involved, there is an infinite number of characters (Fourier modes), but it has been shown that truncations of tensor sums preserve global and local symmetries~\cite{meurice2019examples,meurice2020discrete}, see Ref.~\cite{meurice2020tensor} for a recent review. The reformulation of lattice gauge theories was initially developed in collaboration with condensed-matter researchers~\cite{liu2013exact}, extending the method of Ref.~\cite{xie2012coarse}. The original motivations include clean configuration-space coarse-graining~\cite{meurice2013accurate} and absence of sign problem in presence of a non-zero chemical potential~\cite{denbleyker2014controlling} and a non-zero $\theta$ term~\cite{kawauchi2016phase,nakayama2022phase}. Transfer matrix methods also connect to the  Hamiltonian approach and quantum simulation~\cite{zou2014progress, bazavov2015gauge, zhang2018quantum, meurice2021theoretical}. There has been a considerable effort to extend this approach to models with fermions~\cite{shimizu2014grassmann,shimizu2014critical,takeda2015grassmann,sakai2017higher}, 
scalars~\cite{kadoh2018tensor,kadoh2019tensor,adachi2020anisotropic,akiyama2020tensor}, and other models~\cite{kuramashi2019three,asaduzzaman2020tensor,bloch2021tensor} in various dimensions. Furthermore, other tensor methods that were developed earlier in condensed-matter physics  led to the use of quantum-information tools in approaching QFT problems~\cite{banuls2020review,banuls2020simulating}. Moreover, near-term quantum-simulation algorithms might benefit from a combined approach of variational algorithms and tensor networks. The power of tensor networks could be utilized by splitting large quantum systems by small subsystems~\cite{yuan2021quantum}, and QFTs could be natural targets.\footnote{These activities are detailed in a Snowmass whitepaper~\cite{meurice2022tensor} focused on the tensor approach to QFTss.}
\item[--] \textit{Light-Front quantization.} Hamitonian QFT need not be formulated on fixed time-slices.  In the light-cone quantization approach, fields are quantized along the light-cone $x^+ \equiv t+z$~\cite{mannheim2021comparing}. This is shown to result in a smaller number of physical degrees of freedom compared with the canonical equal-time quantization. The reason is that the sum of occupancies in a Fock state is upper bounded in the light-cone quantization, since there is no possibility for an infinite number of left- and right-moving massive particles which can give rise to a net finite momentum. Such an approach, that is related to the single-particle quantization scheme mentioned above, puts quantum simulation of QFTs on a similar footing with the quantum simulation of quantum chemistry~\cite{liu2020quantum}. Nonetheless, subtleties associated with the zero mode of massless fields and with UV-IR mixing during renormalization~\cite{wilson1994nonperturbative}
complicate the scheme and require careful treatment, see e,g., Refs.~\cite{chabysheva2009zero,martinovivc2020vacuum,chabysheva2022tadpoles} for related progress. A digitization amenable to use on quantum computers for the SU(3) LGT was constructed recently~\cite{kreshchuk2021light,kreshchuk2021simulating,kreshchuk2022quantum}. The light-cone formulation is well suited for calculating the properties of relativistic bound states, while the applications to the scattering problem are emerging~\cite{du2021quantum}.  
\item[--] \textit{Quantum Link Models and qubit regularization.} 
There exist approaches for simulating QFTs without the need for an infinite-dimensional local Hilbert space. Such approaches to QFTs are well known in the condensed-matter literature and were brought to particle physics through the quantum-link approach via the idea of D-theory~\cite{brower1999qcd,brower2004d}.   It is argued that almost all quantum field theories can be obtained in this approach by formulating a lattice field theory with a finite-dimensional Hilbert space, when one introduces a fictitious space dimension so that the infinite local Hilbert space is built up as a direct product of fixed-size Hilbert space on each site in the new direction. The advantage of this method is that the Hamiltonian that needs to be simulated is \emph{local} in this extended space, so the quantum circuits that implement it are potentially simpler. 
Recent work shows how the O(3) model with a theta vacuum can be studied in this framework~\cite{caspar2022asymptotic}. It has also been shown that the extra dimension may not be unnecessary in some cases~\cite{buser2020state}: the low-energy sector of Hamiltonians tuned to a quantum critical point are described by the QFT. Intuitively, at such a critical point the continuum local Hilbert space describes the state of the lattice system over a region of size given by the correlation length of the system, and so can be infinite as this length diverges. The trick, of course, is to find the correct quantum critical point, and these may not exist for theories of interest. Interestingly, however, properties like asymptotic freedom can arise~\cite{bhattacharya2021qubit,alexandru2021universality} in this approach, and, often, with smaller resources, one can get an EFTs that can be completed in the ultraviolet by continuum perturbation theory. A systematic way to explore the qubit-regularization approach within the quantum-link framework was discussed recently in Ref.~\cite{liu2022qubit}. Future work needs to extend this approach to gauge theories and fermionic theories.
\item[--] \textit{Matrix models.} Dimensional reduction can be used to map gauge theories to quantum-mechanical models, while preserving some of the interesting non-perturbative dynamics and structure of the parent QFT. With their much smaller Hilbert spaces, these models are interesting physics targets for near-term simulations of gauge theories, complementary to approaches based on digitization or other truncations of lattice gauge theories with small numbers of sites. For a simple example, the reduction of 1+1D QED with massive charged matter on a small spatial circle leads to a quantum-mechanical rotor model which realizes the same ’t Hooft anomalies and tunneling processes as the parent Schwinger model. These properties are associated with slow dynamics that can be seen in analog simulations on a single Rydberg atom~\cite{shen2021simulating}. In the context of ordinary 4D Yang-Mills theories, reduction on a small spatial torus maps the gauge theory to a matrix quantum-mechanics model with calculable Hamiltonian~\cite{Luscher:1982ma}, and the low-lying spectrum of the matrix model can accurately reproduce the spectrum of the gauge theory obtained in Euclidean lattice simulations~\cite{van2001qcd}. Thus quantum simulations of matrix models may provide interesting insights to nontrivial dynamics of 4D gauge theories on near-term hardware. Related matrix models are also of broader theoretical interest, arising for example in non-perturbative formulations of string theory~\cite{banks1997m} and other models of quantum gravity~\cite{banks2006towards,anninos2020notes}, see Suppl.~\ref{suppl:gravity}. Such models could also serve as a natural target for developing quantum algorithms and simulations~\cite{gharibyan2021toward,buser2021quantum}.
\end{itemize}

As global and local symmetries played a crucial role in the establishment of the SM, it takes a special effort to reconsider these questions 
in the context of Hamiltonian formulation of discretized/digitized approximations of QFTs. In the irrep-basis formulation of lattice gauge theories, a cutoff on the number of irreps retained respects the gauge symmetry except at the cutoff. On the other hand, arbitrary digitizations and truncations in the group-element basis may only respect a subset of symmetries or none. In tensor reformulations of lattice models~\cite{meurice2020tensor}, it has been shown that truncations of tensor sums preserve global and local symmetries~\cite{meurice2019examples,meurice2020discrete}. More generally, symmetries usually define a continuum limit that has universal aspects. Reaching this limit in the most efficient and economical way, rather than closeness to a specific lattice model with lattice artefact, must be considered the ultimate goal.

The questions of gauge invariance and noise-robust implementations of Gauss's laws for Abelian~\cite{kaplan2020gauss,unmuth2019gauge,meurice2020discrete,bender2020gauge} and non-Abelian~\cite{raychowdhury2020solving,davoudi2021search} local symmetries have received considerable attention in recent years. Even if the digitized and truncated formulation of a gauge theory provides a (nearly) gauge-invariant Hamiltonian, simulating the system under that Hamiltonian may break the symmetries. Due to the condition of Gauss's law that is a constraint on the Hilbert space, even if the simulation is launched in the physical sector of the theory, imperfections in the simulation algorithm or in quantum hardware can drive the system out of the physical subspace. For example, Trotterized evolution in a digital simulation may introduce errors that do not respect the symmetries, as does an inaccurate engineering of the dynamics in an analog simulation. Coupling to the environment likely involves gauge-symmetry violating terms too. 

There are two currently-known approaches to gauge-invariant simulations, assuming that the simulation starts in the gauge-invariant sector but may evolve to other sectors due to hardware or algorithmic imperfections. One is to adopt a formulation which is fully or partially gauge invariant by construction. Examples include purely-fermionic formulation of gauge theories coupled to matter in 1+1 D with certain boundary conditions, where a gauge transformation and the application of Gauss's laws fully constrains the gauge degrees of freedom, leaving only fermions which can now interact non-locally~\cite{hamer1997series,sala2018variational,davoudi2021search}. Another example is the loop-string-hadron formulation of the SU(2) LGT in d+1 D~\cite{raychowdhury2020loop}, which incorporates non-Abelian Gauss's laws by construction but leaves an Abelian constraint to be satisfied locally. The second approach is an active protection of the symmetries as the system is evolved in the simulator. Examples include adding a penalty term to the Hamiltonian proportional to the (square of) Gauss's law operator to suppress the leakage to the unphysical Hilbert space~\cite{banerjee2012atomic,hauke2013quantum,banerjee2013atomic,marcos2013superconducting, rajput2021hybridized}, performing random rotations during evolution with the Gauss's law operator to average out the symmetry violation~\cite{tran2021faster,lamm2020suppressing}, adding to the Hamiltonian the Gauss's law operator with properly-chosen coefficients to separate out different Gauss's law sectors in the spectrum~\cite{halimeh2021gauge,halimeh2022gauge} or similar techniques~\cite{halimeh2021stabilizing,van2021suppressing}, using classical noise proportional to the Gauss's law operator to suppress gauge-symmetry violation via a Zeno effect~\cite{stannigel2014constrained}, a similar quantum approach in which quantum control is used to dynamically decouple unphysical sectors during the evolution~\cite{kasper2020non}, and in a more gate-based setting, using controlled operations to disallow unphysical transitions between basis states~\cite{ciavarella2021trailhead}. As a verification step, one could also use oracles in the quantum circuit to detect Gauss's law violations and discard the result~\cite{stryker2019oracles,raychowdhury2020solving}.

More research is needed to clarify the importance of symmetry-protected simulations and whether they will be more resource-efficient in general compared with non-protected counterparts. For example, the measure of merit should be the closeness to the exact evolution, and if a non-protected algorithm has a faster approach to the exact limits, it should be taken as the method of the choice. Furthermore, it is not clear that suppressing errors that are associated with transitions to the unphysical sectors will reduce the total error, as demonstrated in Ref.~\cite{nguyen2021digital}. Furthermore, the incoherent noise appears to be the dominant source of simulation error in the NISQ hardware, and symmetry-protection protocols need to be generalized to address such errors too. Last but not least, it may be that the gauge-theory simulation are robust to small gauge-violating errors in the simulator, as demonstrated for several quantum link models (QLMs) in Refs.~\cite{halimeh2020fate,van2021reliability}, so an active symmetry enforcement with a resource overhead may not be necessary after all. As the field moves toward selecting the best theoretical formulations of gauge theories for quantum simulation, all these questions need to be thoroughly addressed. 

A major part of the development of conventional lattice gauge theory over the past few decades has been to quantify and mitigate systematic errors, such as those associated with discretization, finite volume, excited-state effects on ground-state properties, and quark-mass inputs (if set away from physical values for computational expediency). EFTs played a major role in these efforts as they allowed to construct improved actions and observables, and to find reasonable extrapolation forms for certain quantities. Furthermore, studying finite-volume effects offered a powerful methodology to access few-hadron scattering amplitudes that otherwise would not have been accessible with the lattice-QCD technique~\cite{luscher1986volume,luscher1991two,bulava2022hadron,briceno2018scattering,davoudi2018path,hansen2019lattice}. It is conceivable that such a program will continue to grow in the Hamiltonian-simulation era too. In fact, investigations of infinite-volume and continuum limits of certain QFTs have emerged in recent years~\cite{briceno2021role,carena2021lattice}. Besides new systematic uncertainties encountered in digital quantum simulation, such as digitization of the time variable, and the truncation errors in bosonic theories due to limited qubit resources, constraints such as space discretization in lattice formulations, and the finite extents of time and space remain in effect. Strategies to quantify and extrapolate them away may be rather different in a Hamiltonian-simulation setting, but reliance on EFTs and improvement programs may prove useful in this context too. As a result, more theoretical research is needed to address the question of systematic uncertainties and their quantification for quantum simulation of QFTs.

\subsubsection{Algorithmic research for digital quantum computing and resource analysis}
\noindent
The digital approach to quantum simulation offers controlled ways to prepare, evolve, and measure the states of a quantum system. Importantly, digital algorithms can be generally analyzed rigorously and their asymptotic or exact resource requirement can be bounded given a desired accuracy. Furthermore, a range of error-correction and error-mitigation techniques applies to digital simulations. It is important to invest in designing, analyzing, and improving suitable quantum-simulation algorithms for HEP, and particularly for QFTs of interest. In the context of quantum simulation of physical models in general, and QFTs in particular, this section reviews the basic elements of a digital approach to simulation and recent advancements. Remaining open questions in simulation algorithms, and in understanding their resource scaling will be further discussed.

In a digital simulation, the system's evolution is broken to simpler implementable unitaries, and the way such a digitization is performed defines the simulation algorithm. Implementable in this context means unitaries for which a decomposition exists that is composed of a number of elementary quantum gates that is polynomial in terms of problem size and error tolerance. For the purpose of this section, we focus on the qubits as the quantum-information units and a common choice of universal set of single- and two-qubit gates as units of quantum processing (more general entangling gates can be used as elementary gates as well). More general choices, e.g., higher-dimensional qudits or customized gates, will be discussed later in the context of analog and hybrid approaches to quantum simulation. The algorithm's figure of merit depends on resources that need to be minimized. In the near-term computing model, qubit resources are scarce and entangling gates are lower in fidelity than the single-qubit gates. Therefore, computations that require the least number of ancillary qubits and entangling gates are desired. In the fault-tolerant era of quantum computing, the overhead is associated with error correction. Fault-tolerant implementation of non-Clifford gates such as the T gate is known to be more resource intensive than Clifford gates for error-correcting codes such as surface codes. Therefore, it is the T-gate count that needs to be minimized in the far term.

The most popular simulation algorithms to date are product formulas, which are based on Trotter-Suzuki decomposition of the time-evolution operator~\cite{suzuki1991general}. For example, for a Hamiltonian of the form $H=\sum_{l=1}^\Gamma H^{(l)}$, where the different $H^{(l)}$ do not commute with one another, the operator $\left[\prod_{l=1}^\Gamma e^{-it H^{(l)}/r}\right]^r$ approximates $e^{-iHt}$ up to an error that scales as $\mathcal{O}(t^2/r)$ for positive integer $r$ and real parameter $t>0$. Higher-order formulas can be constructed to enable more accurate simulations but at the cost of increasing the circuit depth~\cite{wiebe2010higher, childs2021theory}. As a result, if the Hamiltonian is a sum of local or semi-local terms, system's evolution can be implemented in polynomial time~\cite{lloyd1996universal}. With no ancillary overhead and simpler implementation, the product formulas may remain the simulation algorithm of choice in the near term. There has been a great deal of progress in developing other simulation algorithms such as Taylor series expansion and linear combination of unitaries~\cite{childs2012hamiltonian, berry2015simulating}, quantum signal processing~\cite{low2017optimal}, qubitization and block encodings~\cite{low2019hamiltonian, chakraborty2018power}, singular-value transformation~\cite{gilyen2019quantum}, off-diagonal Hamiltonian expansion~\cite{kalev2021quantum} and hybrid algorithms~\cite{rajput2021hybridized}, which generally perform more optimally asymptotically, but often involve significant (if scaling poly-logarithmically) ancillary qubits and more complex circuit implementations. On the other hand, better analytical approaches~\cite{childs2021theory}, taking advantage of system's locality and conservation laws~\cite{su2021nearly} or inputting information about the initial state~\cite{csahinouglu2021hamiltonian,hatomura2022state,yi2021spectral,zhao2021hamiltonian}, and empirical analysis of the performance in select cases~\cite{childs2019nearly,stetina2020simulating,nguyen2021digital} have resulted in considerably tighter bounds on product-formula errors in recent years. Research in the quantum-algorithm community continues to improve the current simulation schemes and to devise new strategies. QFT simulations will be a prime application of optimized algorithms given their significant resource requirement. 

In the context of QFTs, among the first thorough algorithmic approaches to quantum simulation is the seminal work by Jordan, Lee, and Preskill~\cite{jordan2011quantum,jordan2012quantum,jordan2014quantum} which sets up an evaluation of the scattering S-matrix in an interacting field theory. This work demonstrates three primary tasks in quantum simulation: i) initial-state preparation amounting to preparing scattering wavepackets, ii) time evolution involving an adiabatic approach of the system to a fully interacting theory and evolving back to isolated wavepackets, and finally iii) final-state measurement amounting to identifying quantities that can be optimally measured and processed to access information about the scattering amplitude, without the need for costly full state tomography. While this algorithms shows the true advantage of a quantum computer, that is to enable a direct evaluation of real-time cross sections in QFTs, its resource requirement will likely prohibit its implementation for even small systems for the foreseeable future. 
Research is in progress to devise and benchmark less resource-intensive approaches to the scattering problem in the near term, including proposals for variational approaches~\cite{liu2021towards}, obtaining phase shifts in prototype spin models via time delay~\cite{gustafson2021real}, or the use of L\"uscher's method~\cite{luscher1986volume,luscher1991two} in extracting low-energy few-body scattering parameters from energy spectra of particles in a finite volume, as is done in the conventional lattice-QCD program~\cite{bulava2022hadron,briceno2018scattering,davoudi2018path,hansen2019lattice}. A complete resource analysis of scattering problems in gauge theories, including QCD, is still lacking and the problem is complicated by the absence of fully developed and efficient state-preparation algorithms for a range of non-trivial states in gauge theories, from the interacting vacuum to the scattering wavepackets of confined hadrons. Promising progress is reported in recent years on state preparation in scalar~\cite{jordan2011quantum,klco2020minimally,bagherimehrab2021nearly}, fermionic~\cite{jordan2014quantum,moosavian2019site}, and gauge field theories~\cite{ciavarella2021preparation}, and it is plausible that customary state-preparation techniques such as adiabatic state preparation~\cite{chakraborty2020digital}, projective cooling~\cite{lee2020projected,gustafson2020projective}, and tensor-network-inspired ansatzes~\cite{moosavian2018faster}, combined with new customized QFT algorithms can lead to further progress in the coming years.

A fundamental element of any quantum-simulation algorithm for QFTs is the implementation of the time-evolution operator, given the formulation and basis states chosen and the encoding adopted for the various degrees of freedom. While the development of efficient algorithms for scalar field theory and Abelian and non-Abelian gauge theories has led to valuable detailed analyses of qubit and gate requirements for various formulations of a range of models~\cite{jordan2011quantum,byrnes2006simulating,shaw2020quantum,ciavarella2021trailhead,kan2021lattice}, it is not clear that the devised algorithms have the most optimal scaling and are suitable for near- and intermediate-term computations. The situation has strong parallels in the classical-computing world, where many initial algorithms, while they generated interest and guided the developments, were replaced by increasingly faster and more resource-efficient algorithms. It also became clear that asymptotic scalings were not necessarily the best guide to accurate resource estimates, and identifying prefactors and benchmarking algorithms in pursuit of learning their empirical performance were essential in advancing computational sciences and their applications. Similar trend is expected in the realm of quantum simulation. For example, it is becoming clear that simulating non-Abelian gauge theories in the irrep basis suffers from costly compilation of non-commutative algebra of the group, necessitating many rounds of accurate synthesis of non-trivial functions as the system evolves~\cite{kan2021lattice}. Both near- and far-term circuit implementations of such dynamical phases are costly and introduce non-negligible overhead to the simulation. For example, for the $SU(2)$ and $SU(3)$ lattice gauge theories in the irrep basis, when approximating the time evolution operator via Trotterization for a maximum error $\epsilon$ for an arbitrary state, the gate count scales as $\propto d\Lambda t^{3/2}(L/a)^{3d/2}\epsilon^{-1/2}$~\cite{tong2021provably,shaw2020quantum}, although logarithmic corrections may prove important~\cite{kan2021lattice}. Here, $d$ denotes the dimensionality of space, $\Lambda$ is the gauge-field truncation in the irrep basis, $L$ is the spatial extent of a cubic lattice, and $a$ denotes the lattice spacing. For an accuracy goal $\epsilon = 10^{-8}$ and a lattice with tens of sites along each spatial direction, as proposed in Ref.~\cite{kan2021lattice}, the simulation requires hundreds of thousand to hundreds of million qubits and of the order of $10^{50}$ T gates and more. Given the unaccounted-for theoretical uncertainties, acceptable values of $\epsilon$ could be orders-of-magnitude larger. A tighter bound could be obtained by considering state-dependent errors~\cite{csahinouglu2021hamiltonian,hatomura2022state}. As the community pursues better approaches to simulating gauge theories, questions such as suitable formulations and practical, and perhaps more customized, implementations given the simulating hardware must be addressed. Hybrid digital-analog algorithms may prove valuable in scaling down the cost, but more work is needed to understand their time complexity, as discussed in the next subsection.

The algorithm and its performance is closely dependent upon the encoding of the degrees of freedom, and this encoding is related to the choice of formulation as discussed in the previous subsection. This choice, in particular, impacts the error-bound analysis of the simulation algorithm by systematically accounting for the truncation errors in bosonic field theories. The first attempts at numerically investigating the truncation errors in scalar field theory and Abelian and non-Abelian gauge theories in both irrep and group-element basis demonstrate the exponentially fast convergence of low-energy observables to the exact values~\cite{klco2019digitization,hackett2019digitizing,davoudi2021search,ciavarella2021trailhead}, but fails to reach this exponential scaling until larger truncation cut offs are used for high-energy and long-time-evolved quantities~\cite{davoudi2021search}. There exists analytical approaches to understand this exponential convergence in certain problems~\cite{somma2015quantum,macridin2018digital,macridin2018electron,tong2021provably}, and a first analysis of error bounds in product formulas considering this exponential convergence in non-interacting scalar field theory and the SU(2) LGT in the irrep basis has appeared~\cite{tong2021provably}. Deriving analytical bounds for evolution under truncated Hamiltonians is generally difficult but is an important step toward more reliable error-bound analysis of algorithms given the formulation used for the QFT. Even in qubit-regularized and QLM approaches to recovering the continuum QFTs, which do not require extrapolation in the dimensionality of the on-site Hilbert space, the rate of convergence to such continuum limits in connection to resources required needs to be examined thoroughly in future investigations.

Besides the issue of quantifying truncation errors and convergence rate toward the continuum limit, one needs to analyze the best available encoding strategies for both bosonic and fermionic degree of freedom, and such a question must be addressed in the context of the QFT formulation used and the simulation algorithm adopted. Among proposed bosonic encodings are the standard binary encoding with which an integer $b$ is encoded in $\sim\log_2 b$ qubits, Gray encoding with the same qubit count as the binary case, and the unary encoding with $b$ qubits~\cite{sawaya2020resource}. There is even more diversity with the fermionic encodings. While fermionic on-site Hilbert space is finite dimensional, implementing the Fermi statistics can be costly, leading to maximally non-local interactions among the qubits through Jordan-Wigner fermion-to-spin transformation~\cite{jordan1928paulische}, or more local but qubit-resource-intensive encodings~\cite{bravyi2002fermionic,verstraete2005mapping, ball2005fermions, whitfield2016local, steudtner2019quantum,setia2019superfast, chen2020exact, chien2020custom, derby2021compact}. In fact, the construction of many local mappings resembles the structure of lattice gauge theories, in which the interactions are locally mediated by the ancillary degrees of freedom but local ``Gauss's law'' like constraints must be imposed on the Hilbert space of main and ancillary qubits, requiring robust implementations to ensure the constraints are not violated and the Fermi statistics remains intact. Symmetry-protection strategies outlined in the previous section can, therefore, be of value in fermionic simulations with local mappings to qubits. There has not been sufficient research on the performance of various bosonic and fermionic encodings in gauge-theory simulations but limited results have appeared for select models and formulations~\cite{mathis2020toward,kreshchuk2022quantum}. The outcome of such analyses will not only help with deciding the optimal choice of formulation and encoding, but can also lead to the development of new and better customized encodings for QFT applications.

Finally, dedicated algorithms are needed for a range of quantities of interest in high-energy physics, beyond the scattering amplitudes, and such algorithms must specify concretely the state-preparation and (partial) state-tomography techniques that are efficient and tailored to the goal of the problem. For example, obtaining the hadron tensor for computing hadron's structure functions requires matrix elements of space-time separated quark-level currents inside a proton and standard quantum algorithms for evaluating correlation functions exist, see e.g., Ref.~\cite{wecker2015solving}. Nonetheless, questions regarding how to prepare the hadron state to a given accuracy, and how to estimate the error associated with state preparation along with that in correlator measurement, need to be fully addressed. Besides the energy spectrum, particle-density distribution, and equal-time and out-of-time correlation functions, entanglement measures such as entanglement entropy and entanglement spectrum will be of critical value in our understanding the phases of matter in and out of equilibrium, and how thermalization, fragmentation, and hadronization occur in collider experiments. Research is needed to assess the applicability and efficiency of recently-proposed methods in entanglement tomography, e.g., using random measurements and classical shadows~\cite{aaronson2019shadow,huang2020predicting,kokail2021entanglement,kokail2021quantum}, for QFTs, particularly in QCD-like theories with confined and composite asymptotic final states.

From this overview of the field and the outstanding problems, it is obvious that algorithmic research will continue to constitute a major component of the field of quantum simulating QFTs. Progress relies on well-equipped field theories that can combine theory inputs with algorithmic needs, along with close collaboration with quantum-algorithm experts and digital-hardware developers, as discussed in more details in Supp.~\ref{suppl:codevelopment}.

\subsubsection{Analog and hybrid approaches to quantum simulation of QFTs}
\noindent
In an analog approach to quantum simulation, the Hamiltonian of a theoretical model is mapped onto the Hamiltonian of an actual physical system engineered in a laboratory, usually as tabletop experiments. Often large Hilbert spaces can be encoded, e.g., the occupations of thousands of sites of an optical lattice by cold Rubidium atoms, but there are only a few ``knobs" that can be turned, e.g., the optical lattice spacing, depth of the potential, etc. Consequently, this approach could be suitable for studies of universal properties or continuum limits of models in situations where the correlation lengths are large. Research in analog quantum simulation of QFTs requires understanding the underlying physics of the simulator. For example, to determine if the capabilities of the hardware can be matched with the features of the target Hamiltonian, one needs to learn what the native or naturally implementable interactions are in the simulator, and what characteristics of the simulator can be tuned easily and what features are harder to modify. A dedicated Supplemental Section (Supp.~\ref{suppl:analog}) reviews the status of the state-of-the-art atomic, optical, molecular, and solid-state analog platforms for quantum simulation, and their prospect for simulating QFTs. Here, we will focus on opportunities and challenges of simulating quantum fields in an analog manner. Furthermore, the need for hybrid strategies which combine the benefits of the digital and analog schemes in one setting will be motivated, and will be further elaborated in later Supplemental Sections.

Interacting scalar field theories, due to their equivalence to coupled many-body quantum-harmonic oscillators appear more natural in systems that have an effective description in terms of coupled harmonic oscillators, such as in superconducting circuits~\cite{kjaergaard2020superconducting,krantz2019quantum}. In fact, the first proposals for quantum simulating sine-Gordon models have been put forward in recent years~\cite{roy2019quantum,roy2021quantum}. In contrast, lattice gauge theories are coupled fermionic-bosonic models that often do not have a simple harmonic-oscillator representation for the gauge bosons. The Schwinger model in the limit of large bosonic occupation can be approximated by a spin-harmonic-oscillator system, for which an analog-simulation proposal has been developed in the context of trapped-ion systems with phonons as the bosonic degrees of freedom~\cite{yang2016analog}. Similarly, the Abelian Higgs model is proposed to be studied using superconducting microwave cavities that implement a boson-based variational quantum algorithm~\cite{zhang2021simulating}. Purely fermionic as well as simple QLM formulations of the Schwinger-model, on the other hand, involve only interacting spins and have received interest in the context of trapped-ion~\cite{hauke2013quantum,davoudi2020towards,andrade2021engineering}, polar molecules~\cite{luo2020framework}, cold-atom~\cite{banerjee2012atomic,wiese2013ultracold,zohar2012simulating,zohar2011confinement,zohar2013simulating,zohar2013cold,rico2014tensor,kasper2016schwinger,mil2020scalable} including with Rydberg arrays~\cite{surace2020lattice}, and superconducting-circuit~\cite{marcos2013superconducting} simulators. In fact, the largest-scale analog simulations of the Schwinger model have been enabled in recent years within the QLM description~\cite{yang2020observation}. The situation with non-Abelian LGTs is less developed. The gauge degrees of freedom in Schwinger-boson and loop-string-hadron formulations of SU(2) LGT admit a harmonic-oscillator description, but the number of local oscillators and the complex nature of interactions among them do not have a natural analog in current simulators. A few proposals exist for simulating the SU(2) LGT in atomic simulators~\cite{zohar2013cold,banerjee2013atomic,dasgupta2022cold}, including encoding non-Abelian Gauss's laws as natural angular-momentum conservation laws in atomic collisions~\cite{zohar2013quantum}, but these proposals have not yet been implemented in experiment.

Going to dimensions higher than 1+1 D presents a bigger challenge, as both the electric and magnetic Hamiltonian must be simultaneously implemented in the simulator. When working in the irrep (electric-field) basis, the magnetic (or plaquette) Hamiltonian is complicated and requires interactions among at least four degrees of freedom, which are intrinsically harder to engineer in any of the current analog quantum simulators. This motivates the need for dual-variable bases, in which the magnetic Hamiltonian is more easily implemented, however, this leads to more complex electric interactions that may not be naturally implementable in the simulator. Engineering a highly accurate minimal building block of an Abelian or non-Abelian LGT in 2+1 dimensions will be a milestone for the field in the coming years. Scaling the system while maintaining the fidelity will then be the next challenge to overcome. To achieve this goal, extensive research is needed in surveying and analyzing theoretically a range of plausible analog-simulation platforms. This challenge also motivates the need to search for the most optimal formulations of QFTs, noting the fact that depending on the digital and analog nature of the simulation, and the type of the architecture used, different optimal formulations may be found.

The engineering challenge associated with implementing a continuous and fully analog evolution of gauge theories further motivates hybrid approaches to the simulation. One one hand, digital quantum computation allows simulating arbitrary local or semi-local Hamiltonians efficiently, providing an approximation to various unitaries with an error that can generally be bounded systematically. On the other hand, analog quantum simulation can only simulate certain systems whose degrees of freedom and interactions are similar to those of the simulator, but provides a more natural and resource-efficient approach to time evolution. Combining the benefits of each mode of the simulator, one can allow some degree of digitization in the simulation so that the engineering of the approximate dynamics is facilitated compared with the analog simulator. On the other hand, it could be the case that the simulator offers access and control of certain degrees of freedom that can encode more naturally the degrees of freedom of the target theory, or that there are intrinsic interactions among these degrees of freedom that can allow a more extended set of quantum gates to be devised. In this case, it is reasonable to devise digital simulations that are augmented by some analog building blocks. As an example, certain QFTs are shown to benefit from a hybrid approach in trapped-ion platforms, where phonons can encode bosonic fields and participate in the dynamics actively~\cite{davoudi2021towards,casanova2011quantum}. As another example, the quantum simulation of LGTs can be digitized such that plaquette interactions involving four- or higher-body terms arise as an effective interaction when a series of two-body local interactions are implemented with the use of mediating ancillary qubits~\cite{zohar2017digital,zohar2017digital,bender2018digital}. Furthermore, higher-dimensional qudits might prove useful~\cite{wang2020qudits} in encoding multi-component fermions or scalar fields~\cite{gustafson2021prospects,kurkcuoglu2021quantum}. A great deal of research is anticipated to uncover the true power of quantum simulators for QFTs when such flexibility in the choice of simulation mode is allowed. 

Another method to encode information in quantum computing is based on continuous variables (CVs), such as the position and momentum of a particle or the quadratures of an electromagnetic field. CV quantum computing was first proposed by Lloyd and Braunstein~\cite{lloyd1999quantum} who presented necessary and sufficient conditions for constructing a universal quantum computer over CVs. Unlike discrete-variable quantum computing, the basic unit of information in CV quantum computing is a quantum system with an infinite-dimensional Hilbert space called qumode. CV quantum computing is a natural platform to study continuous quantum systems such as QFTs. One can also use qumodes to encode qubits by using encoding schemes such as the one proposed by Gottesman, Kitaev, and Preskill~\cite{gottesman2001encoding} or using coherent states~\cite{ralph2003quantum}. Additionally, hybrid protocols that leverage the advantages of both qubits and qumodes have been proposed~\cite{andersen2015hybrid}. Recent experimental breakthroughs, such as the demonstration of quantum advantage using Gaussian Boson Sampling~\cite{zhong2020quantum}, and the introduction of a programmable photonic quantum computer~\cite{arrazola2021quantum}, have sparked a growing interest in the development of CV quantum algorithms. The first CV quantum algorithm that studied the quantum simulation of a QFT was proposed in Ref.~\cite{marshall2015quantum} where scattering amplitudes in a scalar bosonic QFT were calculated, and later extended to scalar electrodynamics~\cite{yeter2018quantum}. For the ground and excited states of a QFT, the quantum imaginary-time evolution~\cite{motta2020determining} algorithm adapted to CV substrates was recently proposed~\cite{yeter2021quantum}. Energy levels in a $\phi^4$ QFT were calculated on a quantum simulator, and by developing similar algorithms, one can also study finite-temperature systems~\cite{sun2021quantum}. More research is needed to estimate the required resources for quantum simulations of QFTs using CV quantum hardware. Such investigations will need to include quantum error-correction schemes, a subject which is still in its infancy for CVs but early adaptations of known discrete-variable codes for CV quantum computing have emerged in recent years~\cite{wu2021continuous, fukui2017analog, tzitrin2020progress}.

More generally, in order for the analog and hybrid approaches to present a realistic path toward reliable simulations of the target theory, two complementary research areas must be developed. First, analytical or empirical analysis of the errors due to inaccurate engineered dynamics must be systematically performed but this generally is a non-trivial task. It is also important to come up with error-correction and error-mitigation strategies that do not rely on the digitization of the evolution or qubit encodings. In the absence of such protocols, it is still conceivable that a set of criteria can be determined, such as the degree of local and global symmetry violations or inconsistencies in different observables, to put into test the result of the simulation. However, full confidence in the simulation result may not be possible without independent implementations of the same problem on different platforms, particularly as the simulation sizes grow beyond classical limits. Finally, an important task the HEP community will take on in the coming years is to investigate which problems are more suitable for implementation in analog simulators given potential inaccuracies and errors. There may be features that require larger Hilbert spaces to be simulated than plausible in near-term digital devices, but are less prone to imperfections in the simulator. Those qualitative features are likely to be related to studies of phases and phase transitions of matter. Therefore, there may be a unique opportunity for HEP physicists over the coming decade to identify the quantum-advantage case that may be enabled by analog or hybrid simulators.

\subsubsection{Illuminating the path: Implementation and benchmark}
In the current NISQ era of computing, the number of basic quantum-computing units is limited and the gate errors put constraints on the depth of the quantum circuits. Similarly, analog simulators are limited in size and quantum control, and face noise and coupling to the environment. Nevertheless, the building blocks of the unitary evolution necessary to study the simplest models can be developed, tested, and optimized with existing NISQ devices. The general idea is to start with simple models in low dimensions (e.g., $Z_2$ LGT, the Schwinger model, etc.) and progressively increase the symmetries and dimensions. An approximately similar sequence of models, sometimes called the ``Kogut ladder" or ``Kogut sequence" ~\cite{kogut1979introduction,kogut1983lattice} was proposed in the development of classical-computational methods, and it is clear that it can also be of value in the quantum-simulation era~\cite{meurice2020tensor}. In fact, from the very first quantum-simulation experiment of a gauge theory in a small trapped-ion quantum computer in 2016~\cite{martinez2016real}, the number of implementations and benchmarks on quantum hardware continued to substantially grow. The simulations of the Schwinger model~\cite{klco2018quantum,kokail2019self,lu2019simulations,nguyen2021digital,xu20213+} have gradually improved compared with the early demonstrations, and there exist now simulation results for small non-Abelian gauge theories as well, including discrete $D_N$ group~\cite{alam2021quantum}, and SU(2)~\cite{klco20202,atas20212} and SU(3)~\cite{ciavarella2021trailhead} LGT in 2+1 D. What these demonstrations show is that increasing the number of qubits in the hardware without comparable increase in gate fidelities is not of value as QFT circuits are not only qubit-resource intensive but also extremely deep. 

Analog quantum simulators have also been used to study the dynamics of LGTs. These range from simulating small building blocks of a LGT, such as a link starting and ending at the fermion sites using two-component ultracold atoms in double-well potentials~\cite{schweizer2019floquet} and other variants~\cite{gorg2019realization} for the $Z_2$ LGT, and using inter-species spin-changing collisions in an atomic mixture in an optical lattice for the U(1) LGT~\cite{mil2019realizing}. Larger-scale demonstrations of gauge-invariant dynamics in the spin-$1/2$ QLM of the U(1) LGT in 1+1 D have been made possible in defect-free arrays of bosonic atoms in an optical superlattice with up to 71 sites~\cite{yang2020observation}, along with the first demonstration of gauge-theory thermalization in this model~\cite{zhou2021thermalization}. Realization of gauge theories in 2+1 D~\cite{zohar2012simulating,zohar2011confinement,zohar2013simulating,tagliacozzo2013optical,ott2021scalable,gonzalez2022hardware} and higher dimensions and of the non-Abelian gauge-theory dynamics~\cite{tagliacozzo2013simulation,zohar2013quantum,zohar2013cold,banerjee2013atomic,dasgupta2022cold,gonzalez2022hardware} in analog quantum simulators will mark an important next step but these yet need more realistic proposals, as discussed before.

The simulation experiments and implementations, enabled either through access to cloud-based quantum processors by companies or via university collaborations, are a critical component of the quantum-simulation program for two reasons. First, they allow hands-on experience with quantum hardware and fill the gap between theory/algorithm and the simulation, and hence guide the developments toward more realistic and hardware-efficient proposals for experiment. Second, they allow the hardware developers to become familiar with the unique problems presented in QFTs and engage in a co-development process where dedicated QFT simulators could perhaps be considered and designed. This co-design process is further discussed in a dedicated Supplemental Section (see Supp.~\ref{suppl:codevelopment}).

\subsubsection{Combining classical computing with quantum simulation
\label{supp:QFT-hybrid}}
In the near or even far term, it may be beneficial to incorporate classical-computing methods in quantum simulation so that quantum resources can be allocated more efficiently. This can enable simulations that otherwise would not be possible. For example, the spectrum of an interacting quantum many-body system can be determined with a quantum computer via known phase-estimation algorithms, which are nonetheless costly and not suited for the present NISQ hardware. As an alternative, the variational principle of quantum mechanics can be taken advantage of to find the lowest-energy eigenvalues by inputting a parametric initial state and evolving it to non-trivial final states. Measuring the Hamiltonian matrix element and using classical optimizers can then allow putting rigorous and ideally tight upper bounds on energies. Variations of such variational quantum algorithms (VQAs)~\cite{peruzzo2014variational,mcclean2016theory}, including quantum approximate optimization algorithms (QAOAs)~\cite{farhi2014quantum}, have been developed in recent years and have been applied to a range of problems~\cite{tilly2021variational}. In the context of QFTs, the low-lying spectrum of lattice Schwinger model~\cite{klco2018quantum,lu2019simulations,kokail2019self}, and a first VQA applied to the SU(2) LGT coupled to fermions in 1+1 D has appeared in recent years~\cite{atas20212}. Such a variational approach can also be adopted to find good approximations to the eigenstates of the system, which in turn can inform conventional lattice field theory calculations that need good interpolating operators for the states~\cite{avkhadiev2020accelerating}. They have also been suggested to facilitate the scattering problem on a quantum computer~\cite{liu2021towards}.

Another area worth exploring is to determine if the conventional lattice-field-theory program can be accelerated with quantum-computing routines. Here, questions that need to be answered include: Can quantum processors be useful in enhancing importance sampling of quantum configurations, especially when sign or signal-to-noise problems are encountered, or when a critical slowing down halts the simulations
toward the continuum limit of lattice field theories? Can quantum platforms speed up inversion of poorly-conditioned large matrices~\cite{harrow2009quantum}, enhance semidefinite programming for construction of optimal field interpolating operators or improving the non-perturbative bootstrap in conformal field theories~\cite{simmons2015semidefinite} (which can benefit from quantum advantages from Gibbs sampling~\cite{bao2019quantum,brandao2017quantum}), or incorporate more economically the factorial growth of the number of contributions in nuclear correlation functions? Initial investigations along these lines have started~\cite{avkhadiev2020accelerating, chang2019quantum} and will likely continue as the fault-tolerant quantum processors become closer to reality.

Another interesting path is to use the input from conventional lattice field theory to accelerate quantum simulation of QFTs. An example is to leave the computationally demanding task of time evolution to the quantum simulator, but learn the density matrix of the initial state using conventional lattice field theory methods, and encode this information in the quantum simulator to avoid the costly state-preparation step, as proposed and tested in Refs.~\cite{harmalkar2020quantum,gustafson2021toward}. Unfortunately, if the initial state is hard to compute classically, for example in situations where a sign or signal-to-noise problem is present, the protocol is faced with the usual issues, but can otherwise be valuable. Tensor-network inspired state-preparation techniques can also be taken advantage of in this context. The final way in which classical resources will play an important role is in refining the classical-quantum boundary beyond which practical quantum advantage occurs in HEP~\cite{gustafson2021large}.

More hybrid classical-quantum simulation protocols, in the spirit of those described in this section, can offer lower quantum-resource requirements compared with full quantum simulations, and can therefore speed up the progress in addressing the physics drives of this program.

\
\

\noindent
In summary, a quantum-simulation program in lattice field theory is starting to form. Its development over the next decade will be guided by many insights from the development and growth of the conventional lattice field theory program. It will rely on the existing and forthcoming advancements in quantum algorithm and hardware, but equally importantly on new proposals and algorithms tailored to the unique features of QFTs, and particularly gauge field theories. QIS-literate quantum field theorists will be the key to these advances, but they can accomplish much more by collaborations and exchanges with the quantum-information science and technology community. As a result, quantum simulating QFTs will be a highly interdisciplinary area of research in HEP over the next decade, and is expected to combine theory, algorithm, and hardware research in exciting new ways.
\vspace{0.5 cm}

\section{Underlying simulations:  Simulating effective field theories}
\label{suppl:EFT}
\noindent
An effective field theory is a field theory that reproduces a given underlying field theory in a particular kinematic regime~\cite{weinberg1979phenomenological}. Examples are four-Fermi effective theory~\cite{fermi1934tentativo}  which describes weak interactions of elementary or composite particles with momentum transfers below the electroweak symmetry-breaking scales, hadronic and nuclear effective theories~\cite{weinberg1990nuclear,rho1991exchange,weinberg1990nuclear,ordonez1996two,weinberg1992three,bedaque1998nucleon,kaplan1998new,bedaque1998effective,kaplan1998two,van1999effective,birse1999renormalisation} which describe interactions of hadrons and nucleons with low momentum transfers, and Soft-Collinear Effective Theory~\cite{bauer2000summing,bauer2001effective,bauer2001invariant,bauer2002soft} which describes the interactions of partons collinear to each others, or the interactions of collinear and soft partons. Since an EFT must reproduce the same physics as the underlying theory in the regime where it applies, relevant information about short-distance physics must still be contained in the EFT. The generation of EFTs follows a renormalization-group matching procedure with which irrelevant high-scale degrees of freedom are integrated out but their effect is encoded in the coefficients of the generated operators and their scale dependence. These coefficients are typically called Wilson coefficients, or in the context of hadronic and nuclear EFTs, commonly named as low-energy constants (LECs). When the matching occurs at energy scales far above the hadronic scale, these coefficients can be calculated perturbatively, given that the strong interaction is asymptotically free. The LECs of the hadronic and nuclear EFTs can only be constrained by matching to experiment or via a non-perturbative evaluation within the underlying QCD theory using the lattice-QCD method.

The separation of the  overall dynamics into  short-distance contributions and long-distance effects provides many advantages. First, short-distance contributions at high scale are often reliably calculable in perturbation theory. The remaining long-distance physics is then described by the dynamics of EFTs, which in many cases are easier to compute compared to the full theory. This is because when expressed in terms of the effective low-energy degrees of freedom, the interactions are often simpler, and new symmetries may manifest themselves. Furthermore, the EFT has to describe the dynamics only at long distances, and so the EFT needs to be simulated over a much smaller energy range than the full theory.

Since simulating the underlying QFTs at a wide range of energies amounts to enormous computational cost, and that we do not know the valid QFTs of nature at a high scale (assuming these are QFTs), the renormalization-group matching and the emergence of effective descriptions of interactions will continue to be a valuable approach in handling problems in HEP in the quantum-simulation era. Quantum simulations have been studied for several different EFT applications: within SCET, for parton-shower physics, for parton distribution functions, and within hadronic and nuclear EFTs. These applications come with distinct features, as will be described in this supplemental material.

\subsubsection{Simulations within Soft-Collinear Effective Theory}
Most cross sections of interest at high-energy colliders such as the LHC are dominated by events that contain only a relatively small number of hard jets. A jet of particles is a collection of particles that have a small invariant mass relative to one another. Jets, therefore, contain a collection of energetic particles that are moving in the same direction (they are said to be collinear with respect to each other). Jets can interact with one another through the exchange of soft particles, which do not raise the invariant mass of any of the jets significantly. The long-distance dynamics that describes the interactions between collinear particles within a jet, and with soft particles that can mediate long-range interactions between the jets, is described by matrix elements in SCET~\cite{bauer2000summing,bauer2001effective,bauer2001invariant,bauer2002soft}. On the other hand, the short-distance physics describing the initial production of the jets is contained in the Wilson coefficients. The full dynamics of such cross sections are, therefore, described by perturbatively-calculable short-distance coefficients and non-perturbative long-distance matrix elements of SCET operators. 

The collinear dynamics within a given jet in SCET are the same as those of the full theory, albeit with a much smaller dynamical range required. Thus, the same techniques as those developed for the quantum simulation of full QCD using Hamiltonian lattice-gauge-theory techniques are applicable in this case, see Suppl.~\ref{suppl:QFT}. The important simplification, however, is that the dynamical range has to be much smaller, therefore reducing significantly the number of degrees of freedom required for a reliable simulation. The first quantum-simulation algorithms for SCET dynamics has been developed in Ref.~\cite{bauer2021simulating}, and will continue to be improved in the coming years.

The soft dynamics of SCET, which is required to calculate the non-perturbative soft matrix elements, is dramatically simpler compared with the dynamics of the full theory. This is because in the soft sector of SCET, collinear degrees of freedom are integrated out, leaving only static color sources in the theory. Furthermore, interactions with soft fermions are power suppressed in SCET, such that the soft dynamics is described by those of a pure gauge theory in the presence of static Wilson lines. Since the upper energy range of the soft theory is about three orders of magnitude below that required for a typical full theory simulation, a dynamical lattice-gauge-theory simulation of this soft theory requires a factor of $10^9$ fewer lattice points than the corresponding simulation in the full theory. This leads to a dramatically smaller resource requirement, such that \emph{ab initio} non-perturbative calculations of soft SCET matrix elements seem feasible on quantum devices in the not-so-distant future. A simplified version of this theory, namely a scalar field theory interacting with Wilson lines was studied in Ref.~\cite{bauer2021simulating}, along with necessary quantum circuits and a small-scale simulation on an IBMQ quantum device. The quantum simulation of a scalar field theory is by now very well studied~\cite{jordan2011quantum,klco2019digitization,macridin2018electron,macridin2018digital,somma2015quantum}, and the addition of Wilson lines is relatively straightforward. The progress in this problem is correlated with that in simulating lattice gauge theories in 3+1 D, which is discussed in Suppl.~\ref{suppl:QFT}.

\subsubsection{Simulating parton showers}
Collider events typically contain a very large number of final-state particles. This can be traced back to the fact that emissions can happen over a very large range in energies, giving rise to logarithmic enhancements of emission cross sections, and particle emission at low relative transverse momenta does not involve a small coupling constant. For this reason, describing exclusive events with a large number of individual particles can not be achieved using the full underlying theory. As already mentioned above, most of the particles in typical collider events are grouped into a relatively small number of jets, such that most emissions are either collinear or soft relative to the particles produced in the underlying hard interaction. In the collinear limit, one can show that the emission is dominated by single amplitudes, such that quantum-interference effects become subdominant and the emission can be described by a probability~\cite{particle2020review,sjostrand2006pythia,bahr2008herwig++,gleisberg2009event,bauer2007event}. In the soft limit, emission can still be described at leading order by an emission probability if one takes the limit of a large number of colors ($N_c \to \infty$). The radiation of collinear and soft particles can therefore be described in a probabilistic manner, using a Markov-Chain algorithm. 

The probabilistic nature of a Markov-Chain algorithm makes including quantum-interference effects challenging. Quantum-interference effects that can be present are effects at subleading orders in $1/N_c$, and interference between amplitudes with different intermediate particles and different internal kinematics. Initial studies have emerged in recent years to formulate the parton-shower description using quantum-simulation methods. For example, a quantum algorithm has been developed in Ref.~\cite{nachman2021quantum} to reproduce the regular parton shower while by computing all possible amplitudes at the same time, it can be constructed to include quantum-interference effects. In particular, it was shown in a toy model that this quantum parton shower was able to include quantum-interference effects arising from different intermediate particles with an exponential improvement in efficiency compared to known classical algorithms. 

More work is required to develop a full parton-shower algorithm for the SM, and to find ways to include the most relevant quantum-interference effects, such as color interference. One can also look for other implementation strategies of the quantum parton shower such as those proposed in Refs.~\cite{Williams:2021lvr,macaluso2021quantum}.

\subsubsection{Simulating parton distribution functions}
Parton distribution functions are a crucial non-perturbative ingredient in any prediction for hadron colliders. In rough terms, they describe the probability to find a parton with a given momentum fraction (or other properties for generalized distributions) inside a hadron. PDFs can be combined with partonic scattering processes to make predictions for hadronic scattering cross sections.

PDFs are defined by the Fourier transform of a matrix element of an operator containing two quark fields separated by in the light-like direction, evaluated between a hadronic state. A Wilson line is required when forming the matrix elements to make the product of two fermion fields at different locations gauge invariant. The fact that the two quark fields are separated by a light-like direction makes the computations of this matrix element difficult when using traditional lattice-QCD techniques. This is because lattice-QCD calculations are performed in Euclidean spacetime to avoid a sign problem. While several techniques have been put forward to allow the calculation of PDFs in lattice QCD, with successful results in several cases~\cite{aglietti1998model,liu2000parton,ji2013parton,chambers2017nucleon,ma2018extracting}, a precise determination of PDFs, and particularly the Bjorken-$x$ dependent distributions at small and large values of $x$, remains challenging. Such determinations are further complicated when PDFs are desired for atomic nuclei used in high-energy collider experiments. Quantum computers can, in principle, compute directly the forward light-cone matrix element relevant for PDFs using a first-principles lattice-QCD framework. 

Several proposals have been put forward in recent years to demonstrate how PDFs can be accessed on a quantum computer~\cite{lamm2020parton,echevarria2021quantum,li2021partonic,mueller2020deeply,qian2021solving}. One important aspect of the matrix element relevant for PDFs is the gauge structure, in particular the Wilson line required for gauge invariance. One may forego the complications in simulating the gauge structure and calculate the simpler hadronic tensor on a quantum computer, then extract the PDF using perturbative information for the partonic scattering~\cite{lamm2020parton}. This is a similar strategy as to traditional extractions of parton distributions from experimental measurements. However, by using  the results for hadronic tensor obtained on a quantum computer rather than using experimental measurements, one has the ability to turn certain  contributions off to make the extraction of particular PDFs easier. Alternatively, the Wilson line can be explicitly constructed using plaquette operators or fermion hopping terms, allowing for an estimate of the full quantum computation of the PDF~\cite{echevarria2021quantum}. Finally, a PDF calculation in the Nambu–Jona-Lasinio (NJL) model was performed in Ref.~\cite{li2021partonic} using a variational Ansatz for the proton state, and following Ref.~\cite{pedernales2014efficient} for the correlation function.

At this stage, it is not clear what the realistic computational-resource requirements are for computing PDFs and hadronic tensor to given accuracy, as the complete algorithms, including that needed for preparation of hadronic states in QCD on a quantum computer, are either non-existing or premature. Over the next decade, theoretical and algorithmic research will improve upon these initial analyses.

\subsubsection{Simulations within hadronic and nuclear EFTs
\label{suppl:EFT-Nuclear}}

Low-energy EFTs of nuclear interactions are important for the HEP mission as they provide a consistent framework to describe the nuclear targets used in high-energy experiments. For example, accurate simulations of these theories in systems of many nucleons are needed to compute semi-exclusive neutrino-nucleus cross sections that are needed for event analysis of long-baseline neutrino experiments but are difficult to measure directly in the laboratory (see Supp.~\ref{suppl:neutrino} for more details). They are also important to evaluate nuclear matrix elements required to analyze experiments looking for neutrinoless double-$\beta$ decay~\cite{engel2017status,cirigliano2022neutrinoless}, and for direct detection of potential dark-matter particles~\cite{baudis2013signatures,andreoli2019quantum}.

At low energies, these interactions contain both two- and three-nucleon forces and are often non-local at higher orders~\cite{machleidt2011chiral}. Simulation of these theories in a general single-particle basis, like the harmonic-oscillator basis, is structurally very similar to a quantum chemistry calculation, given the non-relativistic nature of interactions. Therefore, quantum-simulation techniques developed there can be ported, upon necessary modifications, to the nuclear case (see e.g.~Refs.~\cite{bauer2020quantum,mcardle2020quantum,cao2019quantum} for recent reviews on the progress there). A few main differences are the presence of two additional fermionic species (to account for different isospin components of the nucleon), the presence of three- (and higher-) nucleon interactions, which in the most general case can lead to Hamiltonian composed by $\mathcal{O}(N^6)$ distinct terms for a $N$-qubit system, as well as the presence of pion-exchange interactions that in the static approximation, lead to long-range two-nucleon potentials. This puts a lower bound of the same order on the number of quantum gates needed to simulate real-time evolution, a requirement to extract inelastic cross-sections. Elastic cross sections could be constrained by ground-state calculations alone and this could be performed with possibly much lower quantum resources using variational methods, at the price of requiring a number of measurements scaling as $\mathcal{O}(N^6)$ (see, e.g., Ref.~\cite{cerezo2021variational} for a review of quantum-variational techniques). A substantial reduction in the number of measurements can be obtained in a number of ways: grouping operators into commuting families that can be measured at the same time (see, e.g., Ref.~\cite{yen2020measuring}), extending algorithms that estimate the two-body fermionic reduced density matrix (see, e.g., Ref.~\cite{bonet2020nearly}) to the three-body reduced density matrix, exploring tensor factorization schemes (see, e.g., Ref.~\cite{huggins2021efficient}) or adopting randomized-measurement strategies~\cite{huang2020predicting,garcia2021learning}. To date, a number of variational calculations of small nuclei have been carried out on quantum hardware using simplified low-order interactions and small models spaces with encouraging results~\cite{dumitrescu2018cloud,lu2019simulations,stetcu2021variational}. Different methods for the direct extraction of elastic matrix elements have been also tested on the simple problem of deuteron photo-disintegration~\cite{roggero2020preparation}.

Quantum simulations of ground-state properties of nuclear targets could become an important application for future quantum technologies but the accuracy achieved by classical methods for these problems sets a very high bar for quantum advantage~\cite{hergert2020guided}. Dynamical properties of nuclei like inelastic cross sections are instead much more challenging to compute classically, especially for semi-exclusive scattering in medium- and large-mass nuclei, and quantum simulations have the potential of being impactful already on smaller-scale problems. Improving the efficiency of real-time dynamics will require additional techniques, some of which can also be beneficial for ground-state calculations. A crucial aspect to consider it the choice of encoding for the fermionic degrees of freedom into qubits. Different mappings have, in fact, different requirements in terms of the number of qubits for a given number of nucleons $A$ and of single-particle states $N$ but also lead to spin representations of fermionic operators of different weight (the number of qubits they act non-trivially on). A conceptually simple and very common choice is the Jordan-Wigner mapping which requires $N$ qubits and $\mathcal{O}(N)$ weight for fermionic operators~\cite{jordan1928paulische}. The large weight induced by this mapping is a major obstacle in the optimization of quantum-simulation algorithms. Alternative schemes have been proposed to overcome the difficulty, for example, the Bravyi-Kitaev mapping requiring $N$ qubits but only $\mathcal{O}(\log(N))$ weight for fermionic operators~\cite{bravyi2002fermionic}, as well as recent semi-local mappings with fixed weight~\cite{verstraete2005mapping, ball2005fermions, whitfield2016local, steudtner2019quantum,setia2019superfast, chen2020exact, chien2020custom, derby2021compact}. Some work has already started to explore the relative benefits of different fermionic mappings in nuclear ground-state simulations~\cite{stetcu2021variational} and an important direction in the future will be to extend these studies to dynamics as well as exploring alternative fermionic mappings that also feature error-correction properties at the expense of requiring a larger number of qubits~\cite{setia2018bravyi} or auxiliary fermion methods~\cite{ball2005fermions,verstraete2005mapping,steudtner2018fermion}. 

Another important research direction in the field is the improvement of the approximation to the time evolution operator which usually constitutes  a dominant contribution to the overall resource cost of algorithms that calculate nuclear cross sections (see, e.g., Ref.~\cite{roggero2019dynamic}). A number of alternatives to the traditional Trotter-Suzuki decomposition have been proposed in the last few years, like the Taylor-series method~\cite{berry2015simulating}, quantum signal processing~\cite{low2017optimal} and quantum stochastic drift protocol (QDRIFT)~\cite{campbell2019random} among others. An initial comparison of the gate cost incurred by some of these in simulations of pionless EFT have been carried out in Ref.~\cite{roggero2020quantum}. For more general nuclear EFTs, which generate Hamiltonians with a large number of terms, randomized schemes like QDRIFT can provide important gains in gate requirements as their cost does not depend directly on the number of terms in the Hamiltonian but only on its norm. Finally, alternative methods for inelastic nuclear cross sections that do not directly require the implementation of time evolution but instead use a quantum device to estimate appropriate Chebyshev moments have been proposed~\cite{roggero2020spectral}. As some of the more advanced time-evolution schemes adopt already a Chebyshev expansion~\cite{low2017optimal}, techniques of this kind could help in reducing the gate cost of simulations of inclusive cross sections.

Finally, addressing questions regarding the composition of the interior of neutron stars will not only shed light on our understanding of the phase diagram of strong interactions and the nature of the densest form of matter known in the cosmos, but also impacts the analysis of the gravitational-wave emission from merging neutron stars. \emph{Ab initio} many-body calculations based in the underlying few-body hadronic interactions are computationally challenging, but are crucial in discerning the role of non-nucleonic degrees of freedom in the equation of state of neutron stars. Hadronic EFTs are generalized forms of nuclear EFTs, and provide a systematic framework for describing interactions among hadronic degrees of freedom including baryons. These include chiral perturbation theory~\cite{leutwyler1994foundations,gasser1984chiral,scherer2003introduction} and baryon chiral effective field theory~\cite{bernard2008chiral,becher1999baryon,jenkins1991baryon}. Quantum computing the many-body problem with the hadronic EFT Hamiltonian can evade the intrinsic sign problem in fermionic simulations, and is similar in nature to the computations described in this section for many-body nuclear systems.

Note that first-principles QCD-based simulation of matter will still be necessary to complement this program. First, they give direct access to the phase diagram of QCD without any assumption about the presence of hadronic degrees of freedom. They can, therefore, investigate the existence of the conjectured exotic quark phases~\cite{kogut2003phases}. Moreover, QCD-based determination of the few-hadron interactions will be needed to constrain the hadronic EFT at a range of energies and densities, which is experimentally challenging for short-lived exotic hadronic states. A quantum-computing-based lattice-QCD program in the few-hadron sector can, therefore, be matched to a quantum-computing-based nuclear and hypernuclear structure program via the EFTs (see e.g., Ref.~\cite{lu2019simulations} for a first example), similar to the matching program currently promoted using conventional methods~\cite{barnea2015effective,contessi2017ground,bansal2018pion}. The output of such efforts will subsequently impact research in HEP, including problems in the intensity and cosmic frontiers.
\clearpage

\section{Simulator requirements:  Analog simulators}
\label{suppl:analog}
\noindent
Functional quantum simulators today can be found based on superconducting qubits, trapped ions, and neutral atoms, but other platforms rooted in molecular, optical, and solid-state quantum systems are being developed and explored for simulation application. Devices optimized for programmable quantum simulation have been advanced, particularly those based on trapped-ion and neutral-atom modalities. The requirements for HEP simulations place a high priority on the development of simulators of large scale that exhibit good quantum coherence and high-quality readout, but perhaps more importantly significant amount of control beyond what is customary in the past for quantum simulation of simpler spin systems. While access to superconducting qubit and trapped-ion systems that operate in a digital mode have become available both commercially and through DOE facilities and programs for some time, it is only recently that several new modalities are reaching sufficient maturity to offer cloud-accessible hardware for analog or hybrid analog-digital simulations. In the following, examples of analog-simulation platforms with potential for simulating HEP models will be reviewed and the challenges ahead will be discussed. 

\subsubsection{Cold neutral atoms}
Systems of neutral atoms have been used as analog quantum simulators since the first production of quantum degenerate Bose and Fermi-gases~\cite{anderson1995observation,demarco1999onset,ketterle2008making,hadzibabic2002two}, 
building on fundamental advances in laser- and evaporative-cooling and leveraging the ability to trap and manipulate atoms with the optical dipole force derived from intense laser light. These devices, particularly utilizing purpose-built optical lattices, have been used to probe quantum phases of many models in condensed-matter physics~\cite{hofstetter2018quantum}, 
explore fundamental phenomena in many-body dynamics~\cite{langen2015ultracold} 
and many other topics. A major advance in these systems came with the advent of the quantum gas microscope (QGM)~\cite{bakr2009quantum,sherson2010single,cheuk2015quantum}, 
which for the first time permitted very large systems of neutral atoms to be placed in a single many-body state through evaporative cooling, state preparation, and measurement. In many cases, QGMs can be used with atomic isotopes with both strong and weak collisional effects as well as tunable by external fields~\cite{chin2010feshbach}. 
The single-particle dynamics is also highly configurable, with a range of optical lattices with controllable geometry, dynamical couplings, and floquet engineering~\cite{schafer2020tools} 
available. QGMs have since been constructed with both bosonic and fermionic isotopes, and much of the benefit of these systems has been devoted to quantum-degenerate-gas phenomena where the indistinguishability of particles plays a key role. Many of the systems studied in this way exhibit complex quantum phenomena intractable with classical numerical simulation, particularly those centered on strongly-correlated fermionic systems, and emerging studies on (non-Landau-Ginzburg) topological matter. To date, very few highly-programmable platforms of this type have emerged, and QGMs have been engineered physically from the ground up to support study of a particular quantum simulation. New modalities~\cite{wu2021concise}, however, promise to accelerate the rate at which a given physical platform can be retasked onto new problems.      

More recently, cold atoms held in independently movable optical tweezers and driven by laser light into Rydberg states have been developed as a platform for quantum information processing~\cite{endres2016atom}. 
In these configurations, atoms are prepared singly near a motional ground state in movable optical tweezers, and arranged into a defect-free initial register. Relying only on laser cooling, they achieve an order-of-magnitude higher repetition rate than QGMs, and provide the run-time flexibility of nearly arbitrary geometry. Information is encoded in the internal state of the atom, and strong interactions are introduced on demand using programmed optical coupling to an excited Rydberg state. These flexible systems have advanced quickly in the last five years, demonstrating long coherence times and   
high state-resolved detection fidelities~\cite{madjarov2020high}, elementary gates~\cite{levine2018high,levine2019parallel}, 
and recently the capability of moving atoms dynamically while preserving their coherence~\cite{bluvstein2021quantum}. 
In the simplest configurations, neutral-atom platforms of this type realize an effective spin interaction with a Hamiltonian of the form $H=\sum_{ij} V_{ij}\sigma^z_i \sigma^z_j+\sum_i (\Omega_i\sigma^x_i+\Delta_i\sigma^z_i)$, where $\Omega_i$ represents the strength of a local optical driving field coupling atoms into an excited Rydberg state, $\Delta_i$ is its detuning from resonance, and $V_{ij}=C/|\vec{r}_i-\vec{r}_j|^6$ is the strength of dipole-dipole interactions decaying quickly with distance for atoms positioned by tweezers at $\vec{r}_{i,j}$. The constant $C$ is characteristic of the isotope and Rydberg state chosen in the implementation. In many cases, systems of cold atoms can be understood approximately from Rydberg blockade physics, where it is assumed that no two atoms within a blockade radius (determined from the condition that $V_{ij}=\Omega_i=\Omega_j$) simultaneously transition into the excited state. This is one simple basis, for example, for programming network-graph analysis~\cite{pichler2018quantum} 
into an analog simulator using the arbitrarity of tweezer positions $\vec{r}_i$---for ``geometric'' (unit-disk) graphs, nodes correspond to atoms and edges represent closeness as determined by the blockade distance. In the last year, neutral atom devices utilizing Rydberg interactions have reached sufficient maturity to enable first demonstrations of spin-liquid states~\cite{semeghini2021probing}, probed quench dynamics of systems of unprecedented scale~\cite{bluvstein2021controlling}, extended modalities to gates with shuttled qubits~\cite{bluvstein2021quantum}, and utilized long-ranged and anisotropic interactions~\cite{scholl2021microwave}. Neutral-atom Rydberg systems can represent information using multiple internal states, and newly utilized isotopes of atoms have exhibited extremely long coherence times due to their special level-structure~\cite{madjarov2020high}.  A number of commercial providers have already, or are near completion of, structured cloud-access to these platforms. 

Emerging platforms are likely to use modalities that extend beyond fixed geometry, qubits, and also beyond single- and two-qubit gates~\cite{saffman2010quantum}, to dynamically reconfigurable connectivity~\cite{bluvstein2021quantum}, qudits and multilevel gate implementations~\cite{saffman2010quantum}, and efficient multi-qubit entangling gates~\cite{khazali2020fast}.

\subsubsection{Trapped ions}
In trapped-ion systems, atomic ions are confined through electromagnetic fields, where the balance between their Coulomb repulsion and the external confinement produces a well-defined crystal, typically in 1D or 2D in space. Each atomic ion stores an effective spin in the same atomic states used as the most advanced frequency standards, so their idle coherence is extremely long~\cite{Wang2017}. The spins are manipulated through either optical or magnetic fields. By coupling off-resonantly the spin degrees of freedom to the motional excitations of the ion crystals, a tunable long-range Ising interaction between all pairs is achieved. This Ising Hamiltonian engineered is $H=\sum_{ij} J_{ij}\sigma^x_i \sigma^x_j+\sum_i B\sigma^z_i$, where the native spin-spin interaction can be described by an approximate power law $J_{ij} \sim J_0/|i-j|^\alpha$, with $0.5 \lesssim \alpha \lesssim 2$, as well as an effective magnetic field that can be induced via applying Stark shifts on the ions. Trapped-ion systems have been used to simulate spin models exhibiting such Ising Hamiltonians in up to a hundred spins \cite{Monroe2021,Britton2012,Zhang2017}. These experiments take advantage of only a global pair of Raman laser beams to engineer the effective Hamiltonian, and because the phonon is virtual in the process of coupling the spins, the scheme used is rather insensitive to the phonon occupation~\cite{Monroe2021}. On the other hand, being a second-order process in the strength of the native spin-phonon coupling, contributions from the first-order processes present a source of error in the Hamiltonian engineering. In practice, the coherence times observed in the experiment, given this error source and the environmental and implementation noise, is of the order of a few inverse Ising coupling. Such simulations are, therefore, suitable for problems with fast time dynamics, such as the evolution of quantum-many body systems after a quench~\cite{jurcevic2014quasiparticle,richerme2014non,Jurcevic2017direct,Tan2021domain}.

Simulating QFTs requires more capabilities to be introduced to the scheme above. As a first requirement, a similar level of control as in the digital trapped-ion computers can be employed in the analog systems such that individual addressing of the ions with one or multiple pairs of Raman beams will be possible. With the trapped-ion interaction graph being fully connected, a wide range of possible spin-spin Hamiltonians can be engineered in any dimension~\cite{Porras2004effective}, including a Heisenberg Hamiltonian $H=\sum_{ij} [J^{(xx)}_{ij}\sigma^x_i \sigma^x_j+J^{(yy)}_{ij}\sigma^y_i \sigma^y_j+J^{(zz)}_{ij}\sigma^z_i \sigma^z_j]+\sum_i J_i^{(1)}\sigma^z_i$ with independently tunable coefficients, which is of relevance to simulating the lattice Schwinger model~\cite{davoudi2020towards}. Furthermore, the three-spin Hamiltonians of the form $H=\sum_{ijk} J^{(3)}_{ijk}[\sigma^+_i \sigma^+_j \sigma^+_k + \sigma^-_i \sigma^-_j \sigma^-_k]+\sum_{ij} J^{(2)}_{ij}\sigma^+_i \sigma^-_j+\sum_i J_i^{(1)}\sigma^z_i$ can, in principle, be generated with tunable coefficients, relevant for simulating simple QLMs among other models~\cite{Bermudez2009competing,andrade2021engineering}. In fact, $N$-spin interactions have now been proposed in trapped-ion systems with exciting possibilities for quantum-simulating various models of relevance to HEP, and for simplifying digital circuits~\cite{katz2022n}. Moreover, spin-phonon interactions have been taken advantage of in proposals for simulating the Schwinger model in an analog fashion~\cite{yang2016analog}, and when combined with phonon-phonon gates can enable a hybrid analog-digital approach to simulating the same model and other coupled fermionic-bosonic field theories~\cite{davoudi2021towards,casanova2011quantum}, among other models~\cite{lamata2014efficient,mezzacapo2012digital}. Further, the availability of multiple atomic states to encode higher-dimensional spin systems~\cite{low2020practical,senko2015realization} adds more flexibility to the analog simulator trapped-ion toolbox. Many of these less conventional but more enabling schemes have yet to be tested in experiment, particularly when sizable simulations are desired. Later in this section, the challenges and prospects for going beyond simple QFTs and toward gauge-theory simulations of the Standard Model are discussed.

\subsubsection{Other simulating platforms}
Other atomic, optical, molecular, and solid-state simulating platforms are quickly advancing~\cite{altman2021quantum}. Among the notable platforms with potential prospects for simulating models of relevance to HEP are: i) Laser-cooled polar molecules~\cite{carr2009cold,gadway2016strongly} which combine the strong electric dipolar interaction with one to hundreds of internal states with transitions at convenient frequencies. Progress is being made to address challenges such as cooling the system to the many-body ground state and addressability at the single-molecule level~\cite{liu2019molecular,anderegg2019optical}, ii) Cavity quantum electrodynamics~\cite{vaidya2018tunable,noh2016quantum,hartmann2016quantum,welte2018photon} which enables coupling of distant atoms mediated by the exchange of the photon, or when viewed as many-body systems of cavity photons, the strong photon-photon interactions mediated by atoms, with the possibility of achieving programmable interactions and emergent geometries~\cite{periwal2021programmable}. Future developments will tackle the challenge of combining strong atom-light coupling with local tunability of the interaction of individual atoms with the cavity. iii) Superconducting circuits~\cite{kjaergaard2020superconducting,krantz2019quantum}, which are the leading solid-state quantum-simulation platform, and are among the most geometrically flexible simulators. They operate both in a digital mode, see Supp.~\ref{suppl:digital}, and in an analog mode, by taking advantage of linear and nonlinear couplings between their elements, i.e., cavity photons and artificial two(multi)-level atoms, to tailor interactions. Hyperbolic lattices using circuit QED~\cite{blais2021circuit,schmidt2013circuit} are being developed~\cite{kollar2019hyperbolic}, which present  an potential platform for simulation of quantum effects in curved spaces~\cite{boettcher2020quantum}. iv) Dopants in semiconductors such as in Silicon, which provide fermionic degrees of freedom encoded in conduction-band electrons that populate the array, and nuclear-spin degree of freedom of the dopant sites. Long-range Coulomb interactions makes possible the quantum simulation of many-body problems modeled by an extended Fermi-Hubbard Hamiltonian~\cite{wang2021quantum}. Effective control of tunable parameters in a donor-based simulators~\cite{harvey2017coherent,he2019two} is achieved through controlled placement of donor atoms with atomic-scale precision~\cite{wyrick2019atom,wang2020atomic}, presenting further opportunities for quantum simulation in these platforms in the upcoming years.

\subsubsection{Challenges and needs of HEP simulations}
While the systems described in this section are highly versatile, and in some cases programmable, as exhibited by a large number of published proposals for novel implementations of physics models, mapping the native interactions to a target model requires a subtle co-design process between hardware providers and simulation builders, see Supp.~\ref{suppl:codevelopment}. In analog simulators, this consists of matching both the target Hilbert space and dynamical evolution by tailoring programming degrees-of-freedom available in different devices. Since the native architecture and the target system often do not share the same fundamental symmetries or connectivity of interactions, computing resources are inevitably lost to encoding overheads. As newer modalities come online with efficient multi-qubit operations, new opportunities to mitigate these overheads will likely arise. Furthermore, in systems that present fermionic and bosonic degrees of freedom, as in certain cold-atom simulators, the issues with encoding fermionic statistics and the need for low truncation of bosonic modes can be circumvented.

When it comes to simulating gauge-field theories of relevance to the SM, the existing simulators are still far away from presenting the essential capabilities and it is conceivable that the ultimate solution will be a hybrid approach: native and more versatile set of operations are used but the evolution is digitized to avoid the need for the challenging task of simultaneously applying increasingly-large number of terms in the local Hamiltonian of gauge-field theories toward higher dimensions and more complex groups. While trapped-ion and cold-atom platforms have been more extensively explored in the context of quantum simulation of gauge theories, it is yet to be known if the other platforms briefly described in this section will provide unique opportunities for this task. Intuitively, one may consider molecules as more natural candidates for encoding certain non-Abelian gauge theories, or take the superconducting cavities with hyperbolic interaction graphs as natural candidates for simulating physical models of relevance to HEP in curved spacetime. Nonetheless, such connections must be carefully examined and developed. While theoretical research in illuminating these questions must be pursued, these platforms will first need to continue to show their capability in simulating simpler and more standard benchmark models before being considered seriously as candidates for HEP applications.
\vspace{0.5 cm}

\section{Simulator requirements:  Digital computers}
\label{suppl:digital}
\noindent
Digital quantum computing implements algorithms as sequences of universal gate operations on the underlying qubit architecture, based on, e.g., superconducting systems, cavity QED, neutral atoms, or trapped ions.  Once a native set of gates and connectivity constraints are identified for different systems, high-level algorithms can be compiled and targeted for a variety of hardware.  Today, a number of platforms have reached commercial maturity, and an ecosystem has developed around cloud access for these devices.  Current technologies are limited today in scale below the 100-qubit level, and exhibit noise levels that prevent gate sequences in excess of typically tens of gates per qubit before the likelihood of error is of order unity. A number of metrics have been proposed for characterizing performance of digital quantum-computing devices, most prominently the quantum volume~\cite{cross2019validating} and the algorithmic qubit count~\cite{algorithmic-qubit}. Both are related to the qubit count and the viable (single- and two-qubit-) gate depth, taking into account the resource overhead consumed by overcoming limitations of qubit connectivity. A wide variety of mechanisms can be used to access commercial and non-commercial digital computing platforms, ranging from cloud-service providers who work in concert with commercial hardware vendors to government-sponsored programs like DOE’s quantum testbeds, in which low-level hardware access to trapped-ion and superconducting qubit modalities can be made to introduce specialized control. Gate sequences can be specified in a growing number of high-level quantum-computing languages, often in a hardware-agnostic way, and be provided to vendors for remote execution.  While early software was developed to run isolated single-instance algorithms, some support now exists for more sophisticated techniques, including hybrid methods like VQE~\cite{peruzzo2014variational,mcclean2016theory,tilly2021variational} and QAOA~\cite{farhi2014quantum} applied to gate-based algorithms.

Trapped-ion quantum computers provide high numbers of algorithmic qubits. The qubits exhibit substantially long coherence times (minutes), and the single-qubit and two-qubit gates can be operated with $>99.96\%$ and 98.5\%--99.3\% fidelities, respectively, and with state-preparation and measurement error $<0.5$. Ion chains with tens of functional qubits are achieved in current systems, and importantly, are all mutually connected. This feature reduces circuit depths as the qubits do not need to be swapped and placed in proximity of each other to enable a gate between them. The scaling of trapped-ion digital quantum computers will follow two predictable paths~\cite{Brown2016}. First, the control of large ion crystals will be limited by the dense motional (phonon) modes that mediate their interactions. This will demand that ion crystals be broken into spatially separated modules, with the quantum connections provided either by shuttling ions in space between the modules~\cite{Kielpinski2002,Pino2021demonstration} or by photonic interconnects between nodes~\cite{Monroe2014}. Notably, this latter method for modular expansion allows full connectivity, even to scale. Second, as the number of qubits grows, the limits of quantum-gate fidelity will begin to determine the achievable circuit depths. At this point, it will become necessary to employ error-correction encoding to further the coherent quantum evolution. Error-correcting codes with trapped ions are particularly efficient, owing to the full connectivity and natively low errors in trapped-ion systems~\cite{Egan2021,ryan2021realization,Postler2021}. 

Another digital quantum-computing platform of particular importance is Superconducting electronics-based Quantum Processing Units (SC QPUs). Being semiconductor systems, SC QPUs can leverage extremely high purity solid-state materials and sophisticated material-processing techniques to produce QPUs with coherence times $\sim 100$’s $\mu$s, high single-qubit, 99.8\%--99.96\%, and two-qubit, 98\%--99.6\%, gate fidelities~\cite{qubit-aqt}, and to realize chips of varied qubit counts from a few to close to a hundred. These QPUs offer the capability to execute proof-of-concept quantum algorithms that move the boundaries of producing and utilizing quantum entanglement on demand. Such capabilities are available through commercial platforms and DOE testbeds and are increasingly being used by scientists and engineers to explore novel quantum simulations of physical systems, development of benchmarking and error-mitigation protocols that feedback into improved SC QPU performance, and many other algorithmic implementations. Imperfections in control systems, QPU design, electromagnetic environment, and materials are the culprits of sub-optimal QPU operation performance. SC QPUs developers are constantly engaged in attempting both incremental progress and innovation to attack these imperfections.

Another closely-related architecture, colloquially known as a 3D quantum processor, utilizes superconducting cavities to store quantum information encoded in the vast number of levels ($N$) available from each single harmonic mode. Using a multitude of levels, as opposed to a qubit that uses only two, effectively realizes $2\log(N)$ all-to-all connected qubits per harmonic mode. Moreover, these $2\log(N)$ `virtual' qubits are natively entangled which alleviates the necessity of many sequential conditional gates to create a maximally-entangled state as is required by the more traditional hardware. Beyond the direct encoding of quantum information, superconducting cavities are compatible with modern error-protection strategies and highly-efficient error-correction protocols, which makes this platform a promising quantum-computing platform. Conveniently, the workhorse in this hardware implementation, known as ancilla, is a conventional transmon qubit, which allows one to control the quantum processor using ordinary microwave pulses. Perhaps un-intuitively, the `limited' coherence time of the transmon does not factor into the coherence time of the quantum hardware provided the transmon coherence time exceeds the required gate duration which is easily achievable. The cavity coherence time, however, does affect computational performance, which is the motivation of the Superconducting Quantum Materials and Systems Research Center at Fermilab to use the typical superconducting radiofrequency cavities that are known for their world-record relaxation times. Ongoing R\&D is focusing on the demonstration of record-high coherence of cavities coupled to qubits. Multi-cell cavities can implement powerful QPUs, with qudit-based and all-to-all connection among multiple radio-frequency modes with high quality factors.

While offering rich and novel opportunities, new programming paradigms, compilation, and transcription processes will be necessary to understand how best to utilize these devices in HEP applications. For the purpose of simulating gauge-theories of the Standard Model, for example, it may be best to work with the formulations and encodings that retain the locality of interactions, both among bosons and among fermions and bosons, depending on the connectivity pattern inherent to the hardware architecture used, in order to minimize costly swap operations.

As the scale is increased in many platforms, considerable new opportunities will arise to implement encodings of logical qubits that are resilient against gate, initialization, and readout errors. While proof-of-concept studies of error-correction with different encodings have been completed on various platforms, see e.g., Refs.~\cite{egan2021fault,ryan2021realization,krinner2021realizing,heeres2017implementing,fluhmann2019encoding,nguyen2021demonstration}, no system to date has sufficient resource to meaningfully utilize it in real-world algorithms. Through deep user interaction, co-design, algorithm innovation, and continued improvement in digital-computing hardware, the field is moving toward accelerating the timeline toward universal fault-tolerant quantum computation. Though it is likely that analog quantum simulation and non-error-corrected algorithms represent the majority of near-term applications in HEP, an important thread of research will need to be devoted to the anticipation of larger-scale devices with error-correction schemes. For the purpose of quantum simulation, any model Hamiltonian with local and semi-local interactions can be decomposed into smaller units of time evolution through the Trotter-Suzuki expansion~\cite{lloyd1996universal} and efficiently decomposed into universal gate operations. Other simulation algorithms have proven costly for present-day hardware but can be considered in the fault-tolerant and large-scale era of quantum computing.

\vspace{0.5 cm}

\section{Simulator requirements:  Simulation in the NISQ era}
\label{suppl:NISQ}
\noindent
Noisy Intermediate-Scale Quantum era refers to an era of quantum computing where noisy non-error-corrected operations are performed on devices with $\sim$50--100 qubits~\cite{preskill2018quantum}. Since the qubit and gate overhead of error correction is substantial, and given that the gate errors are not yet in the fault-tolerant regime, the circuit depths beyond a few hundreds to a few thousand gates will push the limits of sustained quantum coherence, and the noise affecting the computation must be mitigated in certain ways. While the full power of universal quantum computing may not be reached for decades to come, it is still imperative to not dismiss the possibilities provided by the NISQ hardware, and to take advantage of available devices with various capabilities and capacities. The reason is that a quantum device capable of generating and manipulating an entangled state of a hundred qubits, even though for a short period of time and subject to noise, pushes the limits of classical simulation. It may be possible to find algorithms that are more resilient to noise, along with features of the simulated theory that are more robust to imperfections in the implementation, so that useful predictions can be made with the NISQ device, especially as targeted and effective noise-mitigation methods are developed and applied. Furthermore, pure algorithmic developments can lead to theoretical scalings that sometimes prove pessimistic, while empirical scalings revealed via NISQ-hardware implementation may turn out to be far more promising. Since the quantum-hardware design is at a stage where the winning candidates are still to be known, algorithm benchmarks given system architecture, connectivity graph, degrees of freedom, and other parameters can inform the next design choices.

\subsubsection{NISQ algorithms and strategies}
There are a few promising approaches that can produce meaningful results even with shallow circuits and without active error-correction sequences in the NISQ era. One well-formulated approach involves utilizing hybrid algorithms that divide classical and quantum resources such that only steps which require probing a large combinatorial space are executed on qubits. The variational quantum eigensolver~\cite{peruzzo2014variational,mcclean2016theory} and quantum approximate optimization algorithms~\cite{farhi2014quantum} employ this hybrid approach. These algorithms have been widely studied for their application in combinatorial optimization, semi-definite programming, and other contexts~\cite{tilly2021variational}, including finding ground-state energy of physical systems, see e.g., Refs.~\cite{peruzzo2014variational,kandala2017hardware,dumitrescu2018cloud}. Similar techniques have gained attention for direct simulation of field theories, including finding the lowest-lying energy spectra of low-dimensional Abelian and non-Abelian lattice gauge theories~\cite{klco2018quantum,lu2019simulations,kokail2019self,atas20212} and formulating costly scattering problems in terms of NISQ variational algorithms~\cite{liu2021towards}.

Another approach to increasing the computational power of shallow-circuit quantum algorithms for the NISQ hardware involves increasing the size of the operational Hilbert space by using three-level systems, i.e., qutrits, or even higher dimensional logic. This type of encoding increases the connectivity of the quantum network and reduces the number of gates required for a given computational task. As an example of the value of a four-level qudit, the four spin-isospin state of the nucleon (spin-half neutron and proton) can be encoded in a qudit and the spin-dependent dynamics of nuclear systems can be more efficiently encoded and simulated, as demonstrated in Ref.~\cite{holland2020optimal}. This example also highlights another approach to NISQ computing, that is to encode the portion of system's dynamics that is harder to track with classical methods on a quantum processor, and combine the results when possible with the components that can be evaluated with more ease on a classical computer, in this case the evolution in position space of the the nucleons. More examples of such a hybrid classical-quantum simulation approach for QFTs are provided in Supp.~\ref{supp:QFT-hybrid}, highlighting strategies that are in play to maximally gain from limited quantum power of NISQ hardware by porting more tractable computations to classical hardware. 

The translation of a quantum algorithm into specific gates is not unique and various approaches can differ significantly in their actual quantum-resource requirements. For example, the original quantum algorithm for computing scattering amplitudes in a scalar QFT~\cite{jordan2011quantum,jordan2012quantum} uses an algorithm by Kitaev and Webb~\cite{kitaev2008wavefunction} for Gaussian-state preparation which generates scattering wavepackets.  This algorithm requires polynomial qubits and gates, but on NISQ devices, the required resources can be daunting, hence needing more optimal Gaussian-state generations~\cite{bauer2021practical}. Additionally, various techniques have been proposed to design minimal and maximally expressive quantum circuits for the NISQ
era (e.g., Refs.~\cite{sim2019expressibility,funcke2021dimensional}) and to split large quantum circuits into small circuits executable on NISQ devices (e.g., Ref.~\cite{peng2020simulating}).
Finally, gauge-invariant simulation of gauge theories may save qubit resources but locality of interactions may be lost. Many digital simulations of lattice gauge-theory simulations on the NISQ hardware to date~\cite{martinez2016real,klco2018quantum,nguyen2021digital,klco20202,ARahman:2021ktn,ciavarella2021trailhead} have been enabled by fully (when possible) or partially removing the redundant degrees of freedom (via appropriate gauge transformation or by solving the Gauss's law) or by imposing symmetries. This simplified simulation can also reduce leakage to the unphysical sector of the gauge theories due to algorithm and hardware errors. Taking full advantage of such strategies for simulating the gauge theories of the SM, while paying attention to the resulting time complexity of the simulation, will be essential in the resource-limited era of quantum computing.

\subsubsection{Error mitigation for NISQ computing}
Given that in the NISQ era operations are noisy and the system is not error corrected, error mitigation is required to make sense of the measurements. There are two general types of errors that can affect a quantum simulation. The first is called readout error, and is responsible for the fact that the measurement of a qubit does not reproduce the amplitude of the qubit before the measurement. The reason for such an error is often due to decoherence that happens on the time scale required to measure a qubit. The second is called a gate error, and arises due to imperfect implementation of quantum operators (gates). On current publically-accessible machines, readout errors typically dominate, with error rates between 1 and 10\%, and the largest gate errors occur in entangling operators, such as the CNOT operation, with a typical error rate slightly below 1\%. However, since gate errors accumulate with the number of gates in an algorithm, gate errors become dominant for longer circuits.

There are several techniques that have been proposed over the years to mitigate these errors. Readout errors can be measured relatively easily by preparing a system in a given state and recording the measurement of this state. This results in a readout-error matrix. Note that given that the number of possible states is exponential in the number of qubits, the size of this matrix, and therefore the computing resources to determine it, scale exponentially with the number of qubits. However, given this matrix, one can invert it using a variety of techniques~\cite{kandala2017hardware,song201710,chen1904detector,geller2020efficient,nachman2021categorizing,wei2020verifying,hamilton2020scalable,maciejewski2020mitigation,nachman2020unfolding,maciejewski2020mitigation}, where the last method is directly related to techniques developed for detector unfolding in HEP collider physics. This corrects a given measurement on average. Recently a technique using active readout error correction, where each qubit to be measured is encoded into multiple qubits using repetition or hamming codes, have been developed~\cite{gunther2021improving,hicks2022active}. This allows for readout-error correction for each individual measurement. Examples of other approaches to readout errors are discussed in Refs.~\cite{kandala2018extending,funcke2020measurement,nation2021scalable}.

Gate errors depend on the details of their implementation in the quantum simulator, and most of the commercial systems available do not provide enough details to  directly mitigate the errors introduced. One often distinguishes between stochastic and  coherent errors. Coherent errors preserve state purity. They are typically small mis-calibrations in control parameters. Coherent errors usually produce similar errors in consecutive executions of a quantum circuit and lead to a systematic bias in the output. Stochastic errors can be understood as either coherent errors with randomly varying control parameters or as processes that entangle the system with its environment. Stochastic errors can often be modeled as depolarizing noise. A method for converting coherent errors into incoherent errors is randomized compiling~\cite{wallman2016noise,li2017efficient,cai2019constructing,cai2020mitigating}. The dominant gate error is typically found in gates that introduce entanglement between qubits, and a common universal gate introducing such entanglement is the CNOT gate. 

Among the common noise-mitigation techniques for CNOT errors (but which can be also used for other gate errors) are zero-noise extrapolation (ZNE) and its variants~\cite{li2017efficient,temme2017error,kandala2019error,klco2018quantum,he2020zero,sun2021mitigating,bauer2021computationally}. ZNE magnifies the noise in the system either by changing the gate operation time or by replacing each CNOT gate by a larger odd number of CNOT gates. Since applying two CNOT gates in direct succession amounts to an identity operation (${\rm CNOT}^2 = \mathbbm{1}$), this does not change the circuit in the noiseless limit, but it increases the amount of noise. This allows to measure the result of the circuit at different noise levels, which can be used to extrapolate the measurement to the noiseless limit. For the dominant depolarizing-error channel, one can prove that this techniques reduces the noise level in a quantifiable way, with the amount depending on the order of the extrapolation used~\cite{he2020zero}. Other techniques have been proposed which either require knowledge of the noise model or use circuits to measure it~\cite{endo2018practical,endo2019mitigating,otten2019recovering,otten2019accounting,strikis2021learning,urbanek2021mitigating}. 
Another interesting idea is to use the symmetries of the simulated system to control and mitigate errors~\cite{bonet2018low,otten2019noise}. An interesting question that requires more study is whether gauge invariance in gauge-field theories can be used to detect and mitigate errors in the quantum simulators~\cite{stryker2019oracles,raychowdhury2020solving}.

While noise mitigation is not a requirement that is specific to simulations for HEP applications, the HEP community might be able to contribute to this important problem. For decades, high-energy physicists have mastered the science of distinguishing tiny elusive signals from dominant background noise, and of sheltering sensitive particle detectors and readout instruments from environmental noise. It is conceivable that further noise-suppression techniques at the level of the hardware, and noise-mitigation schemes at the level of algorithms, can emerge from communications and collaborations among the two communities in the comping decade(s).

\subsubsection{Quantum annealers}
Quantum annealers are non-universal quantum optimizer machines.
Annealers~\cite{johnson2011quantum} with more than 5000 qubits 
and associated simulators (with white noise) are accessible via cloud, and their performance is being benchmarked across a wide range of problems, from commercial optimization problems, to finance, to applied mathematics, biology, materials, nuclear physics, and HEP. One niche target for quantum annealers is to find optimal or near-optimal solutions to NP-hard problems, although their ultimate performance relative to classical competitors is unknown.
In HEP research, the types of problems being addressed include  track reconstruction in the analysis of collider data, e.g. Refs.~\cite{Mott:2017xdb,Zlokapa:2019tkn,bapst2020pattern} and real-time simulations of elementary components required for QFTs~\cite{ARahman:2021ktn,Illa:2022jqb}.
Annealers use time-dependent adiabatic switching to 
minimize the quadratic unconstrained binary optimization (QUBO) cost function, or mappings to an Ising model, constructed to address the problem of interest, including eigenstates and energies, real-time dynamics using the Feynman Clock algorithm, and  problems that are NP-Hard (or NP-complete), such as max-cut problems. 

The quantum devices have extensive hardware controls (e.g., controlling parameters of the annealing process) that require higher-level expertise and allocations of device time to gain experience with. The D-Wave systems have a low bar for entry to new users, with extensive documentation available online, including tutorials.   
Algorithms that have been developed and integrated into HEP computational efforts, such as lattice QCD, have been found to be valuable to working with the annealing systems~\cite{ARahman:2021ktn,Illa:2022jqb}. Beyond optimizations for classical NP-Hard problems, a challenge currently faced in addressing classes of HEP problems is the dimensionality of their QUBO representations. The need for binary-string representations of complex numbers defining wavefunctions provides a challenge to the scale of problems that can be embedded on the device.  With next-generation annealers having increased qubit connectivity and more qubits, larger problem instances can be addressed, but  paths for tackling problems at scale (beyond  pre-conditioning) remain to be established.

\subsubsection{Benchmarking for HEP}
In the context of quantum simulation of HEP, the principles of locality and unitarity that played a crucial role in the establishment of the SM also guarantees that dynamics of QFTs of relevance to nature can be factorized into unitary building blocks that act on a finite Hilbert space independent of the size of the system. This is the central argument for quantum advantage in quantum simulation of physical systems~\cite{lloyd1996universal}. Developing, optimizing, and testing these building blocks is a near-term task that can be started with existing NISQ devices following the roadmap discussed in Sec.~\ref{suppl:QFT}. 
In this context, error mitigation, device comparisons, and benchmarking are essential for making progress.

In particular, it is important to introduce metrics to assess the progress 
made and compare devices. This is an essential aspect of the hardware and software developments that allows to match applications to platforms that best implement them. Indices have been proposed to compare mitigated real-time evolution with the exact Trotter evolution for different quantum devices~\cite{gustafson2019benchmarking}. Furthermore, sophisticated methods such as cycle benchmarking~\cite{erhard2019characterizing} have been developed to characterize the systematic errors of NISQ devices and applied to the quantum Ising model~\cite{yeter2022measuring} with a specific IBMQ machine.

The HEP community will continue to develop and apply new NISQ-tailored simulation algorithms in the coming years. It would be interesting to see what the limit of NISQ computing is for HEP applications and whether one can anticipate any quantum advantage in HEP in the NISQ era of quantum computing.

\vspace{0.5 cm}

\section{Simulator requirements:  Software and compilers}
\label{suppl:software}
\noindent
Early investments by DOE HEP have enabled research into algorithms and applications for the study of HEP problems on the NISQ devices. These early explorations have been critical to understanding how quantum computers may be used to solve challenging physics problems. Experiences with these early science explorations have demonstrated the need to expand the programmability of quantum-computing devices for testing new and creative quantum algorithms.

\subsubsection{Current state of software and compilers}
A variety of low-level quantum-computing programming languages akin to assembler-level programming models in classical computing have been developed over the past decade, and several of those have been used recently to program actual hardware devices. They are low-level programming languages that work well for knowledgeable experts. Widely used software systems include the IBM’s Qiskit~\cite{qiskit}, Xanadu’s Strawberry Fields~\cite{killoran2019strawberry}, Google’s Cirq~\cite{cirq_developers_2021_5182845}, and Microsoft’s $Q\#$ language~\cite{svore2018q}. 

Transpilers and early low-level compilers are being developed to optimize the user-generated operations---generally referred to as circuits---into a shorter and more efficient set that can be translated to pulses and other fundamental operations needed to execute on quantum hardware. Transpilation tools utilize multiple pathways to optimize performance utilizing a wide variety of methods, including graph optimization~\cite{sivarajah2020t}, unitary synthesis~\cite{younis2021qfast}, and ZX-Calculus~\cite{van2020zx} to name a few. Many of these tools also incorporate hardware-specific knowledge, such as topologies and error rates to deliver the best performance possible. Full error correction is simply not feasible for near-term quantum computers due to limits in scale and the levels of noise. Nearly all scientific simulations require (mostly) non-automated application of error-mitigation approaches to achieve reliable results~\cite{endo2021hybrid,he2020zero,urbanek2021mitigating,kandala2019error,huggins2021virtual}.

To ease the programmability for quantum computers, higher-level programming frameworks will be needed. Some efforts have led to OpenFermion~\cite{mcclean2020openfermion} for materials and chemical sciences and Pennylane~\cite{bergholm2018pennylane} that is primarily focused on machine learning. Within the DOE, efforts are underway in the Office of Advanced Scientific Computing Research to develop a software toolkit that includes higher-level programming models~\cite{nguyen2021composable}. No higher-level programming models suitable for HEP are currently available, though homegrown tools are being developed at various institutions.

\subsubsection{HEP programming needs}
To advance HEP research on quantum computers, and enable the broader HEP community to readily participate, models and approaches for programming quantum computers need to mature to a level in which abstractions and library-based methods can be adopted to expedite programming and ensure portability.\footnote{See a Snowmass whitepaper on ``Quantum Computing Systems and Software for High-energy Physics Research'' ~\cite{humble2022snowmass} for further discussions.} This will require the development of programming languages, potentially domain-specific languages, that can readily express the discretization and complex interactions, such as the need to describe the coupling of fermions and bosons and open quantum systems, provides key building blocks, such as non-trivial arithmetic subroutines that are prevalent in quantum simulating the dynamics of non-Abelian lattice gauge theories, or large (sparse) matrix inversion for expediting Monte-Carlo-based routines present in the conventional lattice-gauge-theory program. Considering programmability needs for digital simulations, it should be noted that there is a significant overlap with other science domains, including quantum chemistry and materials, that should be exploited. 

In addition to programming languages and libraries, the HEP research community would benefit from efficient compilers that can take the programs written in high-level programming languages to efficient low-level code that can be run efficiently on quantum-computing hardware. Inter-operability of software is a key feature to ensure low barriers for HEP users to access across different systems, and avoid being restricted to a single, potentially proprietary software pathway. Open-source software and tools will be essential for broad access, and tools that enable researchers to debug, verify, and validate quantum-computer programs and results will be needed. Last but not least, a hybrid computing model involving tightly-coupled classical and quantum hardware could be explored to enable the usual offloading of special-purpose calculations to a QPU~\cite{XACC,nguyen2021quasimo}.

Much of the discussion in this section has so far focused on digital quantum computing, along the lines of the field’s usage of classical computers. However, one could also utilize quantum computers as well-controlled analog systems, as discussed in Supp.~\ref{suppl:analog}. For example, some recent work has utilized the cross-resonance effect of a superconducting quantum computer to replace sequences of one-and two-qubit gate operations~\cite{babukhin2020hybrid, gonzalez2021digital}. This very different quantum-computing paradigm will require that users to effectively engage across the quantum-computing stack, and to have access to software layers that generate pulses needed to drive the interactions of relevance to HEP applications, such as those of lattice gauge theories.

\clearpage

\section{Quantum ecosystem: Hardware co-design and accessibility}
\label{suppl:codevelopment}
\noindent
While it is in principle possible to construct gate-based quantum-computing algorithms without detailed knowledge of the underlying hardware, the resource constraints of today’s systems imply severe limitations on the scale of computations, and imply a need for hardware-aware efficient implementations and designs that take advantage of a detailed understanding of system interactions and connectivity, as well as error and decoherence mechanisms. This favors tightly-coupled collaborative models for executing science programs, in which hardware providers and domain experts in HEP applications co-develop scientific agendas. This requirement also motivate the need for accessible and adjustable devices such that HEP and other domain scientists can use to benchmark algorithms and test new ideas.

\subsubsection{Co-design models and special-purpose devices}
In the early days of the conventional lattice field theory program, research groups at a number of institutions around the world started to successfully design, build, and run classical-computing hardware dedicated to QCD simulations. One such development in the U.S. became the precursor to the so called QCD-on-a-chip (QCDOC) machine, that via technology transfer to IBM became the underlying design of successful Blue-Gene series of high-performance supercomputers~\cite{boyle2005overview,boyle2004qcdoc}. A similar approach might prove valuable in the design and deployment of quantum hardware for lattice-gauge-theory problems. In fact, such a co-design program is even more relevant in the quantum-computing era as the simulating hardware is itself a physical system used for a physics application. For example, as mentioned in Supp.~\ref{suppl:analog}, Rydberg arrays offer the user a degree of flexibility in the design of the architecture of atom configurations hence model interactions, trapped ions offer native bosonic degrees of freedom, i.e., phonons excitations that can be used to encode more efficiently the bosonic fields and the interactions beyond those possible with only qubits, and cavity QED hyperbolic lattices can be engineered to simulate physics in curved geometries, along with many other examples. In the next 5--10 years, scientific resources should be employed to explore a broad range of co-designed hardware schemes, fostering a close collaboration between the HEP and the QIS communities.  

In order for the co-design process to be effective, the interaction between researchers within institutions that deploy quantum hardware is essential. This allows to find synergies that can lead to collaborations on the design of algorithms, protocols, and even devices, with immediate applications. Among successful examples in the present day is the DOE-funded Advanced Quantum Testbed~\cite{aqt} at the LBNL, where internal and external collaborators and users are engaged in developing benchmarking methods that offer new insights into noise processes and associated mitigation strategies in quantum hardware. This was done through an iterative process of protocol execution feeding back into the external theory teams. To assist in this process, application-specific circuit synthesis, programming sequences, readout, feedback, and noise mitigation were employed for maximizing circuit depth and learning outcomes. Through these and other interlocked collaborations and user interaction, the depth and breadth of science that can be accomplished through novel technology can be advanced, enabling science that goes beyond cloud-based quantum-computing resources.

In addition to DOE programs at the national laboratories, several commercial hardware service providers now offer in-house services to help design efficient hardware-cognizant implementations on demand~\cite{IBM,Quera,Xanadu,QCI,Dwave}. 
This will remove the wall between the users and cloud-based services with preset features and will allow the needs of the domain scientists to be communicated directly to the hardware and software developers. At the university setting, experimental efforts have long been in harmony with theoretical proposals in quantum simulation, and there is a direct relationship between strong theory groups and strong experimental groups at the universities. Such a model will potentially be the key to success in arrangements outside academia as well. For HEP applications, resources and opportunities in co-design efforts will need to be directed toward systematic evaluation of the best methods to simulate Hamiltonians of relevance to HEP efficiently and accurately, design and perform state-preparation and state-tomography in strongly-interacting QFTs, understand the role of decoherence mechanisms, and how best to incorporate symmetries and gauge and scale invariance, or take advantage of them to mitigate or correct the errors. 

\subsubsection{Hardware-access model}
In the rapidly-evolving landscape at the interface between HEP and QIS, it is important for the HEP community to identify optimal ways to have access to state-of-the-art programmable quantum computers or quantum-simulation experiments. In the coming years, the ability to access multiple platforms with a variety of architectures would lead to rapid progress on target problems in HEP, as each platform would be most suitable for particular problems. As already mentioned, this access could be through existing experimental setups, either located in universities and national laboratories or by contracting private companies. At the software and compiler front, the most popular frameworks for programming universal quantum computers such as Qiskit, Cirq, and pyQuil have so far been offered as free and open-source resources. This has been critical for the HEP community, and will be important that this tradition continues as these packages become more advanced and gain commercial value. By making quantum machines available to researchers, companies can ensure competitiveness in the future, which is a practice now pursued by several countries, see e.g., Ref.~\cite{pollet2022france,gougard2021germany,flinders2020uk}. Finally, it is also often necessary to run relatively large simulations on classical computers, as numerical tests and benchmarks in the quantum-simulation problems is a necessity. It is conceivable that these calculations start to require a non-negligible time at High Performance Computing (HPC) facilities, and this requirement should be recognized ahead of time.

One successful model for how accessibility of resources in HEP can be ensured is the USQCD-collaboration model. USQCD~\cite{usqcd} is a meta collaboration of the majority of U.S. lattice gauge theorists. It has benefited from dedicated hardware at Brookhaven National Laboratory, Fermi National Accelerator Laboratory, and Thomas Jefferson National Accelerator Facility, and provides access to these resources for the members via proposal solicitation and a rigorous review process. It also makes community-wide bids for time on leadership-class computing facilities managed by the DOE, which has led to successful acquisition of computing time for large-scale lattice-QCD projects over the years. One could envisage a meta collaboration of U.S. HEP-QIS theorists whose main goal would be to strengthen individual efforts and coordinate their access to different hardware platforms available at national labs or commercial sites. It would also facilitate communications between people developing QC hardware and the needs of the theorists developing algorithms and software for HEP applications. USQCD has also been  successful in bringing up a collaborative algorithm-development team feeding into independent and basically competing individual groups. It can be thought of as a ``virtual U.S. lab'' that could be emulated for the QIS community. It is also important to identify the most pressing needs of the potential users. For these reasons, the communication between the HEP researchers and experimentalists developing new computational platforms need to be enhanced, and these communications can be strengthened by a unifying voice within the HEP community.

\
\

\noindent
In summary, in the quantum-era of computing, progress in quantum simulation of HEP problems will be fostered by close collaborations among HEP and QIS scientists with a range of expertise, leading to a valuable co-design process and potentially to special-purpose platforms. This should be accompanied by direct and broad access to leading hardware and software technologies, and via a community-approved mechanism that can unify, advocate for, and coordinate the accessibility needs of HEP scientists.
\vspace{0.5 cm}

\section{Quantum ecosystem:  Skill sets and workforce development}
\label{suppl:workforce}
\noindent
\noindent
Quantum simulations of processes essential to HEP research objectives require a diverse and inclusive quantum-ready workforce with skills  that extend significantly beyond those traditionally in HEP. This workforce will be distributed and collaboratively trained at universities, national laboratories and even some technology companies. The pipeline for recruitment should start earlier, extending into high school. A complementary and robust portfolio of funding mechanisms, career development opportunities, career paths and mentoring will be required to create and sustain this workforce.

\subsubsection{Quantum workforce for HEP}
Workforce development is a key component in advancing HEP's quantum-simulation objectives. Through the natural re-alignment of some of the existing workforce, and the recruitment and training of junior scientists, the HEP workforce is expected to evolve in a way to enable HEP objectives to be accomplished using quantum simulation.  The intrinsically interdisciplinary nature of quantum simulation for HEP requires growing or collaborating with expertise in QIS, computer science, applied mathematics, statistics, material science, nuclear physics, and more, with scientists, engineers and developers residing in national laboratories, universities, and technology companies.  At this time, this growth and engagement has been underway in HEP for approximately five years, and it will need to continue and be further enhanced to meet HEP scientific  objectives during the next decade and beyond.

The growth of the quantum-ready HEP workforce is occurring in the context of significantly-increased national quantum information science and technology research and development to meet broad-based present and future needs of the nation~\cite{NAP25196}.  When mature, the associated quantum economy is expected to be comparable to the silicon economy, and with a workforce of scale to enable a robust and efficient pipeline from basic research to commodity devices in quantum computing,  communication,  sensing, and more.   It is important to recognize that the growth of the quantum-ready HEP workforce, with skill-sets necessary to meet HEP objectives, will be of significant benefit to the national quantum activity. Historically, HEP scientists have contributed to, or moved into, other areas of research, such as computing, ``big data", and device fabrication, and have shaped the development of those areas in substantial ways.  The same is anticipated in the emerging quantum era.

The present ``Snowmass" process is the first in which QIS is being considered an integral part of HEP research and development, and consequently, estimates of the resources and training required to meet the HEP QIS needs are less sophisticated than in the more mature areas.  This next period is one of research and development in basic science through to advanced technologies. This includes continued calibration, development, resource estimation and engagement, including in the development of a quantum-ready HEP  workforce with successful career opportunities. 

\subsubsection{Required skill-sets and the role of universities, national labs, and private sector}
The skills required for quantum simulations of processes essential to the HEP mission are diverse~\cite{hughes2021assessing}.
Experiences from the last decades of workforce development in lattice QCD, event generators, and other HPC-intensive research areas, have provided models of engagement that are expected to have applicability.
Collaborations of scientists, engineers, and developers will be required to coalesce the skill-sets required to accomplish quantum-simulation objectives.   
These skills include HEP phenomenology, quantum field theory and quantum mechanics, lattice field theory, high-performance computing, statistical analysis, experimental design and optimization, machine learning and artificial intelligence, software stack development, quantum- and classical-computing algorithms, quantum-circuit design, implementation and optimization, and more.
Integration with quantum hardware development through HEP co-design efforts will further broaden this skill-set. 
The education and training programs for these skills are distributed within  universities, national laboratories and the private sector. Therefore, coordination between these sectors is required, and all have an important role to play. For instance, one expects universities to provide basic to advanced education in physics, including quantum mechanics, quantum field theory, and particle theory, along with basic to advanced education in statistics and applied mathematics.  
National Laboratories  continue to develop the talent produced by universities and technology companies, and can assemble significant  teams to address  mission-specific objectives.
Technology companies, in part, perform focused research and development to produce commodity products.  For quantum simulation, this includes developing quantum computers and quantum devices, developing robust software stacks, and Application Programming Interfaces.

These educational and training pipelines need to be both strengthened and expanded in the area of QIS.
While Physics departments at universities specialize in education in physics, both experimental and theoretical, an increase in the the advanced offerings in broadly-defined quantum field theory and quantum many-body systems, for instance through courses in HEP, nuclear physics and condensed matter physics, would directly benefit efforts in HEP quantum simulation. It would also be beneficial for universities to integrate quantum-circuit design between physics, engineering, and computer science. Such moves are already underway, but the inclusion of HEP relevant offerings would be beneficial.

Given the scope of the anticipated quantum ecosystem, 
the pipeline for quantum education and training should begin before students reach university.
It would be beneficial for introductory QIS to become part of high-school level science requirements. The QuarkNet program~\cite{quarknet} for developing the future technical workforce stands as an existing example of how the community could integrate HEP-QIS at this level.  
Increased  opportunities for high-school students to work with scientists at universities and national laboratories are expected to be beneficial to quantum-simulation efforts, including in HEP.

The National Quantum Initiative (NQI) centers~\cite{NQI} and QuantiSED programs effectively support the development of junior scientists, including postdoctoral fellows and graduate students.  While Covid-19 has provided a substantial limitation to in-person meetings, workshops and collaborations, the regular meetings and activities have proven valuable in education and training.
More QIS-training and educational opportunities at all career levels are required.
For example, there will be utility in supporting HEP-specific summer schools in quantum simulation, like the recently-started Quantum Computing Internship for Physics Undergraduates Program at FNAL, alongside the more general schools in QIS that are organized by Los Alamos National Laboratory. Programs analogous to the ``lattice-QCD Hackathons" are anticipated to have a future role in training for quantum-simulation research.
Named fellowships are used successfully at labs and universities to attract talent at the postdoctoral and Ph.D. levels for strategic purposes, and such fellowships for HEP quantum simulation would be of value.

\subsubsection{Career development and mentoring}

In order to meet the needs of the nation at all scales, including those for HEP quantum simulation, career development and mentoring 
has to be available at all stages and levels, for those so wishing. 
It is important to emphasize that this area of research did not exist until recently, and further it has a different required skill-set for the participating workforce.  Until recently, the  workforce in this area has come exclusively from transitions from other areas, and the first generation of QIS-HEP scientists is currently being trained. 
Therefore, the landscape of research in HEP-QIS has been changing quite rapidly, but is anticipated to stabilize during the next decade.  There is only anecdotal information available regarding career paths, how one is best positioned for a career in QIS, what courses  students should take and so forth.   As universities introduce new courses, there will be differences between offerings at different institutions.  This makes mentoring by members of our community, at all levels, crucial, as is engagement with those in the national laboratories, the business sector, the legal sector, etc.  This needs to be addressed during the upcoming period. Programs such as the National Science Foundation's ExpandQISE~\cite{ExpandQISE} will be essential in empowering smaller and less-involved institutions with the QIS research.

The next period will be one of experimentation to identify the types of mentoring and career guidance that are effective.
Importantly, one type of institution will no longer be sufficient to impart all of the skills that are required, and collaborative mentoring should be explored and assessed.  Therefore, new and creative ways of educating and training junior, mid-career, and senior scientists, engineers, and developers are to be encouraged and explored.  Importantly, this exploration should include drawing talent from previously under-considered groups.  

In designing/exploring potentially new avenues of developing  this needed HEP-QIS workforce, 
flexibility should be considered so that under-represented sectors of the population have unimpeded and equal access to QIS education and training, unlike previous generations, and are welcomed into the HEP community.
For this inclusive and diverse community to thrive, its members must be supportive of each other and of the broader scientific goals, and behave respectfully, ethically, and honestly. 
Codes of Conduct are now an integral part of formal collaborations in the community, and will  continue to be important in future efforts.
HEP should participate in the on-going national-level discussions on this crucially-important subject.

One of the lessons learned from the recent decades of SciDAC~\cite{scidac} and HPC-oriented projects in HEP is in regard to the career path of senior researchers who start as domain scientists and evolve into specialist algorithm designers and coders for HEP.  This set of talented and essential scientists and developers will typically leave the community to assume  positions in the private sector.
This impacts the workforce, while building connections with the private sector.  
Without identifying more desirable career paths within HEP, this situation may repeat itself in the HEP quantum-simulation era.

Joint positions between laboratories and universities have been successful in sustaining the HEP workforce. In contrast, the use of strategic bridge positions  to strengthen the HEP workforce has been under-utilized. As there is a strategic need to increase the workforce in quantum simulation for HEP, consideration should be given to using joint and bridge positions at universities at the Assistant Professor level (or higher) to attract talent into the program, and also to address the expected in-balance between Ph.D. graduates and postdocs leaving HEP research for a career in the private sector.  The HEP experts in universities are sure to be attracted to technology companies, and efforts should be undertaken to future-proof the HEP program for the long-term health of the community.

\
\

\noindent
In summary, the HEP community is starting to develop a quantum-simulation workforce in order to address important objectives of the HEP scientific mission.
For a diverse and inclusive quantum-ready HEP workforce of  scientists, engineers, and developers to be at the forefront of quantum simulation, 
close collaborations between universities, national laboratories, and technology companies are critical, see also Supp.~\ref{suppl:partnership}.
The types of resources and the nature of engagements that are anticipated to be required for this to be successfully accomplished were outlined in this Suuplemental Section. 

\vspace{0.5 cm}

\section{Quantum ecosystem: Academia, government, industry partnership}
\label{suppl:partnership}
\noindent
Quantum simulation in HEP is a unique research discipline that cannot grow in isolation. When algorithm, software, and hardware developments are concerned, there are multiple players in the field, from universities to national laboratories to private companies, each contributing to the advancements according to the organizations' strategic agendas. It is important for  HEP scientists to acknowledge this entangled web of contributors, and to form partnerships that are long-lasting and mutually beneficial.

\subsubsection{The role of national laboratories in advancing  quantum simulation}
HEP has a long-standing history of driving developments in classical high-performance computing (HPC), an example of which is the emergence of IBM's BG/L,P,Q supercomputers starting from the collaborative efforts of Columbia University, RIKEN, and and Brookhaven National Laboratory (BNL).
An important outcome of this co-design effort and HPC testbed(s) were the technical skills acquired through ``hands-on" experience(s).  Similar collaborative efforts are emerging in HEP in the areas of machine learning and artificial intelligence, as well as quantum computing, the results of which will benefit many scientific and technology communities.

The QuantiSED program(s) through DOE HEP that successfully supports  QIS in HEP includes integrated collaborative efforts including FNAL, LANL, BNL, LBNL  with universities and technology companies, and brings together experts from HEP and other areas, such as computer science, atomic-molecular-optical physics, machine learning, nuclear physics, condensed matter, and QIS, with a significant effort related to quantum simulation. Research groups at the laboratories have skills related to, and are focused on, accomplishing the objectives of the laboratory.
The QuantiSED programs provide a strategic supplement to these skills through a closer coupling to outside experts.
It is expected that such activities will be essential for further developing the HEP-QIS workforce going forward.
It is also expected that the QIS expertise at the national laboratories with a HEP footprint to continue to increase significantly during the next period to meet HEP objectives.
A similar evolution is anticipated in university-based HEP groups, but in a way that complements the growth at the laboratories. This is, in part, to provide an education beneficial to the laboratories.

For HPC-centric HEP objectives, such as those advancing lattice QCD and event generators, 
the SciDAC program has played a critical role, bringing together domain scientists and experts in applied mathematics, computer science, statistics and more through the SciDAC Institutes. These valuable projects are typically jointly supported by DOE HEP and ASCR.  Analogous opportunities for advancing quantum simulation to meet HEP scientific objectives would provide a valuable resource.

\subsubsection{Engagement with Technology Companies}
A robust engagement between universities, national laboratories, and technology companies is essential. There is significant and  growing expertise already in the private sector in all areas of quantum.  To minimize the time from development of an idea, prototyping and subsequent integration into a commodity device, close ties with technology companies are vital.  
In order for HEP to be at the forefront of quantum simulation, engagement with technology companies is critical, and will be mutually beneficial.
This engagement spans areas such as quantum hardware, operating systems, and  application programming interfaces.

A somewhat novel aspect of HEP-QIS is the need to address IP rights within collaborations, which will involve legally-binding agreements.
This represents somewhat of a  ``different culture" for many in the HEP community, and perhaps this may be a conceptual hurdle that has to be overcome as it does collide with the basic concepts of open science. It has to be the case that all developments of importance to HEP research from engagements with technology companies are future-proofed. That is to say that any advance that enables accomplishing one or more objectives of the HEP mission must be able to reside in the community and not be lost behind an IP barrier if a company decides that this line of research is no longer a priority.

\subsubsection{Engagement with other domain sciences, engineering, and computing}
The types of problems that require quantum simulation in HEP are similar in form to those in the other domain sciences. For example, non-equilibrium dynamics in quantum many-body systems is of interest in QIS, condensed matter, nuclear physics, fluid dynamics, fusion, biology, as well as HEP.  
It is also the case that the techniques that are being developed for the simulation of QFTs may also be of relevance for quantum sensing and communications. 
Workshops, meetings and collaborations, as are being enabled by the NQI centers, are crucial to optimal exchange of ideas and translations between languages and notations of these different communities. In order for the communities to not have to reinvent the wheel, it is important that areas of overlap are identified and developments are being transported across seemingly-disjoint disciplines. For example, lessons can be learned from how quantum information impacted and advanced studies of condensed-matter systems over the past decade, since it is likely that the HEP research can be similarly invigorated by the QIS concepts and tools. Finally, HEP will likely be instrumental in motivating and advancing QIS research, thus an integrated HEP-QIS effort will be mutually beneficial in quantum simulation.

\clearpage
\vspace{0.5 cm}

\section{Community endorsement}
\label{suppl:endorsement}
\noindent
\begin{enumerate}
\item{M. Sohaib Alam
\\
\emph{Quantum Artificial Intelligence Laboratory (QuAIL), NASA Ames Research Center, Moffett Field, CA, 94035, USA}

\emph{USRA Research Institute for Advanced Computer Science (RIACS), Mountain View, CA, 94043, USA}}
\item{Mohsen Bagherimehra
\\
\emph{Chemical Physics Theory Group, Department of Chemistry, University of Toronto, Ontario, Canada}

\emph{Institute for Quantum Science and Technology, University of Calgary, Alberta, Canada}
}
\item{Aniruddha Bapat
\\
\emph{Physics Division, Lawrence Berkeley National Laboratory, Berkeley, CA 94720, USA}
}
\item{Douglas H. Beck
\\
\emph{Department of Physics and Illinois Quantum Information Science and Technology (IQUIST) Center, University of Illinois, 1110 W. Green Street, Urbana, IL 61801, USA}
}
\item{Paulo F. Bedaque
\\
\emph{Department of Physics and Maryland Center for Fundamental Physics, University of Maryland, College Park, MD 20742, USA}
}
\item{David Berenstein
\\
\emph{Department of Physics, University of California, Santa Barbara, CA 93106, USA}
}
\item{Joseph Carlson
\\
\emph{Theoretical Division, Los Alamos National Lab, NM 87545, USA}
}
\item{Alessio Celi
\\
\emph{Departament de F\'isica, Universitat Aut\`onoma de Barcelona, E-08193 Bellaterra, Spain}
}
\item{Shailesh Chandrasekharan
\\
\emph{Department of Physics, Duke University, Durham NC, USA} 27708-0305
}
\item{Anthony N. Ciavarella
\\
\emph{InQubator for Quantum Simulation (IQuS), Department of Physics, University of Washington, Seattle, WA 98195}
}
\item{Eleanor Crane
\\
\emph{Cambridge Quantum Deutschland GmbH, Leopoldstrasse 180, 80804 Munich, Germany}
}
\item{Marcello Dalmonte
\\
\emph{The Abdus Salam International Centre for Theoretical
Physics (ICTP), strada Costiera 11, 34151 Trieste, Italy}
}
\item{Andrea Delgado
\\
\emph{Oak Ridge National Laboratory, Oak Ridge, TN, USA}
}
\item{Xi Dong
\\
\emph{Department of Physics, University of California, Santa Barbara, CA 93106, USA}
}
\item{Lena Funcke
\\
\emph{Center for Theoretical Physics, Co-Design Center for
Quantum Advantage, and NSF AI Institute for Artificial Intelligence and
Fundamental Interactions, Massachusetts Institute of Technology, 77
Massachusetts Avenue, Cambridge, MA 02139, USA}
}
\item{Bryce Gadway
\\
\emph{Department of Physics, University of Illinois, 1110 W. Green Street, Urbana, IL 61801, USA}
}
\item{Daniel Gonz\'{a}lez-Cuadra
\\
\emph{Institute for Theoretical Physics, University of Innsbruck, 6020 Innsbruck, Austria}
\\
\emph{Institute for Quantum Optics and Quantum Information of the Austrian Academy of Sciences, 6020 Innsbruck, Austria}
}
\item{Alexey V. Gorshkov
\\
\emph{Joint Quantum Institute and Joint Center for Quantum, Information and Computer Science, NIST/University of Maryland, College Park, Maryland 20742, USA}
}
\item{Erik Gustafson
\\
Fermi National Accelerator Laboratory, Batavia, IL 60510, USA
}
\item{Mohmmad Hafezi
\\
\emph{Joint Quantum Institute, University of Maryland, College Park, MD 20742, USA}
}
\item{Jad C. Halimeh
\\
\emph{Department of Physics and Arnold Sommerfeld Center for Theoretical Physics (ASC), Ludwig-Maximilians-Universit\"at M\"unchen, Theresienstra\ss e 37, D-80333 M\"unchen, Germany}
\\
\emph{Munich Center for Quantum Science and Technology (MCQST), Schellingstra\ss e 4, D-80799 M\"unchen, Germany}
}
\item{Aram Harrow
\\
\emph{Center for Theoretical Physics, Massachusetts Institute of Technology, Cambridge, MA 02139, USA}
}
\item{Philipp Hauke
\\
\emph{INO-CNR BEC Center and Department of Physics, University of Trento, Via Sommarive 14, I-38123 Trento, Italy}
}
\item{Florian Herren
\\
\emph{Fermi National Accelerator Laboratory, Batavia, IL 60510, USA}
}
\item{Ciaran Hughes
\\
\emph{Theoretical Physics Department, CERN, 1211 Geneva 23, Switzerland}
}
\item{Travis S. Humble
\\
\emph{Oak Ridge National Laboratory, Oak Ridge, TN, USA}
}
\item{Joshua Isaacson
\\
\emph{Fermi National Accelerator Laboratory, Batavia, IL 60510, USA}
}
\item{Karl Jansen
\\
\emph{Deutsches Elektronen-Synchrotron DESY, Platanenallee 6, 15738 Zeuthen, Germany}
}
\item{Natalie Klco
\\
\emph{Institute for Quantum Information and Matter and Walter Burke Institute for Theoretical Physics,
California Institute of Technology, Pasadena CA 91125, USA}
}
\item{Michael Kreshchuk
\\
\emph{Physics Division, Lawrence Berkeley National Laboratory, Berkeley, CA 94720, USA}
}
\item{Andreas Kronfeld
\\
\emph{Fermi National Accelerator Laboratory, Batavia, IL 60510, USA}
}
\item{Doga Murat Kurkcuoglu
\\
\emph{Superconducting Quantum Materials and Systems Center (SQMS), Fermi National Accelerator Laboratory, Batavia, IL 60510, USA}
}
\item{Nima Lashkari
\\
\emph{Department of Physics and Astronomy, Purdue University, West Lafayette, IN 47907, USA}
}
\item{Randy Lewis
\\
\emph{Department of Physics and Astronomy, York University, Toronto, ON, M3J 1P3, Canada}
}
\item{Andy C.~Y.~Li
\\
\emph{Fermi National Accelerator Laboratory, Batavia, IL 60510, USA}
}
\item{Norbert M. Linke
\\
\emph{
Joint Quantum Institute and Department of Physics,
University of Maryland, College Park, Maryland 20742, USA}
}
\item{Niklas Mueller
\\
\emph{Department of Physics, Maryland Center for Fundamental Physics, and Joint Quantum INstitute, University of Maryland, College Park, MD 20742, USA}
}
\item{Yasser Omar
\\
\emph{Portuguese Quantum Institute, and Instituto Superior T\'{e}cnico, Universidade de Lisboa, P-1049-001 Lisbon, Portugal}
}
\item{Vincent R. Pascuzzi
\\
\emph{Brookhaven National Laboratory, Upton, NY 11973, USA}
}
\item{Indrakshi Raychowdhury
\\
\emph{Birla Institute of Technology and Science, Pilani, K. K. Birla Goa Campus, Zuarinagar, Goa - 403726, India.}
}
\item{Enrique Rico
\\
\emph{IKERBASQUE, Basque Foundation for Science, Plaza Euskadi 5, 48009 Bilbao, Spain}
\\
\emph{Department of Physical Chemistry, University of the Basque Country UPV/EHU, Apartado 644, 48080 Bilbao, Spain}
\\
\emph{EHU Quantum Center, University of the Basque Country, UPV/EHU, Barrio Sarriena s/n, 48940 Leioa, Biscay, Spain}
}
\item{Ananda Roy
\\
\emph{Department of Physics and Astronomy, Rutgers University, Piscataway, NJ 08854-8019, USA}
}
\item{Alexander Schuckert
\\
\emph{Cambridge Quantum Deutschland GmbH, Leopoldstrasse 180, 80804 Munich, Germany}
}

\item{Hersh Singh
\\
\emph{InQubator for Quantum Simulation (IQuS), Department of Physics, University of Washington, Seattle, Washington 98195-1550, USA}
}
\item{Jesse Stryker
\\
\emph{Department of Physics,= and Maryland Center for Fundamental Physics, University of Maryland, College Park, MD 20742, USA}
}
\item{Federica Maria Surace
\\
\emph{Department of Physics and Institute for Quantum Information and Matter,
California Institute of Technology, Pasadena, California 91125, USA}
}
\item{Shan-Wen Tsai
\\
\emph{Department of Physics and Astronomy, University of California, Riverside, CA 92521, USA}
}
\item{James P. Vary
\\
\emph{Department of Physics and Astronomy, Iowa State University, Ames, IA USA}
}
\item{Wei Xue
\\
\emph{Department of Physics, University of Florida, Gainesville, FL 32611, USA}
}
\item{Torsten V. Zache
\\
\emph{Institute for Theoretical Physics, University of Innsbruck, 6020 Innsbruck, Austria}
\\
\emph{Institute for Quantum Optics and Quantum Information of the Austrian Academy of Sciences, 6020 Innsbruck, Austria}
}
\item{Erez Zohar
\\
\emph{Racah Institute of Physics, The Hebrew University of Jerusalem, Jerusalem 91904, Givat Ram, Israel}
}
\item{Peter Zoller
\\
\emph{Institute for Theoretical Physics, University of Innsbruck, 6020 Innsbruck, Austria}
\\
\emph{Institute for Quantum Optics and Quantum Information of the Austrian Academy of Sciences, 6020 Innsbruck, Austria}
}
\end{enumerate}

\vspace{0.5 cm}

\section{Acknowledgements and funding information}
\label{suppl:acks}
\noindent
We are grateful to the members of the community who endorsed this document as named in Suppl.~\ref{suppl:endorsement}, as well as to Mohsen Bagherimehrab, Aniruddha Bapat, Shailesh Chandrasekharan, Lena Funcke, Jad Halimeh, Aram Harrow, Philipp Hauke, Joshua Isaacson, Karl Jansen, Natalie Klco, Michael Kreshchuk, Andreas Kronfeld, Norbert Linke, Vincent Pascuzzi, Indrakshi Raychowdhury, Enrique Rico Ortega, Ananda Roy, Federica Surace, Wei Xue, Erez Zohar, and Martin Zwierlein for valuable feedback on an earlier draft of this whitepaper.

\vspace{0.2 cm}

Christian Bauer is supported by the U.S. Department of Energy's (DOE's) Office of Science under contract DE-AC02-05CH11231. In particular, support comes from Quantum Information Science Enabled Discovery (QuantISED) for High Energy Physics (KA2401032).

Zohreh Davoudi is supported in part by the U.S. DOE’s Office of Science Early Career Award, under award no. DE-SC0020271, the DOE’s Office of Science, Office of Advanced Scientific Computing Research, Quantum Computing Application Teams program, under fieldwork proposal number ERKJ347, and the Accelerated Research in Quantum Computing program under award DE-SC0020312. She also acknowledges support from National Science Foundation Quantum Leap Challenge Institute for Robust Quantum Simulation under grant OMA-2120757.

A.~Baha~Balantekin is supported in part by the U.S. DOE's Office of Science, Office of High Energy Physics, under award no.~DE-SC0019465. 

Tanmoy Bhattacharya is partly supported by the Los Alamos National Laboratory and the U.S. DOE', Office of Science, Office of High Energy Physics, under Contract with Triad National Security, LLC, Contract Grant no. 89233218CNA000001 to Los Alamos National Laboratory.

Marcela Carena and Ying-Ying Li are supported by the DOE through the Fermilab QuantiSED program in the area of “Intersections of QIS and Theoretical Particle Physics”. Fermilab is operated by Fermi Research Alliance, LLC under contract number DE-AC02-07CH11359 with the United States Department of Energy.

Wibe Albert de Jong was supported by the DOE's Office of Science, Office of Advanced Scientific Computing Research Accelerated Research for Quantum Computing Program under contract no. DE-AC02-05CH11231.

Patrick Draper and Aida El-Khadra acknowledge support from the DOE's Office of Science QuantISED program under an award for the Fermilab Theory Consortium "Intersections of QIS and Theoretical Particle Physics".

The work of Masanori Hanada is partly supported by the Royal Society International Exchanges award IEC/R3/213026. 

The work of Dmitri Kharzeev is supported in part by the U.S. DOE's Office of Science grants no. DE-FG88ER40388 and DE-SC0012704, and Office of Science, National Quantum
Information Science Research Centers, Co-design Center for Quantum Advantage under contract DE-SC0012704.

Junyu Liu is supported in part by International Business Machines (IBM) Quantum through the Chicago Quantum Exchange, and the Pritzker School of Molecular Engineering at the University of Chicago through AFOSR MURI (FA9550-21-1-0209).

Yannick Meurice is supported in part by the U.S. DOE's Office of Science, Office of High Energy Physics QuantISED program, under award no. DE-SC0019139.

Christopher Monroe is supported by the NSF's STAQ program, under award PHY-1818914 and the DOE's Office of Science, Office of High Energy Physics, under award no.~DESC0019380.

Guido Pagano acknowledges support by the DOE's Office of Science, Office of Nuclear Physics, under award no. DE-SC0021143. He is further supported by the NSF CAREER Award (award no. PHY-2144910), the Army Research Office (W911NF21P0003), and the Office of Naval Research (N00014-20-1-2695, N00014-22-1-2282).

The work of Enrico Rinaldi is partly supported by the Royal Society International Exchanges award IEC/R3/213026. He is further supported by Nippon Telegraph and Telephone Corporation (NTT) Research.

Martin Savage is supported in part by the U.S. DOE's
Office of Science, Office of Nuclear Physics, InQubator for Quantum Simulation (IQuS) under award no.~DE-SC0020970.

George Siopsis acknowledges support by the Army Research Office under award W911NF-19-1-0397, the National Science Foundation under award DMS-2012609, and by the Defense Advanced Research Projects Agency (DARPA) Optimization with Noisy Intermediate-Scale Quantum devices (ONISQ) program under award no.~W911NF-20-2-0051.

K\"ubra~Yeter-Aydeniz was supported by MITRE Corporation TechHire Program, approved for public release with case number 21-03848-2.

\section{Statement of Conflict of Interest }
\label{suppl:COI}
\noindent
A number of the authors of this whitepaper have a financial interest in the field of quantum computing and quantum simulation: Nate Gemelke is the Chief Technology Officer of QuEra Computing Inc., Junyu Liu is a scientific advisor for qBraid Corporation, Mikhail Lukin is the co-founder of QuEra Computing Inc., Christopher Monroe is co-founder and chief scientist at IonQ Inc., and K\"ubra~Yeter-Aydeniz is the Lead Quantum Algorithms Specialist at the MITRE Corporation.

\bibliography{bibi}

\end{document}